\documentclass[aps,prx,twocolumn,%superscriptaddress, 
groupedaddress,longbibliography]{revtex4-2}

\usepackage{dcolumn}
\usepackage{url}
\expandafter\let\csname equation*\endcsname\relax
\expandafter\let\csname endequation*\endcsname\relax
\usepackage{amsmath,amssymb}
\usepackage{amsthm}
\usepackage{graphicx,bm,color,hyperref}
\usepackage{mathrsfs}
\usepackage{dsfont}
\usepackage{mathtools}
\usepackage{bbm,bm}
\usepackage{mathdots}
\usepackage{xcolor}
\usepackage{tabularx}
\usepackage{booktabs}
\usepackage{hyperref}
\usepackage{amssymb}
\usepackage{amsfonts}
\usepackage{changes}
\usepackage{subfigure}
\usepackage{xcolor}
\usepackage{float}
\usepackage{cleveref}
\usepackage{enumitem}
\usepackage{braket}
\usepackage{algorithm}
\usepackage{algpseudocode}
\usepackage{longtable}
\usepackage{stmaryrd}
\usepackage{todonotes}

\usepackage[T1]{fontenc}
\usepackage[utf8]{inputenc}
\usepackage{lmodern}

\usepackage{mhchem}

\usepackage{tikz}
\usepackage{adjustbox} % For adjusting the figure size

\usepackage{changes}
\renewcommand{\arraystretch}{1.5} % Adjust this value for more or less spacing

\newtheorem{definition}{Definition}

\crefname{theorem}{Theorem}{Theorems}
\crefname{proposition}{Proposition}{Propositions}
\crefname{definition}{Definition}{Definitions}
\crefname{lemma}{Lemma}{Lemmas}
\crefname{figure}{Figure}{Figures}
\crefname{corollary}{Corollary}{Corollary}
\crefname{conjecture}{Conjecture}{Conjectures}
\crefname{section}{Section}{Sections}
\crefname{appendix}{Appendix}{Appendixes}
\crefname{observation}{Observation}{Observation}
\crefname{remark}{Remark}{Remark}
\crefname{example}{Example}{Examples}
\crefname{equation}{Eq.}{Eqs.}
\crefname{table}{Table}{Tables}
\crefname{theorem}{Theorem}{Theorems}

\newcommand{\GF}{\mathbb{F}_2}

\newcommand{\lrw}{\operatorname{lrw}}

\newcommand{\rank}{\operatorname{rank}}
\newcommand{\cutrk}{\mathrm{cutrk}}
\newcommand{\adj}{\Gamma}

\begin{document}

% \title{Vertex Ordering Optimization for Photonic Graph States with Quantum Emitters}
\title{Generation of Photonic Graph States with minimal number of quantum emitters}

 \author{Konstantinos-Rafail Revis$^{(1,2)}$} \email{ krevis@uni-paderborn.de}
 \author{Nils Tomke Ottink$^{(3)}$}
 \author{Pierre-Emmanuel Emeriau$^{(3)}$}
 \author{Paul Hilaire$^{(4)}$}

\affiliation{
Department of Computer Science, Paderborn University, Warburger Str. 100, 33098, Paderborn, Germany$^{(1)}$\\
Institute for Photonic Quantum Systems (PhoQS), Paderborn University, Warburger Str. 100, 33098 Paderborn, Germany$^{(2)}$\\
Quandela SAS, 7 Rue Léonard de Vinci, 91300 Massy, France$^{(3)}$\\
LTCI, Inria, Télécom Paris, Institut Polytechnique de Paris, Palaiseau, France$^{(4)}$\\
}

\begin{abstract} 
Graph states are a fundamental resource for measurement and fusion-based quantum computing, quantum networks, and sensing. Preparing them in a photonic system deterministically is, in principle, possible, but finding efficient schemes to prepare them was a long-standing problem addressed recently. Additionally, heuristic optimization schemes for reducing the required number of two-qubit gates were developed. However, the problem of reducing the number of emitters by optimizing the emission ordering was not addressed, due to its computational complexity, as it is connected to a well-known NP-hard problem from graph theory, the linear rank width computation. In this work, we focus on developing heuristic polynomial algorithms to reduce the number of emitters required. In total, we propose four distinct algorithms, which demonstrate up to $30\%$ emitter reduction on random graphs. Furthermore, we provide numerical and statistical evidence that the combination of our optimization schemes with the preexisting algorithms for optimizing the two-qubit gates of the preparation protocol can further reduce them by around $20\%$. Finally, we examine the developed algorithms for various useful graph state families, such as graphs useful for measurement-based quantum algorithms, and cluster states and graph codes used for quantum error correction, to determine the performance of each algorithm.

\end{abstract} 

\maketitle

\section{Introduction}
Entanglement is a key resource that distinguishes quantum technologies from their classical counterparts. In quantum computation, communication, and sensing, the ability to generate and distribute multipartite entangled states plays a central role in determining the tasks we can implement efficiently and at scale. Graph states are a class of multipartite entangled states that have a powerful and versatile description \cite{hein_multiparty,hein_graphstates_review,briegel_persistent}. They are gaining increasing attention for realizing measurement-based quantum computation (MBQC) \cite{raussendorf_oneway,mbqc_flow}, fusion-based quantum computation (FBQC) \cite{fbqc}, quantum error correction (QEC) \cite{qec_gen1,qec_gen3,bell_graph_qec, varnava2006loss, bell2023optimizing,  stean1,shor1}, quantum repeaters and quantum networks \cite{azuma_all_photonic,repeaters,repeaters2,repeaters3,repeaters4,repeaters5,repeaters6,repeaters7,repeaters8,repeaters10}, and quantum enhanced sensing \cite{shettell_graph_metrology,sense2}.

The photonic paradigm is the only case where one can have flying qubits, and thus, they are required in quantum networks. However, a particular difficulty in this case is that the photons do not interact in the linear-optical regime and thus we are not equipped with deterministic two-qubit gates for them \cite{browne_rudolph,notwoqubit1}. Even if such approaches are powerful, as they are the basis for FBQC, this non-deterministic nature leads to substantial overheads in multiplexing, feed-forward, ancillary photons, and loss management \cite{fbqc,rhg_3,cashman2025finding}. It is therefore necessary to have deterministic generation schemes that avoid such direct photon-photon entangling gates. A promising route is the emitter-based photonic graph state generation, where matter qubits such as quantum dots \cite{dots1,dots3,russo_graph_generation}, color centers \cite{color1,color2,color3}, and atoms \cite{atom1,atom2} act as quantum emitters that sequentially produce photons while mediating the entanglement between them. Furthermore, for specific graph state families, there are tailored schemes proposed to generate them, such as cluster states \cite{lindner_cluster_strings,gimeno_large_cluster} and repeater graph states \cite{buterakos}. The entangling gates are performed on the emitters, and the resulting quantum correlations are transferred to the photons through emission. This makes the generation process deterministic in principle and shifts the central experimental challenges from the photon-photon interaction to the control of a smaller number of matter qubits. Deterministic cluster states have already been demonstrated experimentally with semiconductor quantum dots \cite{schwartz_cluster, coste2023high, cogan2023deterministic, meng2024deterministic, huet_reconfigurable}, atoms \cite{thomas2022efficient, yang2022sequential}, and even superconducting qubits \cite{besse2020realizing}.

Finding efficient and deterministic generation schemes for photonic graph states was realized through trial and error without warranties of optimality until Ref. \cite{li_algo} provided an algorithm for the deterministic generation of arbitrary photonic graph states with the minimum number of emitters for a fixed photon emission ordering.

Ref. \cite{li_algo} linked the minimum number of required emitters to a graph property of the target photonic state, its cut-rank, and found a constructive algorithm reaching this minimum. 

It reveals that this physical problem has a deep connection with the well-known graph theory problem of linear rank-width \cite{oum_linear_rankwidth,oum_rankwidth,Oum_rank_width_rev}. Additionally, this demonstrates a crucial limitation, since the underlying assumption is that the ordering, so in our case, the emission sequence is fixed and predefined. In this case, the cut-rank and thus the number of emitters can be computed efficiently. However, finding the optimal ordering, so the emission sequence, is NP-hard \cite{rank_width_nphard} since it requires solving the linear rank-width problem \cite{li_algo}, which we will define rigorously later. Therefore, the generation algorithm addresses the problem after this decisive ordering fixing is made, and for large states, this choice becomes even more critical. 

Recent works have also examined complementary aspects of the resource optimization problem. For a given fixed emission ordering, and consequently number of emitters, the emission circuit itself is not unique, and thus the required resources can be further reduced. In Ref. \cite{eva_opt}, they introduced optimization algorithms that substantially reduce the two-qubit operations between the emitters, while in Ref. \cite{ghanbari_framework}, they proposed a cost-aware graphical framework characterizing elementary graph operations that reduce two-qubit gate count and circuit depth. Related efforts include loss-aware preparation strategies and software tools for photonic-state generation \cite{Ghanbari_opt,kaur_loss_aware,lin_graphiq}.

However, by assuming a fixed, predefined emission order, these works leave open the question of how much reordering can systematically reduce the number of emitters relative to a random choice. Here, we treat the emission ordering as a missing preprocessing step in the deterministic photonic graph state generation with quantum emitters. Since the exact optimization is NP-hard, we aim to develop efficient heuristics that find emission orderings that require a low number of emitters. A recurring obstacle we face is that the vertex ordering optimization is hard because of the local minima one can be trapped in. Our algorithms are therefore designed such that they can escape them using various computational and graph-theoretic concepts. Some of the key techniques we are using include spectral ordering via the Fiedler vector \cite{fiedler_algebraic_connectivity}, reverse Cuthill-McKee ordering \cite{cuthill_mckee}, low emitter motivated graph decompositions, and simulated annealing \cite{kirkpatrick_sa}. Local complementation is a cornerstone tool since graphs related by a sequence of them correspond to locally Clifford (LC) equivalent states \cite{van_den_nest_lc,hein_graphstates_review}. We introduce four algorithms, each targeting a different aspect of the problem: one replaces the cut-rank function with a more tractable graph-theoretic proxy, another is deterministic with a fixed number of steps, and another one exploits the fact that only caterpillar graphs and those LC equivalent to them have linear rank-width of one \cite{cater_are_1}.

We then benchmark our algorithms on Erd\H{o}s-R\'enyi random graphs, achieving up to $30\%$ reduction in emitter count compared with a random emission ordering. We further examine two metrics relevant to practical implementation, the total number of two-qubit gates and the total gate count, and identify which algorithms perform best for each. We also show that using our algorithms improves existing two-qubit gate reduction heuristics such as the one from \cite{eva_opt}, yielding at least a $20\%$ further improvement. Finally, we apply our algorithms to graph state families relevant to quantum technologies. Raussendorf-Harrington-Goyal (RHG) lattices \cite{rhg_1,rhg_2,rhg_3} and graph states associated with quantum error correcting codes, \cite{shor_and_gen_graph_rep,gen_graph_rep2,foliation1,foliation2,hgp_code1,hgp_code2,bb_code} examining instances of states with over $400$ photonic qubits.

This work aims not only to address a vital missing point in the endeavor of the photonic graph state generation with emitters, which is the reduction of the latter, but also to indicate that treating this problem as an extra step can also reduce the rest of the resources required for the practical implementation of photonic graph states. The rest of the paper is organized as follows: in Section \ref{sec:prems}, we have the introduction of preliminary notions from graph theory, graph states, and photonic graph states. In Section \ref{sec:algos}, we present in detail the four algorithms we developed. In Section \ref{sec:exp-rand} we present the result of numerical experiments to test the performance of the best version of our algorithms on random graphs. We establish that imposing them as a preprocessing step before heuristic algorithms for the reduction of two-qubit gates can further improve the performance of the latter, and we also examine the effect of each step of our pipeline. In Section \ref{sec:applications}, we test our algorithms on various states relevant for quantum technologies, and we get more insight about their performance, and in Section \ref{sec:conclusion}, we conclude. The code for the implementation of our algorithms, as well as for the data analysis presented in this work, can be found in \cite{github}.

\section{Preliminaries}\label{sec:prems}
In this section, we cover some important concepts for our work. We present the relevant notions from graph theory, then we introduce some basic properties of graph states, and we discuss the photonic graph state generation with quantum emitters. Beyond this, the main objective of this section is to highlight the link between the minimum number of emitters required to generate a photonic graph state and graph theory concepts related to linear vertex ordering and cut-rank of a graph, such as the linear rank-width.

\subsection{Graph theory}

An undirected simple graph $G = (V,E)$ is defined on a finite set of vertices $V = \{0,\dots,n-1\}$ and edges between them $E\subseteq \{(u,v)\in V\times V| u\neq v\}$ \cite{graph_theory}. The \emph{neighborhood} of a vertex $i\in V$ is defined as
\begin{equation}
    \label{eq:prems-neigh}
    N_i = \{j\in V|(i,j)\in E\}.
\end{equation}
The \emph{adjacency matrix} of $G$ (with respect to a fixed ordering of $V$) is the $n \times n$ matrix $\adj = (a_{ij})$ over $\GF$ where $a_{ij} = 1$ if $(i,j) \in E$ and otherwise it is zero.

A key family of graphs that is relevant for us are the caterpillar graphs, and below we provide a definition.
\begin{definition}[Caterpillar graphs \cite{caters_def}]
    A tree $T= (V,E)$ is a \emph{caterpillar}, if the removal of all leaves (vertices of degree one) yields a path, possibly empty or consisting of a single vertex. Equivalently, $T$ is a caterpillar if it contains a path $P = (v_1,\cdots,v_{k})\subseteq V$, referred to as \emph{spine}, such that every vertex $v\in V$ satisfies either $v\in P$ or $v\in N(v_i)$ for a unique $v_i\in P$.
\end{definition}
In simple words, a caterpillar graph is a tree in which all vertices are within distance $1$ of the central path.

A \emph{(linear) vertex ordering} of $G$ is an assignment of each position in a sequence of length $n$ to a vertex, so a labeling of the positions $\{0,\dots,n-1\}$ by vertices of $G$. Formally, this is a bijection.
\begin{equation}
\label{eq:prems:pidef}
  \pi :\{0, \ldots, n-1\} \to V,  
\end{equation}

which we represent as a list $[\pi(0),\cdots, \pi(n-1)]$, where $\pi(i)$ denotes the vertex occupying position $i$. In the context of photonic graph state generation, a vertex ordering corresponds directly to the sequence in which the associated photonic qubits are emitted. For each \emph{cut index} $k \in \{0, \ldots, n-1\}$, the ordering $\pi$ induces a bipartition of $V = V_\pi^{(k)} \sqcup \bar V_\pi^{(k)}$ with
\begin{align}
  V_\pi^{(k)} &= \{\pi(0), \ldots, \pi(k)\}\\
  \bar V_\pi^{(k)} &
  % = \{\pi(k+1), \ldots, \pi(n-1)\}
  = V \backslash V_\pi^{(k)}.  
\end{align}

Let's denote by $\adj[A, B]$ the $|A|\times |B|$ submatrix obtained whose rows are indexed by $A$ and columns by $B$ and let the \emph{cut matrix} at position $k$ be:
\begin{equation}
  \adj[V_\pi^{(k)}, \bar V_\pi^{(k)}] \in \GF^{(k+1) \times (n-k-1)}.    
\end{equation}

It encodes all edges that cross the cut $( V_\pi^{(k)}, \bar  V_\pi^{(k)})$.

We proceed with the definition of the cut-rank of a graph.
\begin{definition}[Cut-Rank \cite{Oum_rank_width_rev}]
Let $G = (V, E)$ be a graph and $S \subseteq V$ a vertex subset with complement $\bar{S} = V \setminus S$. The cut-rank of $S$ is:
\begin{equation}
  \cutrk_G(S) = \rank_{\GF}\bigl( \adj[S, \bar{S}] \bigr).
\end{equation}
% where $\adj[S, \bar{S}]$ is the $|S| \times |\bar{S}|$ submatrix of the adjacency matrix containing rows indexed by $S$ and columns by $\bar{S}$.
\end{definition}
The cut-rank function is a symmetric, submodular function of the partition and ranges from 0, when there are no connections between $S$ and $\bar{S}$, to $\min(|S|,|\bar{S}|)$. By convention, if $S=V$ and thus $\bar S=\emptyset$, we consider that $\cutrk_G(S) = 0$.

\begin{figure*}
    \centering
    \includegraphics[width=1.0\textwidth]{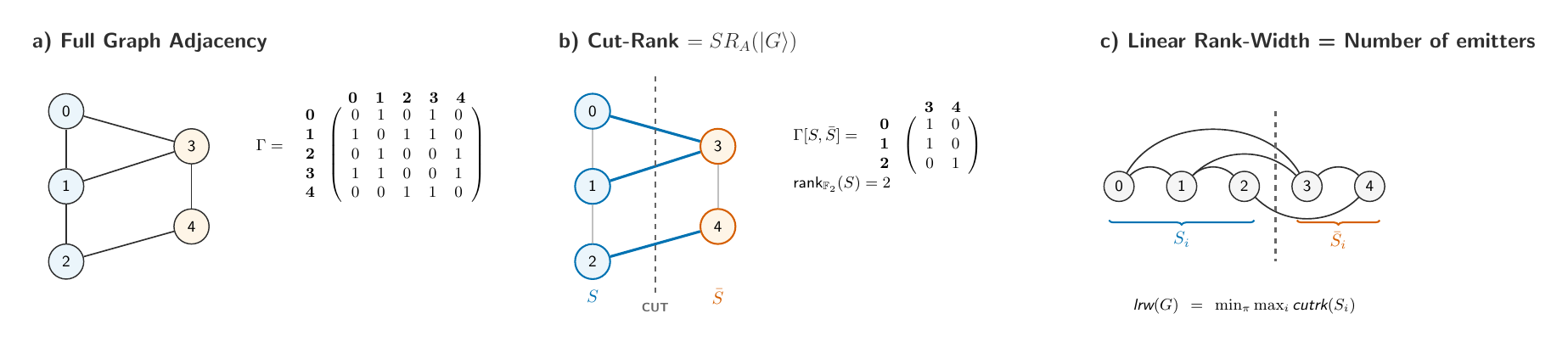}
    \caption{(a) An example of a graph with 5 vertices and its corresponding adjacency matrix. (b) An example of the cutrank of a graph with the corresponding adjacency submatrix. For a specified bipartition, the cut-rank of a graph is equal to the Schmidt Rank, which is an entanglement monotone. (c) The linear rank-width of the graph. In the figure, one of the cuts of the graph is presented. For a given ordering, we must calculate the adjacency submatrix for every cut ranging from $0$ until $n-1$ and then minimize over all possible permutations. The labeling that satisfies the linear rank-width of a graph yields the emission ordering that requires the minimum number of emitters for the quantum dot-based generation of the desired graph state.}
    \label{fig:prems-graph-theory}
\end{figure*}

We now introduce the linear rank-width, which is a central quantity in this work:
\begin{definition}[Linear rank-width and optimal linear layout]\label{def:lrw}
  The \emph{linear rank-width} of $G$, denoted $\lrw(G)$, is
  \begin{equation}
    \lrw(G) =\ \min_{\pi} \max_{0 \le k \le n-1} \cutrk_G\!\bigl(V_\pi^{(k)}\bigr),  
    % \lrw(G) =\ \min_{\pi} \max_{0 \le k \le n-1} \cutrk_G\!\bigl(L_k, R_k\bigr),  
  \end{equation}
where the minimum is taken over all linear vertex orderings $\pi$ of $G$.
We call an \emph{optimal linear layout}, denoted by $\pi_{\text{opt}}$, an ordering which realizes the linear rank-width, i.e., an ordering such that:
$$
\max_{0 \le k \le n-1} \cutrk_G\!\bigl({V_{\pi_{\text{opt}}}^{(k)}}\bigr) = \lrw(G).
$$
\end{definition}
In other words, $\lrw(G)$ is the minimum, over all orderings of the vertices, of the maximum $\GF$-rank of the cut matrices induced by that ordering. 

For the path $P_n$ on $n$ vertices with ordering from left to right, every cut matrix has exactly one row with a single non-zero entry, so each cut has rank one. Hence, $\lrw(P_n) = 1$.
For the complete graph $K_n$, any ordering $\pi$ places some $k$ vertices on the left and $n-k$ on the right. The cut matrix is the all-ones $k \times (n-k)$ matrix, which has $\GF$-rank one. More generally, caterpillar graphs and the ones we can transform to caterpillars \cite{cater_are_1}, have $\lrw$ of one. For instance, path graphs and complete graphs are, respectively, caterpillar and caterpillar-equivalent graphs.

In general, the computational complexity of calculating the $\lrw$ for an arbitrary graph is NP-hard \cite{rank_width_nphard}. There are fixed-parameter-tractable algorithms (FPT) for graphs with bounded linear rank-width and thus the problem can be solved in $f(k)n^{\mathcal{O}(1)}$ where $k$ is the target width, but the dependence on $k$ is typically $2^{2^{\mathcal{O}(k)}}$, making exact FPT algorithms impractical for moderate $k$ \cite{fpt_rank_width}.

\subsection{Graph states}\label{subsec:graph_states}
A link between graph theory and quantum mechanics can be made by considering graph states, which are quantum states characterized by graphs. The following definition focuses on graph states on qubits, but qudit and continuous-variable graph states can also be defined using weighted graphs.

\begin{definition}[Graph state]
A graph state \cite{hein_multiparty,hein_graphstates_review,briegel_persistent} is a class of pure multipartite entangled states defined via a simple graph by assuming that every vertex of the graph corresponds to a qubit prepared in the $\ket{+} = (\ket 0 + \ket 1) / \sqrt 2$ state. Then the edges correspond to the entangling controlled-Z (CZ) operation:
\begin{equation}
    \label{eq:prem-gs-def}
    \ket{G} = \prod_{(u,v)\in E}CZ_{uv}\ket{+}^{\otimes n }.
\end{equation}    
\end{definition}

Graph states are a subclass of stabilizer states. Moreover, any stabilizer state is equivalent to a graph state through local Clifford gates  \cite{van_den_nest_lc}. They can be uniquely defined as the $+1$ common eigenstates of $n$-qubit Pauli stabilizer groups. 

The \emph{stabilizer generators} of a given graph state with $n$ qubits are defined as:
\begin{equation}
    \label{eq:prems-stab-gen}
    g_i = X_i\prod_{k\in N_i} Z_k \quad \forall i \in V,
\end{equation}
with 
$$\forall i \in V, \quad g_i \ket G = + \ket G.$$

% The adjacency matrix  is a useful mathematical quantity to characterize graph states, since the $n$ generators of the state uniquely define the state, and all necessary information about its entanglement properties can be extracted from it. 

\paragraph{Entanglement in graph states.}
A broadly examined entanglement measure for multipartite states is the Schmidt measure \cite{schmidt_measure}, but it is challenging to calculate for general pure states. However, considering the bipartition $(S,\bar S)$, the Schmidt measure in this case equals the Schmidt rank, which for a multipartite qubit system $\mathcal{H}_S\otimes\mathcal{H}_{\bar S}$ is an entanglement monotone:
\begin{equation}
    \label{eq:prems-sschmidt-rank-def}
    SR_S(\ket \psi) = \log_2\rank (\text{tr}_S(\ket{\psi}\bra{\psi}))
\end{equation}
Equation \eqref{eq:prems-sschmidt-rank-def} corresponds to the $\alpha \to 0$ limit of the R\'enyi entropy of the reduced state $\text{tr}_S(\ket{\psi}\bra{\psi})$. Since the entanglement spectrum of a graph state is flat, the Schmidt rank in fact coincides with the R\'enyi entropy of any order $\alpha$ in this case.

As per Refs. \cite{hein_graphstates_review,hein_multiparty}, for the special case of graph states, the Schmidt rank is equal to the cut-rank of the corresponding adjacency matrix: 

$$SR_S(\ket G) = \cutrk_G(S).$$

\paragraph{Local complementation.}

\begin{figure}
    \centering
    \includegraphics[width=1.0\linewidth]{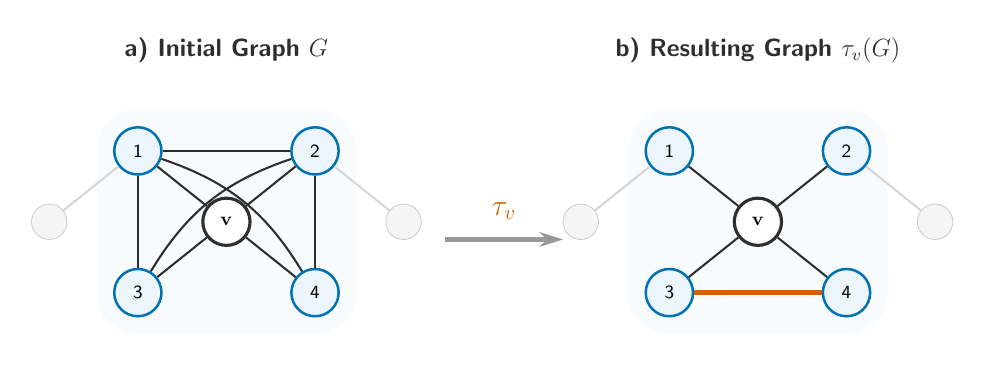}
    \caption{An example of local complementation applied to the vertex $v$ with the neighborhood $N_v = \{1,2,3,4\}$.}
    \label{fig:prems-lc-ex}
\end{figure}

Local complementation \cite{BOUCHET199375} is a concept from graph theory that has a central role in graph states entanglement theory. 
\begin{definition}[Local Complementation]
Let $G = (V, E)$ be a simple undirected graph. The local complementation at a vertex $v \in V$, denoted by $\tau_{v}(G)$, is the operation that transforms $G$ into a new graph $G' = (V, E')$ by replacing the subgraph induced by the neighborhood of $v$ with its complement. Formally, the edge set $E'$ of the transformed graph is given by:

\begin{equation}
    E' = (E \cup E_v) \backslash (E \cap E_v)
\end{equation}
with $E_v = \{(u,w) \in E | u,w \in N_v, u\neq w\}$, all possible edges in the neighborhood of $v$.
\end{definition}
An example of this operation on a graph is given in Figure \ref{fig:prems-lc-ex}: for the local complementation at a vertex $v$, existing edges between the neighbors of $v$ are removed, and if there are no edges between two neighbors of $v$, the local complementation creates it. The local complementation is therefore a graph operation that can modify the total number of edges of a graph. As per Ref. \cite{van_den_nest_lc}, two graph states $\ket{G}$ and $\ket{G'}$ are Locally Clifford (LC) equivalent if and only if a sequence of local complementations exists $\tau_{v_i}(G)\circ\cdots\circ\tau_{v_j}(G)$ that transforms $G$ to $G'$. In terms of quantum gates, we have \cite{hein_graphstates_review,hein_multiparty}

\begin{equation}
    \label{eq:prems-lc-qdef}
    \ket{\tau_{v}(G)} \propto  \sqrt{-iX_v}\prod_{j\in N_v}\sqrt{iZ_j} \ket G.
\end{equation}
Local complementation is essential for finding LC equivalent states with fewer edges. Finding the graph with the minimal number of edges using only local complementation, also called the minimum edge representative (MER) \cite{lc_minimization}, is an important pre-processing step in our work. Finding the exact MER requires finding the total orbit of a graph, which is unrealistic for large graphs. Nevertheless, it can be done for smaller graph instances and Refs. \cite{qubit_orbits,Adcock_2020} and \cite{qutrit_orbits} did it for qubit graph states with fewer than $12$ qubits and qutrit graph states with fewer than $7$ qutrits, respectively.

\begin{figure}
    \centering
    \includegraphics[width=0.8\linewidth]{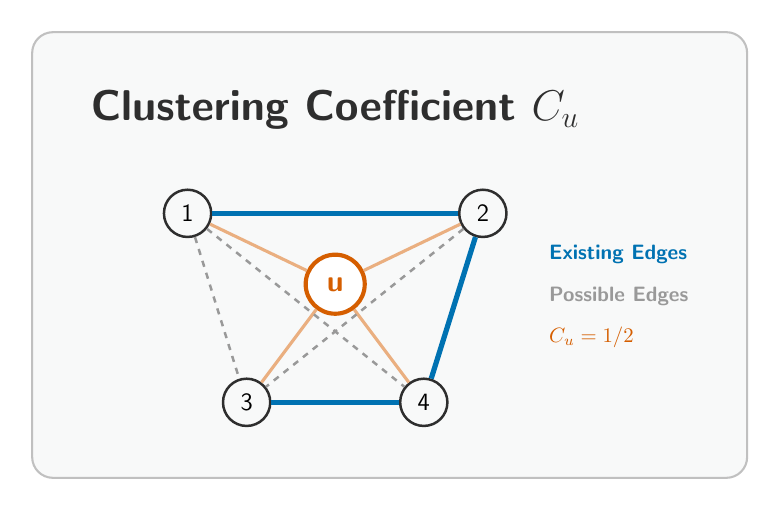}
    \caption{A schematic presentation of the calculation of the clustering coefficient for a graph at the vertex $u$.}
    \label{fig:prems-cc}
\end{figure}

In our case, we employ a simple but rather effective heuristic. For this, we need to introduce the \emph{clustering coefficient}  \cite{Ghanbari_opt}, which quantifies how close the neighborhood of a vertex is to a complete graph. More concretely, we define it as
\begin{equation}
    \label{eq-prems-clust-coef}
    C_u = \frac{2|\{(i,j):j,i\in N_u, (i,j)\in E\}|}{|N_u|(|N_u|-1)}.
\end{equation}
A representative example is given in \ref{fig:prems-cc}. The clustering coefficient is an interesting metric for edge reduction through local complementation: indeed, the local complementation at a vertex $u$ reduces the total number of edges if and only if $C_u>0.5$.
We use it in our edge reduction strategy, which is simply the sequential application to any vertex with $C_v>1/2$ until $C_v\leq1/2$ $\forall v\in V$. This strategy does ensure that a local minimum is obtained, but does not ensure that we reach the global minimum. However, it can be computed efficiently, and usually provides results close to, if not the same as, the actual minimization algorithms, and thus it is preferred.

\subsection{Photonic graph state generation}
\begin{figure*}
    \centering
    \includegraphics[width=0.8\linewidth]{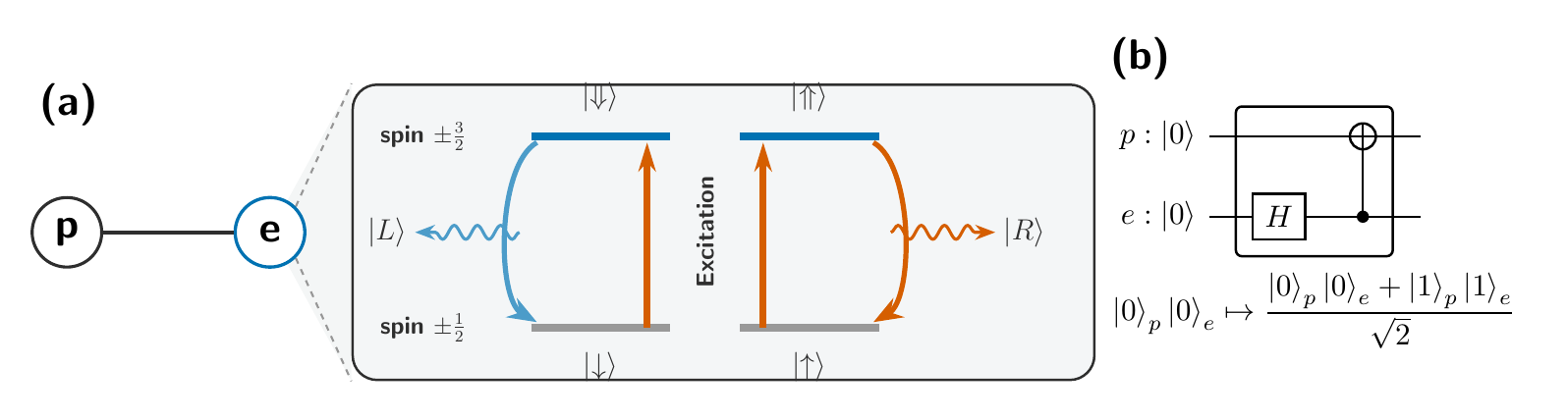}
    \caption{Schematic presentation of photonic graph state generation (a). As \cite{lindner_cluster_strings} proposed an energy level structure, with doubly degenerate ground and excited states with spin $\pm 1/2$ and $\pm 3/2$. Due to the selection rules, polarization of the emitted photon (p), which is either left (L) or right (R), becomes entangled with the spin of the emitter. The process can be repeated, and along with the local gates, any caterpillar or LC equivalent to them in any state can be created using only one emitter. (b) The emission of a single photon is represented as a quantum circuit. Initially the state we have is $\ket{\uparrow}_e+ \ket{\downarrow}_e$. After the excitation we have $\ket{\Uparrow}_e+ \ket{\Downarrow}_e$. Then the emission follows that yields, $\ket{\uparrow}_e\ket{R}_p+ \ket{\downarrow}_e\ket{L}_p = \ket{0}_e\ket{0}_p+ \ket{1}_e\ket{1}_p $.}
    \label{fig:emission}
\end{figure*}

One of the biggest challenges for generating graph states in a photonic setup is that deterministic linear-optical entangling gates do not exist \cite{browne_rudolph,notwoqubit1}.
Therefore, while Eq. \eqref{eq:prem-gs-def} gives a direct construction procedure of the graph, we cannot exploit it deterministically with linear optics.
% Therefore, a direct application of equation \eqref{eq:prem-gs-def} is not possible deterministically with linear optics.
To overcome this obstacle, the use of \emph{quantum emitters} with internal qubits as entanglement interfaces has been proposed, see Figure \ref{fig:emission}(a) and Ref. \cite{lindner_cluster_strings} for a detailed schematic explanation of the process. Given an adequate energy level configuration, these emitters act as a central hub where emitted photons preserve a coherent bond with the emitter's internal state. Utilizing one emitter along with single-qubit operations allows the generation of a graph family known as \emph{caterpillar graphs} as well as any graph that can be locally complemented to them.

The generation of the target state is done sequentially, requiring specific operations to be performed before each photon is released. By manipulating the emitter qubits, we can build entanglement between the photons, using the emitters as ancillas. We can model this as a quantum circuit, where each emission event is a CNOT gate, with the emitter being the control and the photon being the target, initialized in the $\ket 0$ state, as depicted in Figure \ref{fig:emission}(b). This circuit-based approach provides a way to quantify the resources required for the generation of the desired states and to benchmark endeavors of optimizing any of the relevant aspects of the problem. 

\paragraph{Algorithms for deterministic photonic graph state generation.}
Ref. \cite{li_algo} presents an algorithm that determines the minimum number of emitters and the emission circuit to produce the desired photonic graph state, given an imposed photon emission order. The operations on the photons are restricted to the local Clifford group, while the gates on the emitters can also be non-local Clifford gates. Given a graph $G$ and a fixed vertex ordering $\pi$, they showed that the minimum number of quantum emitters to generate the photonic graph state sequentially is equal to the maximum of a function, denoted the \emph{height function}. They defined the height function based on the stabilizer tableau, but we provide here a graph-theoretic definition which is equivalent when restricted to graph states.

\begin{definition}[Height function of a graph state]
    Given a vertex ordering $\pi$, the height function of a graph state $\ket G$ is the function $h_{\pi,G}: \{0, \dots, n-1\} \mapsto\mathbb N$
    \begin{equation}
    h_{\pi, G}(x) := \cutrk_G(V_\pi^{(x)})
    \end{equation}
\end{definition}

The height function is linked to the bipartite entanglement entropy, $S_{V_\pi^{(x)}}$, between the two subsets of qubits $V_\pi^{(x)}$ and its complement $\bar V_\pi^{(x)}$.
% The \emph{height function} $h \colon \{0, 1, \ldots, n\} \to \N$ is defined given an $\mathcal{T}(\pi)$ at step $k$. More specifically, we can define the height function $h(x) = S_A$ to be the bipartite entanglement entropy when a linear graph is arranged in a linear setup and a cut is assumed, dividing it into a subregion $A = \{0,\cdots, x\}$ and its complement.
% % Note that $S_A$ can be any of the R\'enyi entropies. 
% Note that $S_{V_\pi^{(x)}}$ can be any of the R\'enyi entropies. 

Ref. \cite{li_algo} proves that the minimum number of emitters for a given emission ordering $\pi$ can be determined as
\begin{equation}
    n_e(\pi, G) = \max_{0\leq x\leq n-1} h_{\pi, G}(x).
\end{equation}

Therefore, the number of quantum emitters needed to generate a photonic graph state $\ket G$ in a given order is directly connected to a graph theoretic quantity, the cut-rank of the adjacency matrix of the $G$.
% Given an arbitrary ordering $\pi$, the maximum of the height function and the maximum of the cut-rank are:
% \begin{equation}
% \label{eq:prems-h-less-cut}
%     \max_k h(k) \ge \max_k \cutrk_G(L_k, R_k).
% \end{equation}

To synthesize the desired circuit, the algorithm in Ref.  \cite{li_algo} reduces the tableau to a disentangled basis, which is a separable state. The process deconstructs the state via the sequential matter qubit
coupling, resulting in a \emph{time-reversed emission model}. While the algorithm proceeds backward from the final photon $n_p$ to the first $1$, the physical generation circuit is obtained by inverting this sequence, providing the required steps to build the graph state. In this framework, they introduce the time-reversed measurement and absorption. The time-reversed emission is the absorption of a photon by the emitter. This is possible when the stabilizer generator is of the form $g = Z_{p_j}Z_{e_i}$. The CNOT gate used for the emission in the circuit interpretation can also be directly used for the absorption. Applying $CNOT_{e_ip_j}$ transforms the generator into $g = Z_{p_j}I_{e_i}$, which disentangles and absorbs $p_j$.

In case a stabilizer generator cannot be turned into the required form, the photon cannot be absorbed immediately. Instead, a time-reversed measurement must precede the absorption. While a standard $X$-basis measurement removes an emitter from the graph, the time-reversed version reintroduces an emitter. This is a necessary step when the height function obeys $h(p_j)<h(p_{j-1})$. This reversal is realized by applying a Hadamard gate to an emitter followed by a $CNOT_{e_ip_j}$. 

In a follow-up work \cite{eva_opt}, a series of algorithms called Naive, Heuristic1 (Heu-1), and Heuristic2 (Heu-2) were introduced to reduce the CNOT gates between the emitters themselves required for the emission algorithm to work, achieving up to $66\%$ reduction.

An alternative framework for the generation of photonic graph states was presented in Ref. \cite{ghanbari_framework}. The fundamental primitives remain the same. Namely, the sequential state generation, that joint operations between photons are not permitted, as well as how the minimum number of required emitters for a fixed emission order is found. Their addition is the graphical framework they are introducing, allowing them to examine the resources and possible tradeoff between the different kinds of resources required to generate the state.

However, the chosen labeling corresponds to the emission ordering of each photonic qubit. 
The optimal emission order, i.e.,  the one that minimizes the number of quantum emitters is also the one that minimizes the maximum of the height function, i.e., the cut-rank over all linear cuts. This corresponds to the definition of the optimal linear layout (see Def. \ref{def:lrw}), $\pi_{\text{opt}}$:
$$n_e(\pi_{\text{opt}}, G) = \lrw(G).$$

Therefore, for any arbitrary and especially large graph, there is no efficient algorithm to find this optimal sequence unless NP=P. However, developing heuristic algorithms for determining a labeling yielding fewer numbers of emitters than a randomly chosen one is not only possible but also essential, since, as we showcase later in this work, it can reduce the resources needed for the state generation.

\section{Algorithms}\label{sec:algos}

In this section, we present the algorithms we developed for obtaining the ordering such that we obtain the minimum number of emitters. As shown previously, the number of emitters is linked to the maximum of the height function of a given vertex ordering. The optimal ordering leads to a minimum number of emitters, which is equal to the linear rank-width of the target graph state. Therefore, the goal of the four algorithms that we present here is to find an ordering for which the number of quantum emitters approaches the linear rank-width of the graph.

The research space contains $n!$ possible orderings, and the objective is controlled by the worst cut rather than by an average property of the ordering. However, small permutations between wisely chosen vertices of the graph can lead to improvements in the final emitter count and thus can lead to a substantial reduction of the required resources to prepare the state.

\subsection{Simulation pipeline}

\begin{figure*}
    \centering
    \includegraphics[width=1.0\linewidth]{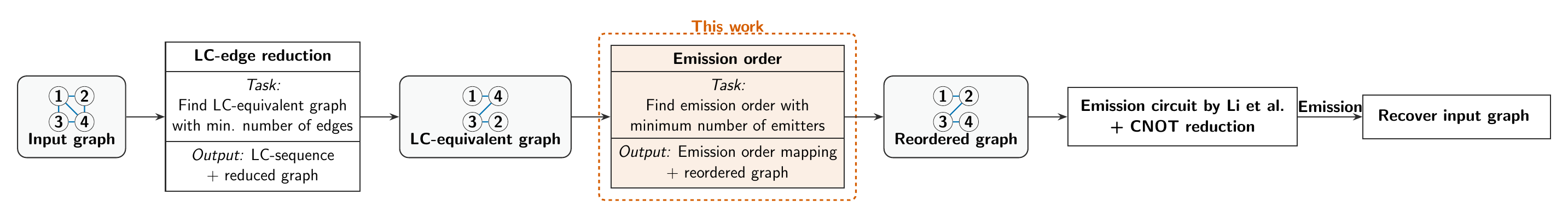}
    \caption{A schematic description of obtaining the emission circuit for an arbitrary graph state. We firstly apply an LC reduction by applying a series of local complementations to every vertex with a clustering coefficient above $1/2$. We then find an emission ordering that reduces the number of emitters ($n_e$) required to generate the circuit. Then, given the reordered graph, we are using the emission algorithm presented by Ref.  \cite{li_algo} and the CNOT reduction algorithms from Ref. \cite{eva_opt}.}
    \label{fig:pipeline}
\end{figure*}

In Figure \ref{fig:pipeline}, we present the complete pipeline required starting from an arbitrary graph with a random labeling and then obtaining the emission circuit and the corresponding requirements. Given an input graph, we first use the greedy edge reduction strategy based on local complementation introduced in the preliminaries (see Subsec. \ref{subsec:graph_states}). After this step, we obtain a local Clifford equivalent graph with fewer edges. This preprocessing step is shown to improve the performance of the full pipeline.  The central part of our work lies in the second step of the pipeline, where we use heuristic algorithms to find a vertex ordering that minimizes the maximum of the height function for this graph. As we discussed previously, this number is connected to the minimum number of emitters required to produce the photonic graph state in Ref. \cite{li_algo}. The final step is to use the algorithms in Refs. \cite{li_algo} or \cite{eva_opt} to find a quantum circuit that produces this photonic graph state.

In the main text, we provide a higher-level description of the algorithms, together with the utility and underlying motivation. The interested reader can find a more detailed presentation of these algorithms, with their pseudocode in Appendix \ref{app:pseudocodes}. There are four heuristics algorithms presented in this work, \texttt{hill\_climbing}, \texttt{height\_function\_sa}, \texttt{path\_clustering} and \texttt{min\_la\_sa}.
All but \texttt{path\_clustering} require an initial ordering. In this section, we present first the algorithm subroutine for initial ordering selection, then a general presentation of a key simulation technique, the simulated annealing, and finally, we proceed with the presentation of the details of our four algorithms.

\subsection{Initial ordering}
% \subsubsection{Phase A and B: Selection of an initial ordering}

A critical decision is the initial ordering of the algorithm. Naively using a random initial ordering can be a drawback. These orderings are not expected to be optimal, but they provide a low cost starting point. For this reason, in Phase A, we resort to proposing three different initializations: the spectral ordering, the reverse Cuthill–McKee ordering, and the minimum-degree ordering, which can all be efficiently computed.

\paragraph{Spectral ordering:} 
The reasoning here is to find a ``direction'' in the graph, along which the vertices close along this direction tend to be densely connected, and thus an ordering that has a reduced height function should order the vertices along that axis. The vector that gives this stretch direction is the Fiedler vector. The calculation of the Fiedler vector requires calculating the \emph{graph Laplacian}, which contains information for both the degree of each vertex as well as the connectivity between the vertices. We provide a formal definition below.

\begin{definition}[Graph Laplacian \cite{graph_theory}]
  The Laplacian of $G$ is $L = D - \adj$, where $D$ is the diagonal matrix with
  $D_{ii} = \deg(i)$.
\end{definition}

The matrix $L$ is symmetric and positive semidefinite by construction, and thus its eigenvalues are real, obeying $0 = \lambda_1 \le \lambda_2 \le \cdots \le \lambda_n$. The eigenvector of the eigenvalue $\lambda_2$ is called \emph{Fiedler vector} \cite{fiedler_algebraic_connectivity}, denoted as $\mathbf{f}$. We obtain the spectral ordering $\pi_{\text{spec}}$ by sorting the vertices according to their corresponding Fiedler vector entries: that is, $\pi_\text{spec}(i)$ gives the vertex whose Fiedler vector entry has the $i$-th smallest value,

\begin{equation}
    \label{eq:pi-spec}
    \pi_\text{spec} = \text{argsort}(\mathbf{f}).
\end{equation}

% Intuitively, this vector illuminates the stretch direction of the graph since vertices close together in $\pi_\text{spec}$ tend to be densely connected.

\paragraph{Reverse Cuthill-McKee} 
We proceed with the \emph{Reverse Cuthill-McKee} (RCM) algorithm \cite{cuthill_mckee}, originally designed to find an ordering $\pi$ that minimizes the bandwidth of sparse matrices, defined as $\max_{(i,j) \in E} |\pi^{-1}(i) - \pi^{-1}(j)|$, where $\pi^{-1}(i)$ is the position of vertex $i$. This algorithm, therefore, produces an ordering in which adjacent vertices remain close to one another. While this is not equivalent to minimizing the linear rank-width, it serves as a useful graph-theoretic proxy, as it can be computed efficiently. The direct Cuthill-McKee algorithm permutes a sparse adjacency matrix as input, which is why we perform the edge reduction step via local complementation beforehand.

The algorithm proceeds by a breadth-first search (BFS) starting from a peripheral vertex, defined as a vertex of maximal eccentricity, where the eccentricity of a vertex $v$ is the greatest distance from $v$ to any other vertex in the graph. Peripheral vertices, therefore, lie at the boundary of the graph rather than at its center, and using one as the starting point ensures that BFS layers expand outward such that the bandwidth is minimized. Within the BFS, neighbors are visited in increasing order of degree. The resulting BFS ordering is then reversed, while the reversed and unreversed orderings are equivalent in terms of bandwidth, the first is the established convention in the literature, as it was shown by \cite{george1971computer} to reduce fill-in when the ordering is then used in sparse matrix factorization.

\paragraph{Minimum-degree ordering}

Our third initialization approach is a greedy one, where each vertex is processed at a time according to the local degree, which is the number of neighbors of a vertex. We start by assigning labels to a vertex with the minimum number of neighbors, and we continue until no vertices are left. In case there are multiple vertices with the same number of neighbors, we pick the one with the most neighbors in the already placed list. Then the corresponding ordering $\pi_{\text{deg}}$ is obtained. The reasoning behind it is that low-degree vertices have fewer connections when placed early, and thus, we do not lead to a big increase in the height function, which is something we want to avoid since we are interested in the maximum rank across the linear cuts and not the average.

\paragraph{Initial ordering selection}
Then we proceed with Phase B, which is the evaluation of each one of the $\mathcal{C} = \{\pi_\text{spec}, \pi_\text{rcm}, \pi_\text{deg}\}$ we calculate the best ordering
\begin{equation}
    \pi^* = \arg\min_{\pi\in\mathcal{C}} n_e(\pi, G).
\end{equation}

We also introduce the \emph{bottleneck cut index}
\begin{equation}
      % k^* = \arg\max_k \cutrk(L_k, R_k)
      k^* = \arg\max_k \cutrk(V_{\pi^{*}}^{(k)}),
\end{equation}
which identifies the cut index at which the height function, and hence the number of required emitters, attains its maximum. Since certifying the global minimum of $\lrw$ requires an exhaustive search over all $n!$ orderings, which is computationally infeasible, we instead rely on local search. We stress the interest of this intuition: rather than perturbing the ordering uniformly at random, we focus perturbations around the bottleneck cut index $k^*$, since this is where a change is most likely to reduce the maximum of the height function. This targeted approach allows us to escape local minima more efficiently and reduce the number of emitters required.

\subsection{Simulated annealing}\label{subsec:sa}

\emph{Simulated annealing} (SA) \cite{kirkpatrick_sa,van1987simulated} is a widely used probabilistic technique for approximating the global optimum of a given function, and we employ it here as well. Let us provide a brief and general overview of the SA approach within the scope of the problem we are addressing in our work. We start with the initial ordering obtained by Phase B, with cost function $w = \max h_{\pi, G}$. At a temperature $T$, we change the labeling of the graph to generate a random neighbor with ordering $\pi'$ and cost $w'$. We then compute $\Delta = w'-w$. If $\Delta\leq 0$, we accept $\pi'$, since it is at least as good as $\pi$. On the contrary, if $\Delta > 0$ we accept the new ordering $\pi'$ with probability $e^{-\Delta/T}$, also known as \emph{Metropolis criterion} \cite{metropolis}. This allows an exploration phase, for which it is required to allow the acceptance of solutions that worsen the cost function, such that after a while, an even better ordering is discovered. The goal of this phase is to avoid being stuck in local minima. We then reduce the temperature $T$ according to a cooling schedule, and we repeat the process.

We referred to the temperature $T$, which plays a central role in escaping local minima since it affects the probability of acceptance in case $\Delta>0$, as well as the cooling schedule, which is how $T$ is changing. Assuming that we are choosing an initial temperature value $T_0 = T_{\text{start}}$, we use a geometric cooling schedule to gradually reduce the temperature. Therefore, at a step $k$ of the process:

\begin{equation}
    T_{k+1} = \alpha \cdot T_k, \quad T_0 = T_{\text{start}}, \quad 0<\alpha<1
\end{equation}
with $\alpha$ the cooling factor. Therefore, the temperature decreases exponentially as:
\begin{equation}
    T_k = T_{\text{start}} \cdot \alpha^k
\end{equation}
and thus it is called a cooling schedule. We also need to fix a temperature value for which the algorithm terminates, denoted as $T_{\text{min}}$, and thus the total number of temperature levels is given by:
\begin{equation}
    K = \left\lceil \frac{\ln(T_{\min} / T_{\text{start}})}{\ln \alpha}
    \right\rceil.
\end{equation}
At each temperature level, $S$ perturbation trials are performed. The specific form of these perturbations differs across the algorithms we developed and thus will be discussed in detail in the corresponding parts.

One can also account for the possibility of premature convergence, which signals that we are trapped inside a local minimum once again. Therefore, in case rather high-quality solutions are required, we have included an optional periodic reheating mechanism in the provided codebase, whereby the temperature is occasionally increased by a reheat factor, temporarily restoring the system's ability to accept uphill moves before geometric cooling resumes. This concludes the presentation of parts of the pipeline, allowing us to proceed with the presentation of the algorithms we developed.

\subsection{\texttt{hill\_climbing}}
\begin{figure}
    \centering
    \includegraphics[width=0.6\linewidth]{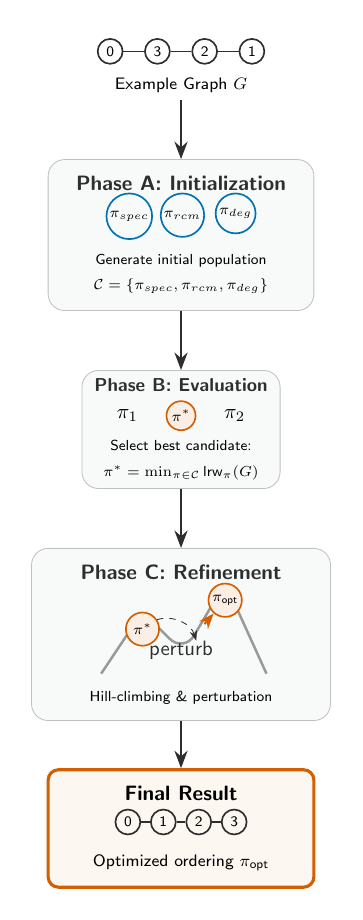}
    \caption{A schematic overview presentation of the steps for the \texttt{hill\_climbing}. In phases A and B, we choose an initial ordering before proceeding with phase C, which is the hill climbing algorithm, where swaps are tried exhaustively in a window around the bottleneck cut and randomly across the ordering, and accepted only if they reduce the width..}
    \label{fig:rank_width_over}
\end{figure}

Let us start by giving a general overview of our algorithm called \texttt{hill\_climbing}. Our approach is summarized in Figure \ref{fig:rank_width_over}. For a detailed presentation of the pseudocode of the algorithm, we refer the reader to Appendix \ref{app-RW}. 

\begin{figure}
    \centering
    \includegraphics[width=0.8\linewidth]{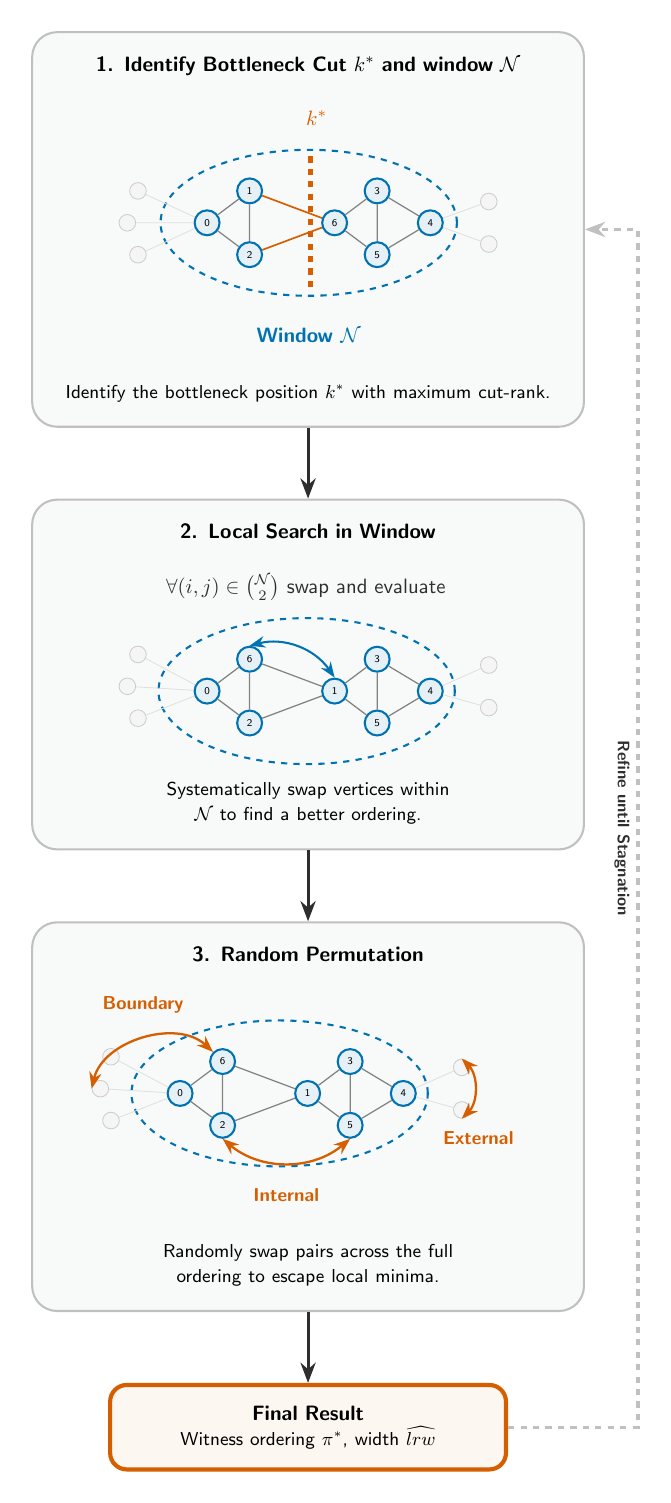}
    \caption{A schematic overview presentation of the Phase C of the \texttt{hill\_climbing}.}
    \label{fig:rank_width_refine}
\end{figure}

% \subsubsection{Phase C: Refinement}
We start by finding an initial ordering based on the previous discussion, then we refine this previous arrangement. We do so by applying a local-search hill-climbing procedure to the best initial ordering candidate to further reduce its width, escaping local optima via random perturbations.
% The final phase is the refinement of the previous arrangement.
In this case, we aim to further reduce the maximum of the height function using a local search. The key observation is that a single swap of two positions $i$ and $j$ in the ordering only affects the cut matrices at indices $\min(i,j)$ through $\max(i,j) - 1$. All other cut matrices remain identical. Therefore, it is natural to focus on perturbing near the bottleneck cut. For this reason, we define the window $\mathcal{N}$ as 
\begin{equation}
    \label{eq:algo-rank-window}
    \mathcal{N} = \{k^* - \beta, \dots, k^* + \beta\} \cap \{0, \dots, n-1\}
\end{equation}
where the offset $\beta$ is predefined at the beginning of the algorithm. The window $\mathcal{N}$ defines the set of vertices that will be exhaustively examined. This is depicted in step 1 of Figure \ref{fig:rank_width_refine}. Then a local search is performed where for every pair $(i,j)\in\binom{\mathcal{N}}{2}$ we swap the vertices, and we evaluate the new maximum of the height function. The neighborhood of a solution $\pi$ consists of all orderings obtainable by swapping one pair of positions.  With $|\mathcal{N}|$ positions, the complete neighborhood has $O(|\mathcal{N}|^2)$ elements. Evaluating each neighbor costs $O(|\mathcal{N}|^2)$ (for $|\mathcal{N}| - 1$ cut-rank computations), so exhaustive neighborhood search has $O(|\mathcal{N}|^4)$ cost per iteration, which is prohibitive for large subset $\mathcal N$. Therefore, the offset $\beta$ must be relatively small. This step is depicted in step 2 of Figure \ref{fig:rank_width_refine}.
However, swapping labels within the window is not sufficient to ensure that we have minimized the width of the ordering, since we may have obtained a local minimum rather than a global one. For this reason, a uniformly random pair of positions $(i, j)$ is selected from the entire ordering, and the swap is accepted only if it reduces the number of emitters. Therefore, there are three possibilities. The first is that we choose two vertices from the window, in which case no improvement is achieved. This possibility is, in reality, rather unlikely for large graphs, since we fix the offset and, in practice, $|\mathcal{N}|\ll n$. The other two possibilities are that either one or both indices do not belong to $\mathcal{N}$, in which case it is possible to improve our results further. We schematically describe this effect in step 3 of Figure \ref{fig:rank_width_refine}. In summary, the key idea of the \texttt{hill\_climbing} algorithm is that we firstly perform an exhaustive search on a small region of the graph, since it is possible to reduce $n_e(\pi, G)$, the value of the height function's maximum for the current ordering. Then we proceed with random permutations over the graph, but an exhaustive search in this case is not possible.

\subsection{\texttt{height\_function\_sa}}

We proceed with our analysis by presenting a new algorithm called \texttt{height\_function\_sa}. As it became apparent with the hill climbing subroutine of \texttt{hill\_climbing} that randomly swapping indices can yield better vertex labeling, as we may escape a local minimum. However, the bottleneck refinement at the last step of \texttt{hill\_climbing} can stagnate when no width reducing move is found. On the contrary, the SA is built to manage to escape such local minima by temporarily accepting moves that worsen the cost function thanks to the Metropolis criterion. Building on this intuition, we also start with an initial ordering candidate selected as previously. The main difference in this case is that we utilize a \emph{simulated annealing} (SA) \cite{kirkpatrick_sa} approach for further refining the initial ordering choice. Such probabilistic algorithms are well established in the literature as yielding strong practical performance, even in the absence of rigorous optimality guarantees \cite{kirkpatrick_sa,van1987simulated}.

The algorithm \texttt{height\_function\_sa} has some key differences compared to the \texttt{hill\_climbing}. The first one concerns how the exploration phase of the two algorithms is terminated. The \texttt{hill\_climbing} algorithm is deterministic in this respect: the number of perturbation iterations is fixed directly and chosen ahead of time as an explicit parameter. The \texttt{height\_function\_sa} algorithm, in contrast, has no such directly specified step count. Instead, termination is governed by the cooling schedule described in Section \ref{subsec:sa}. Once the initial and final temperatures $T_{\text{start}}$ and $T_{\text{min}}$, together with the cooling factor $\alpha$, are fixed, they jointly determine the number of temperature levels $K$, and thus the total number of perturbation trials performed before the algorithm terminates. In other words, while \texttt{hill\_climbing} fixes the exploration budget directly, \texttt{height\_function\_sa} fixes it only indirectly, as a consequence of the chosen temperature bounds and cooling rate.

\begin{figure}
    \centering
    \includegraphics[width=1.0\linewidth]{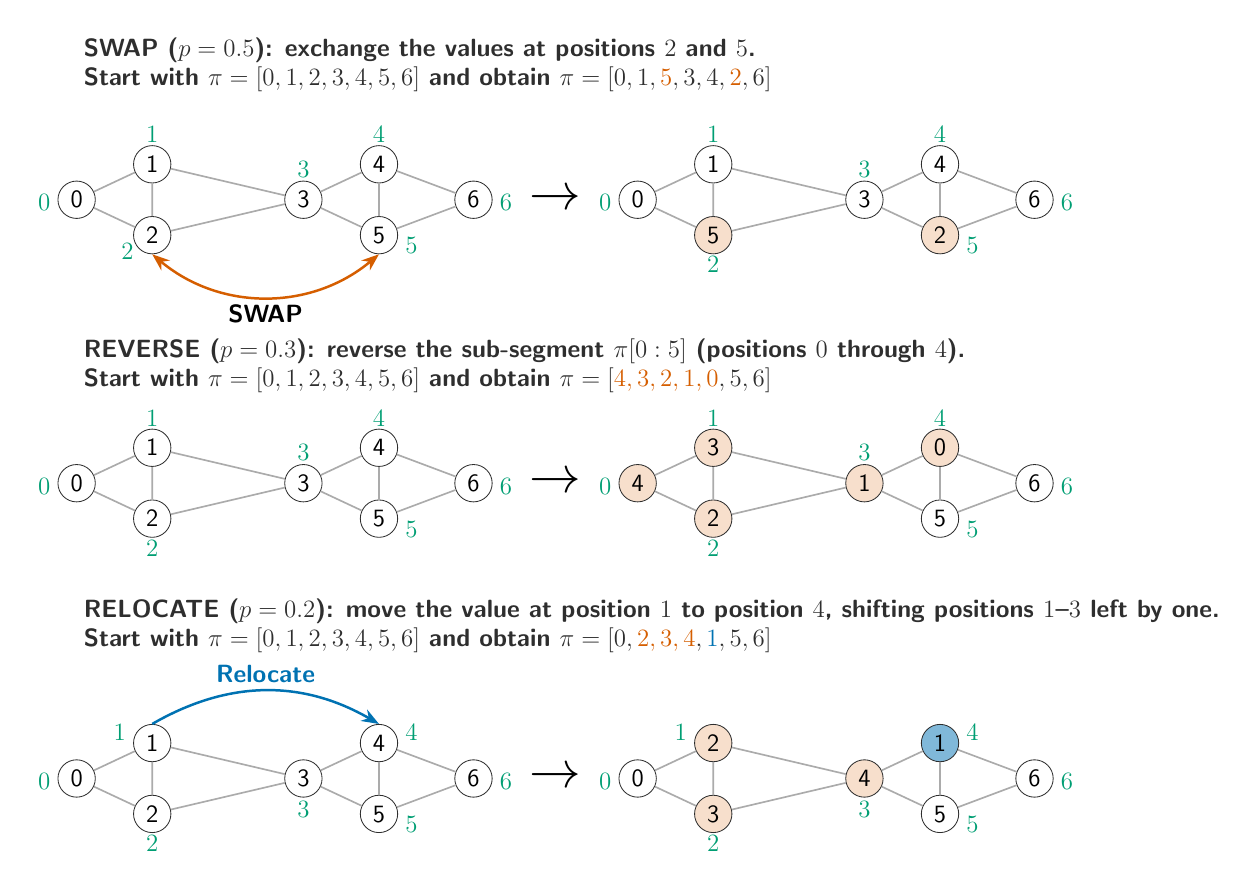}
    \caption{
    An example of the three perturbation moves used for the \texttt{height\_function\_sa}. In the top panel, the \textsc{swap} operation is presented, where we choose two labels uniformly at random, and we swap them. The next operation is the \textsc{reverse} presented in the middle panel, and here we choose to start from the first element, which is zero, and choose the segment of the first 5 elements. In the bottom panel, the \textsc{relocate} operation is presented, where we assume that we randomly chose the label 1 to be moved to the position of the label 4, and at the same time, every label in between is switched one position to the left to fill the gap.}
    \label{fig:ranksa_moves}
\end{figure}

The second key difference is the perturbation moves used. When we discussed the SA technique in Section \ref{subsec:sa}, we briefly mentioned it. However, care must be taken with the type of perturbations one uses. It is always possible to opt for simple swaps between the labels of the vertices, but escaping the local minima and avoiding suboptimal solutions requires having different ways to shuffle the emission ordering. For this reason, we used three distinct perturbation types. The first one is the \textsc{swap}, with probability $p_{\text{SWAP}} = 0.5$, for which we select two positions $i$ and $j$ uniformly at random and we exchange them. An example is presented in the upper panel of Figure \ref{fig:ranksa_moves}. The second one is the \textsc{reverse}, with probability $p_{\text{REV}} = 0.3$, for which we select a random starting position $i$ and a segment length $\ell \in [2, \min(n - i, 8)]$ and then we reverse the subsegment $\pi[i: i + \ell]$. This operation is described in the middle panel of Figure \ref{fig:ranksa_moves}. Finally, the \textsc{relocate} with probability $p_{REL} = 0.2$ removes a vertex at a random source position and reinserts it at a random destination position, shifting all the labels in between one position to the left. An example of this operation is depicted in the bottom panel of Figure \ref{fig:ranksa_moves}. The probabilities for the three cases can be modified, but they should always add up to one. Having different types of vertices reshuffling has two advantages. The first one is that allows us to explore different configurations that can yield fewer emitters faster. The second one is that during the exploration phase, when the temperature is high, it is possible to accept moves that temporarily worsen the cost function. This diversity can lead to a final ordering superior to the one obtainable without such substantial changes to the vertex ordering.

Let us now discuss the implications of each one of the moves separately. The core idea here is that the \textsc{swap} is the cheapest operation to do and provides a local exploration. The \textsc{reverse} aims for neighboring segments and can correct local sub-orderings. We chose the segment length to be 8 at maximum to preserve locality. Finally, \textsc{relocate} is more disruptive since it allows long-range transports of the label, without the overhead of a full random start. After each move, if $\Delta\le0$, the new ordering is accepted, otherwise we accept it with probability $e^{-\Delta/T}$. In case a solution is not accepted, we will restore the ordering. This algorithm is efficient since it scales as $\mathcal{O}(N_{\text{trials}}\cdot n^2)$.

Before concluding this subsection, let us summarize the key differences between the algorithms presented until this point. The first one is that the \texttt{height\_function\_sa} uses three different permutation moves for the exploration phase, while the \texttt{hill\_climbing} uses only swaps between labels. Secondly, by definition a SA approach accepts permutations that can worsen the obtained number of emitters, such that escaping local minima can become easier. On the contrary, the hill-climbing approach only accepts permutations that improve the number of emitters. Having three different perturbation moves is crucial during the exploration since they allow for more impactful distortion in the labeling and thus increase the possibility of obtaining a better labeling. Additionally, we can modify the probabilities of each one of these moves to be used, equipping us with more flexibility during the exploration phase. Last but not least, there is a difference in how this exploration phase is implemented. In the case of the hill-climbing algorithm, we allow for simple swaps with no specific indication of whether we are moving in the correct direction, whereas with simulated annealing, we allow to rapidly escape from plenty of suboptimal local minima. Finally, the cooling schedule of the SA allows us to have a more controlled approach to how the escape from the local minima can be achieved.

\subsection{\texttt{path\_clustering}}
\begin{figure*}
    \centering
    \includegraphics[width=0.8\linewidth]{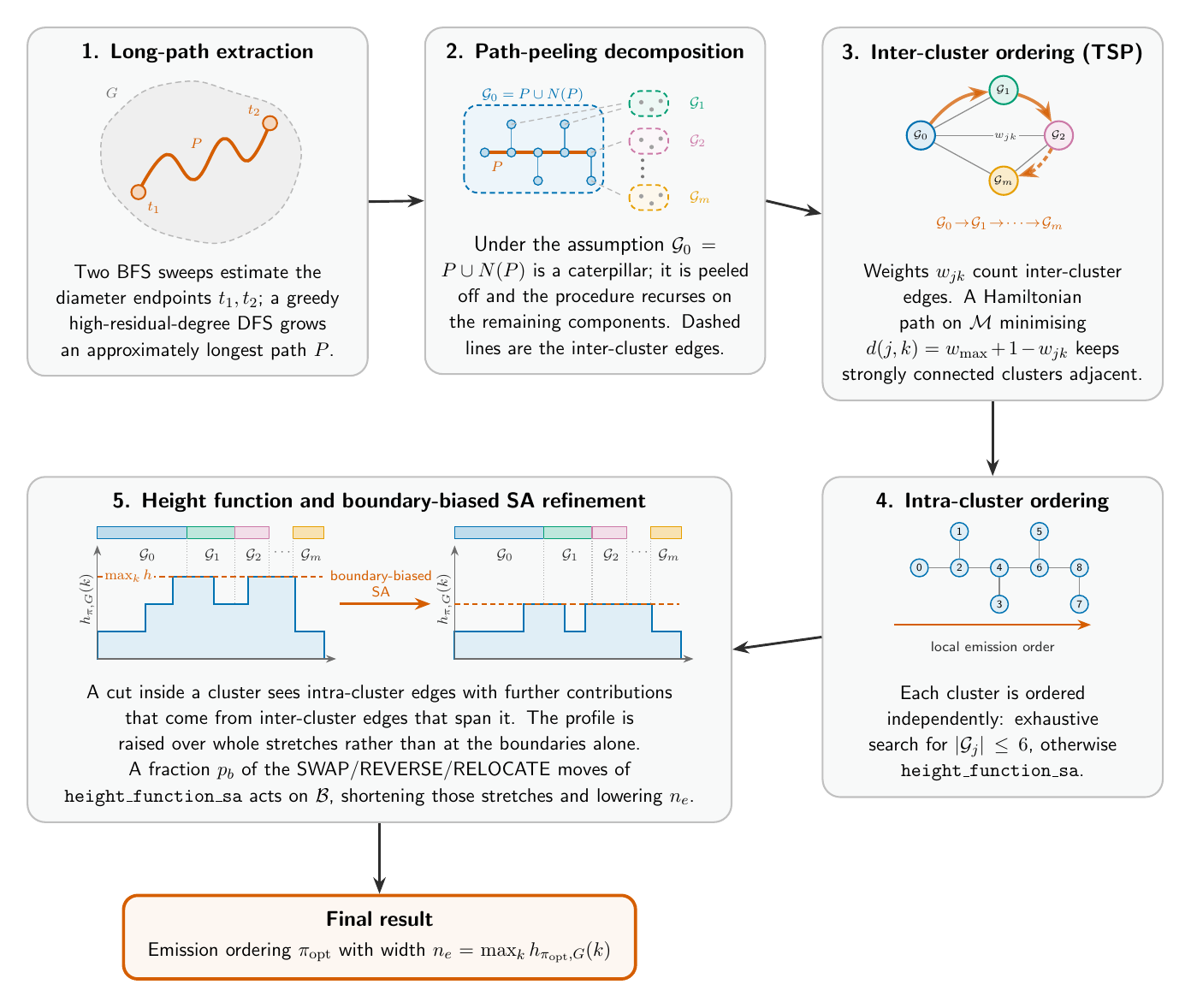}
    \caption{A schematic overview of the \texttt{path\_clustering} algorithm. In panel 1, we start with a long-path extract from the original graph $G$, and we then extract it along with its neighbors, and we recursively continue on the remaining components (panel 2). In step 3, we depict the meta-cluster $\mathcal{M}$, which in principle is a weighted graph, with weights the number of neighbors. We impose an ordering on the $\mathcal{M}$ by solving the traveling salesman problem (TSP). We always start from the first cluster. With orange arrows, we highlight the order in which we are choosing to label $\mathcal{M}$. Until this point, we have not modified the emission ordering. This is done in step 4, where, based on the order of $\mathcal{M}$, we change the emission ordering within each cluster. The last step is the general refinement of the ordering with a boundary-biased SA approach that yields emission orderings that require fewer emitters than the previous step.}
    \label{fig:path_scheme}
\end{figure*}

We proceed with our analysis by introducing the \texttt{path\_clustering} algorithm. There are two main facts taken into consideration for designing this algorithm. The first one is that only the caterpillar states, or the LC equivalent to them, can be prepared using only one emitter. Therefore, it is natural to develop an approach that systematically exploits this property. The second idea is that we use a divide-and-conquer approach rather than only trying to impose an emission ordering optimization over all vertices, as we did up until this point.

Before getting into more details on the algorithm, let us give a general overview of it. We start by decomposing the target graph into caterpillar graphs. We then create a weighted graph, which we call \emph{cluster metagraph}, where each node represents one of those graphs, and the weight of each edge is the number of edges of $G$ between the two clusters. An ordering is then imposed on this graph by solving the Traveling Salesman Problem (TSP) \cite{tsp_nphard}. This ordering imposes which set of vertices will be emitted first, resulting in a preliminary preference on the order we aim to optimize. Using the resulting labeling, the subgraph associated with each cluster is analyzed independently to obtain a local ordering. This ordering is subsequently refined via a simulated annealing run. For the caterpillar decomposition, we first need to find the longest path in a graph. However, we have to note that the longest path problem, where given a graph $G$ and an integer $\ell$, decides whether $G$ contains a simple path of at least $\ell$, is NP-complete \cite{path_npcomplete}. Additionally, the traveling salesman problem is NP-hard \cite{tsp_nphard}. Therefore, we use some heuristics to address both of the problems. An illustration of each step of \texttt{path\_clustering} is given in Figure \ref{fig:path_scheme}, where we illustrate each step of the algorithm on an example.

We begin by describing the approximation employed for the longest path, where we combined two polynomial time heuristics such that we obtain results rather close to the optimal, if not the actual optimal, but with much less computational requirements. The first key element is the \emph{BFS-diameter} heuristic, which approximates the diameter endpoints via two rounds of BFS, each with cost $\mathcal{O}(|V|+|E|)$. Starting from an arbitrary start vertex $s_0$, we compute $t_1$, which is the furthest point from $s_0$, using a BFS search. We then calculate $t_2$, which is the furthest point using BFS. The distance of the pair $(t_1,t_2)$ is approximately the diameter of the graph. For tree graphs, this yields the exact diameter. The second heuristic is based on a \emph{greedy depth-first search path extension} for which, starting from a vertex, we build a path by always extending to an unvisited neighbor with the most remaining unvisited neighbors. This greedy heuristic runs with $\mathcal{O}(|V|+|E|)$ per call. By preferring vertices with high residual degree, we keep future extensions open, and thus we obtain long paths in practice. The complete heuristic used in our case is a combination of the two algorithms, since we combine the BFS-diameter endpoints, high-degree vertices, and random seeds into a multi-start framework. With $k$ random starts and $d$ high-degree seeds, the total number of calls is $2(k+d+2)$. Each call costs $\mathcal{O}(|V|+|E|)$, so the total complexity is $\mathcal{O}\left((k+d)\cdot(|V|+|E|)\right)$.

We proceed with the second step of \texttt{path\_clustering} algorithm. Given a long path $P_1$ of the $G$, the first cluster is constructed by gathering 
the path $P_1$ along with all the adjacent vertices of the path. Then these vertices, called \emph{cluster vertices}, are removed from the graph, and the previous process is repeated until we end with an empty graph or isolated nodes. Let us formally define the path cluster.
\begin{definition}[Path cluster]
Let $P= (v_1,\cdots,v_{\ell})$ be a simple path in $G$ and let $C \subseteq V$ denote the set of vertices already claimed by previously extracted clusters. The \emph{cluster graph} seeded by $P$ with respect to $C$ is the vertex set
\begin{equation}
    V(\mathcal{G}_P)= \{v_1,\ldots,v_{\ell}\}\cup\bigcup_{i=1}^{\ell}\bigl(N(v_i)\setminus C\bigr)
\end{equation}
consisting of the path vertices together with all their unclaimed neighbors, and $\mathcal{G}_P$ denotes the subgraph of $G$ induced by $V(\mathcal{G}_P)$.
\end{definition}
Each vertex is claimed exactly once, so the sum of cluster sizes equals $n$. At each level of recursion, the algorithm for extracting the longest path calls on disjoint components, costing $\mathcal{O}(|V|+|E|)$ in total. Therefore, the number of recursion levels is at most $n$, and thus the total complexity is $\mathcal{O}(n\cdot(|V|+|E|))$. Assuming that the cluster extraction is completed, we can construct the \emph{cluster meta-graph}.
\begin{definition}[Cluster meta-graph]
Given clusters $\mathcal{C} = \{\mathcal{G}_1,\cdots, \mathcal{G}_p\}$ the cluster meta-graph $\mathcal{M}$ has one node per cluster $\mathcal{G}_j$ and edges $(\mathcal{G}_j,\mathcal{G}_k)$ with weight:
\begin{equation*}
w_{jk} = |\{(u,v)\in E(G): u\in\mathcal{G}_j,v\in\mathcal{G}_k\}|.
\end{equation*}
\end{definition}

\begin{definition}[Boundary vertices]
A vertex $v\in\mathcal{G}_j$ is a boundary vertex if it has at least one neighbor in a different cluster $\mathcal{G}_k$ with $j\neq k$. This set is denoted as B.
\end{definition}
The \emph{boundary vertices} are important elements for the algorithm, similar to the bottleneck point discussed in the previous sections. This is because we expect that the local ordering we impose on each cluster graph will yield $\lrw = 1$ or very close to it. However, hen we evaluate the width of the global ordering, such emission ordering can be suboptimal, and a substantial increase in the cut-rank can be obtained. The complexity of computing all the intra-cluster weights requires a single pass of the edges of the initial graph with a vertex to cluster lookup table costing $\mathcal{O}(|E|)$ in time and $\mathcal{O}(|V|)$ space for lookup.

The order in which clusters appear in the global vertex ordering determines which inter-cluster edges cross the cut boundaries. Intuitively, strongly connected clusters will be adjacent in the ordering. This is crucial so that their shared edges are concentrated at a single cut rather than spread across many. Therefore, we formulate it as the TSP on the $\mathcal{M}$ graph. It requires finding a Hamiltonian path through the clusters that minimizes the total distance, which is inversely related to the edge weight. For this, we define the distance matrix. Given a cluster meta-graph $\mathcal{M}$ with weights $\{w_{jk}\}$, we define the distance between the clusters $j$ and $k$ as:
\begin{equation*}
    d(j, k) =
  \begin{cases}
    w_{\max} + 1 - w_{jk} & \text{if } w_{jk} > 0, \\
    w_{\max} + 1           & \text{otherwise},
  \end{cases}
\end{equation*}
with $w_{\max} = \max_{jk}w_{jk}$. The inversion ensures that the clusters sharing many edges have a small distance. When the number of clusters, denoted as $|V_\mathcal{M}|$ is small, in our case, we fix it for $|V_\mathcal{M}|\leq 6$, we solve the TSP exactly by enumerating the $(|V_\mathcal{M}|-1)!$ ordering since the primary cluster is fixed to be zero, which is a feasible choice in practice. In the case where $|V_\mathcal{M}|>6$, solving the TSP problem is not practical, and thus we follow a procedure similar to the SA pipeline used for the \texttt{height\_function\_sa}, with a few modifications. The first one is that for the initial ordering we use the $\pi_{\text{deg}}$, while the only moves allowed during the SA algorithm are the \textsc{swap} moves between the vertices of the metagraph. The rest of the details of the SA algorithms are the same as presented in Section \ref{subsec:sa}. The exact TSP costs $\mathcal{O}(|V_\mathcal{M}|!)$ and the SA-based TSP costs $\mathcal{O}(K_{tsp}\cdot S_{tsp}\cdot |V_\mathcal{M}|)$, with $K_{tsp}$ the number of temperature levels and $S_{tsp}$ the number of steps per temperature.

At this point, we have imposed an ordering on the meta-graph $\mathcal{M}$ which already imposes a preference on the vertices that will be emitted first. Even if it is known that creating the subgraphs for each cluster requires one or few emitter, we need to find the exact ordering to do so. In case there are 6 or fewer vertices inside a cluster, we can perform an exhaustive search. This is something we can not opt for for large clusters, and thus we resort to using the \texttt{height\_function\_sa}. This step illuminates that the \texttt{path\_clustering} algorithm provides an alternative to the initial conditions used for the previous algorithms and a preliminary partial global ordering. 

Bearing this in mind, we proceed with the final step of \texttt{path\_clustering} algorithm, which is a global SA refinement, similarly to \texttt{height\_function\_sa}. The core difference is that we should focus on perturbations on or close to the boundary vertices. Therefore, we introduce the boundary bias $p_b$, with a default value of $0.7$, that controls the number of perturbations happening close to the boundary. The perturbation moves allowed are the same as the \texttt{height\_function\_sa}. The core idea is that we expect the global SA refinement to increase the cut-rank for cuts inside the caterpillar graphs we have from the decomposition, but reduce the cut-rank on the cuts on the boundaries between the clusters, which are much larger. With this refinement step \texttt{path\_clustering} algorithm concludes.

\subsection{\texttt{min\_la\_sa}}

As we have extensively discussed, finding the optimal emission ordering can be mapped to solving the linear rank-width problem, which can not be exactly solved for any general graph. However, there are specific cases where a solution is known exactly, for instance, for the caterpillar graphs. Additionally, this problem is not the only ordering problem in graph theory. Other such problems are the cutwidth minimization problem, the bandwidth minimization problem, or the vertex separation problem, to name a few \cite{linear_layout}. For each one of them, extensive research has been conducted to address cases where they are exactly solvable or at least a fairly good approximation algorithm is built. Therefore, it is natural to address the case in which we are using a different ordering problem as a proxy, and then we calculate the number of emitters and the rest of the metrics we are interested in.

Among the plethora of ordering problems, the minimum linear arrangement problem (minLA) seems to be a promising candidate. The cost function for a linear arrangement of a graph is defined as the sum, over all edges, of the absolute difference between positions assigned to their endpoints,
\begin{equation}
     LA(\pi,G) = \sum_{(u,v)\in E(G)}|\pi^{-1}(u)-\pi^{-1}(v)|
\end{equation}
where we express the inverse map of the bijection given in the equation \eqref{eq:prems:pidef} as $\pi^{-1}: V \xrightarrow{} \{0,\cdots,n-1\}$. The vertex ordering $\pi$ that minimizes this sum is a solution to the minLA problem. A solution to this problem yields an ordering that favors the emission of photons that are adjacent in the graph, close together in time.

Let us assume that our edge reduction scheme reduces any LC-equivalent graph to a caterpillar graph. The caterpillar graph can be decomposed into two subgraph classes, stars and linear graphs. The minLA for a linear graph is an enumeration from one side to the other. For a star graph, the emission order for a single emitter is arbitrary since it is LC equivalent to the fully connected graph. The minLA for a star places the centered vertex $v_c$ as the middle of the arrangement. A star with $\ell$ leaves has a minimum cost of $LA( \pi{\text{opt}},G) = \frac{\ell}{2}(\frac{\ell}{2}+1)$ if $\ell$ is even and $LA( \pi{\text{opt}},G) = (\lfloor\frac{\ell}{2}\rfloor+1)^2$ if $\ell$ is odd \cite{Cohen_Optimal_LA}. The two subgraph minLA solutions are compatible with each other and yield an emission order realizable by a single emitter, since we create the central path with one emitter, and at each position, a dangling bond occurs, the requisite number of additional photons are emitted with Hadamard gates applied to them \cite{Hilaire_Near-deterministic}. 

In combination with edge reduction, this proxy ordering could ideally arrange all graphs that can be prepared with a single emitter. Less is known in general about graphs requiring more than one emitter, except for specific graph families. Computing minLA is NP-hard for general graphs and exactly solvable in $\mathcal{O}(2^n|E|)$ \cite{Garey_minLA_NP-complete,Koren_scale_algo_LA}. However, several approximation techniques can be used, and one of the most prominent is the SA, which we have already presented in this work. We are following the same SA process as before, with the only difference that in the exploration phase, we randomly choose two vertices and we swap their labels.

\section{Numerical Experiments on Random Graphs}\label{sec:exp-rand}
Bearing in mind the four algorithms for optimizing the emission ordering, which we presented in Section \ref{sec:algos}, we turn our focus on their performance on random graphs with a random labeling. More specifically, we test the best version of our algorithms on an extended dataset of random graphs and calculate the number of emitters, emitter CNOTs, and the total gate count that our algorithms yield. In this case, the emission circuit is obtained directly from \cite{li_algo}, and no optimization of the emitter CNOT is implemented. In the next subsection, given the previously obtained emission ordering, we examine whether the imposed optimization can further reduce the number of required CNOT gates in the emission circuit by combining our algorithms with the one presented in \cite{eva_opt}. A detailed comparison of each step of the developed algorithms is given in the Appendix \ref{app:pipeline-exam}.

\subsection{Best case comparison and emitter CNOT reduction}\label{subsec:best-case-and-cnot-red}

We first need to construct a graph state database of random Erd\H{o}s-R\'enyi graphs. For this reason, we have a dataset where we sweep on both the probability of connection $p$ as well as the number of nodes to get a more detailed idea of the performance. Additionally, the metrics we consider are the number of emitters, the number of emitter CNOTs, and the total gate count. The goal is to quantify the improvement one gets if our algorithms are used compared to random emission ordering. Then we aim to compare the developed algorithms across the four relevant metrics we are interested in. For each $p$ and number of nodes, we assume 10 graphs. We compare the results our algorithms yield in comparison to the random labeling they had when we initially created them.

\begin{figure*}
    \centering
    \includegraphics[width=1.0\linewidth]{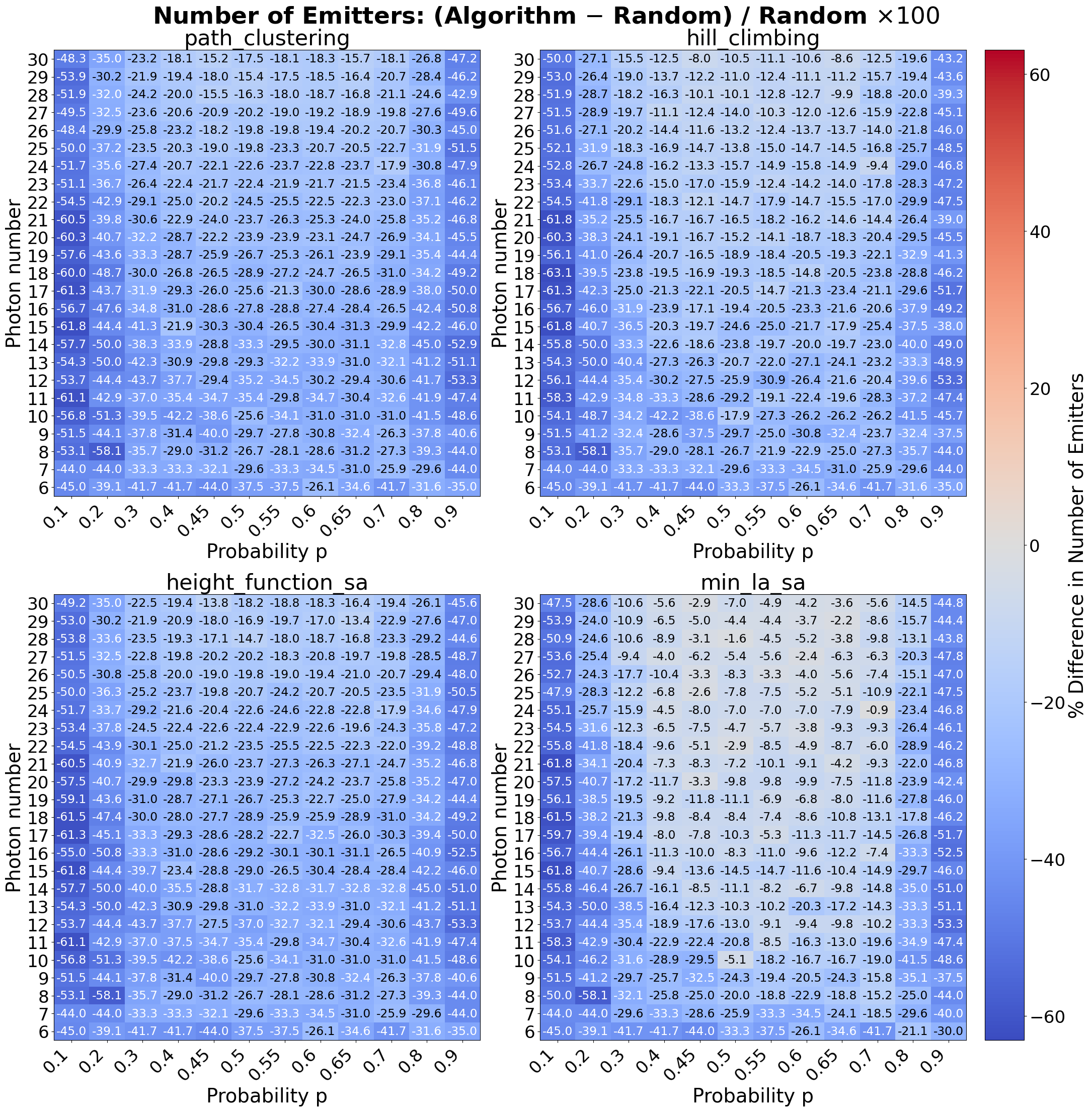}
    \caption{The comparison between the emission ordering given by our algorithms and a random emission, for the number of emitters when they are tested on the dataset of random Erd\H{o}s-R\'enyi graphs.}
    \label{fig:best_case_comp_ems}
\end{figure*}

In Figure \ref{fig:best_case_comp_ems} we have the residual percentage difference between a random emission order and the one each one of our algorithms yields. As it is expected for each one of the cases, the number of emitters is reduced. It is noteworthy that for low-density graphs ($p\in\{0.1,0.2\}$) the improvement is larger than for graphs in the medium regime ($p\in\{0.45,0.5, 0.55\}$). Two important remarks are emerging from this observation. The first one is that the same behavior is found for graphs with high density, which is expected since an edge reduction preprocessing step was imposed before we employed our optimization algorithms. The second one is that it is expected to achieve further reduction in the low-density regime since the states in that case are either trees or close to them. Even though such graphs are generally not caterpillars, and thus cannot necessarily be created with one emitter, tree graph states are known to admit a generation scheme requiring a comparably small number of emitters \cite{buterakos}. This explains why fewer emitters are required in the preparation circuit compared to states with the same number of photons but a medium density for their corresponding graphs.

\begin{figure*}
    \centering
    \includegraphics[width=1.0\linewidth]{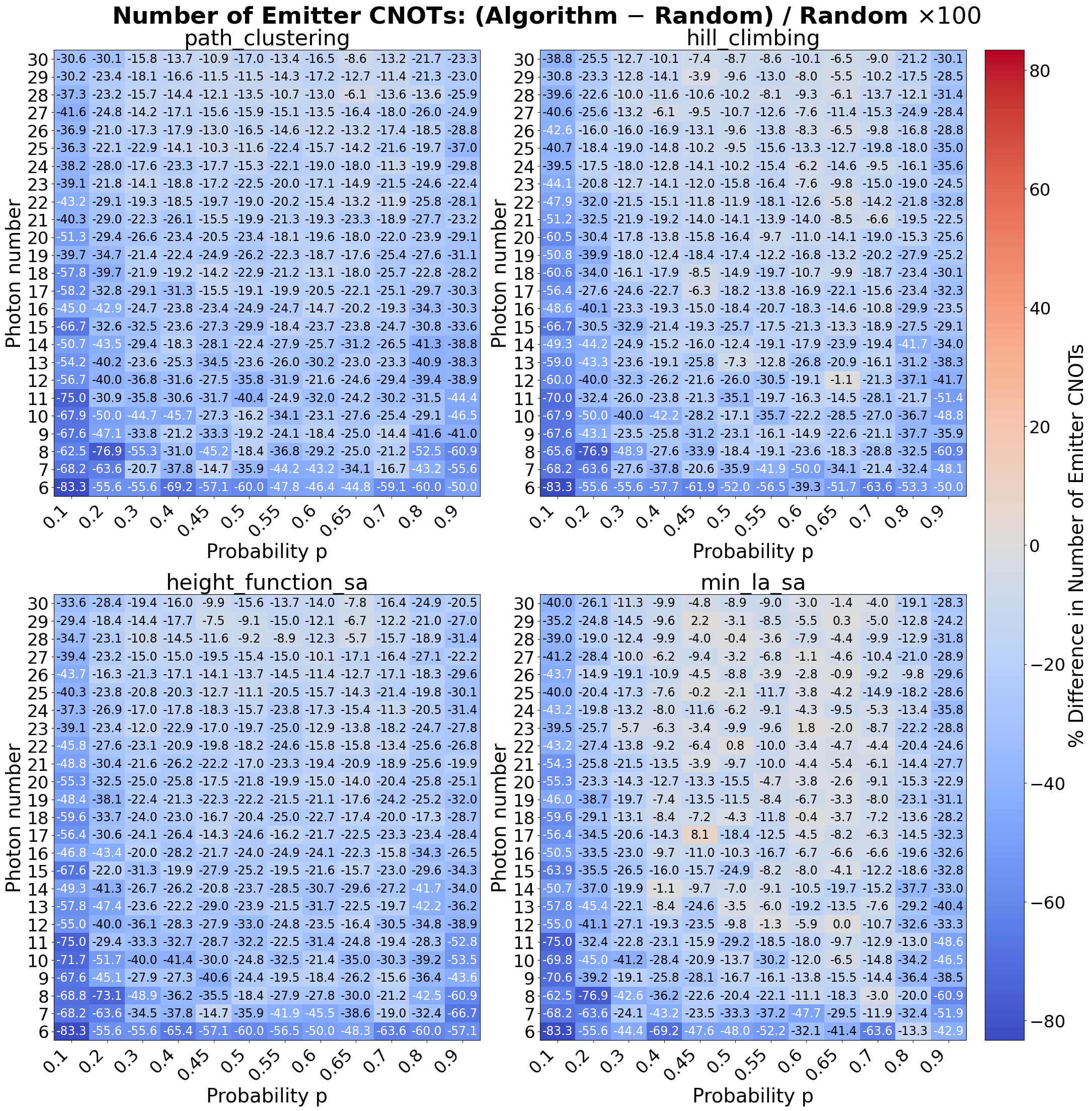}
    \caption{The comparison between the emission ordering given by our algorithms and a random emission, for the emitter CNOTs when they are tested on the dataset of random Erd\H{o}s-R\'enyi graphs.}
    \label{fig:best_case_comp_cnot}
\end{figure*}

\begin{figure*}
    \centering
    \includegraphics[width=1.0\linewidth]{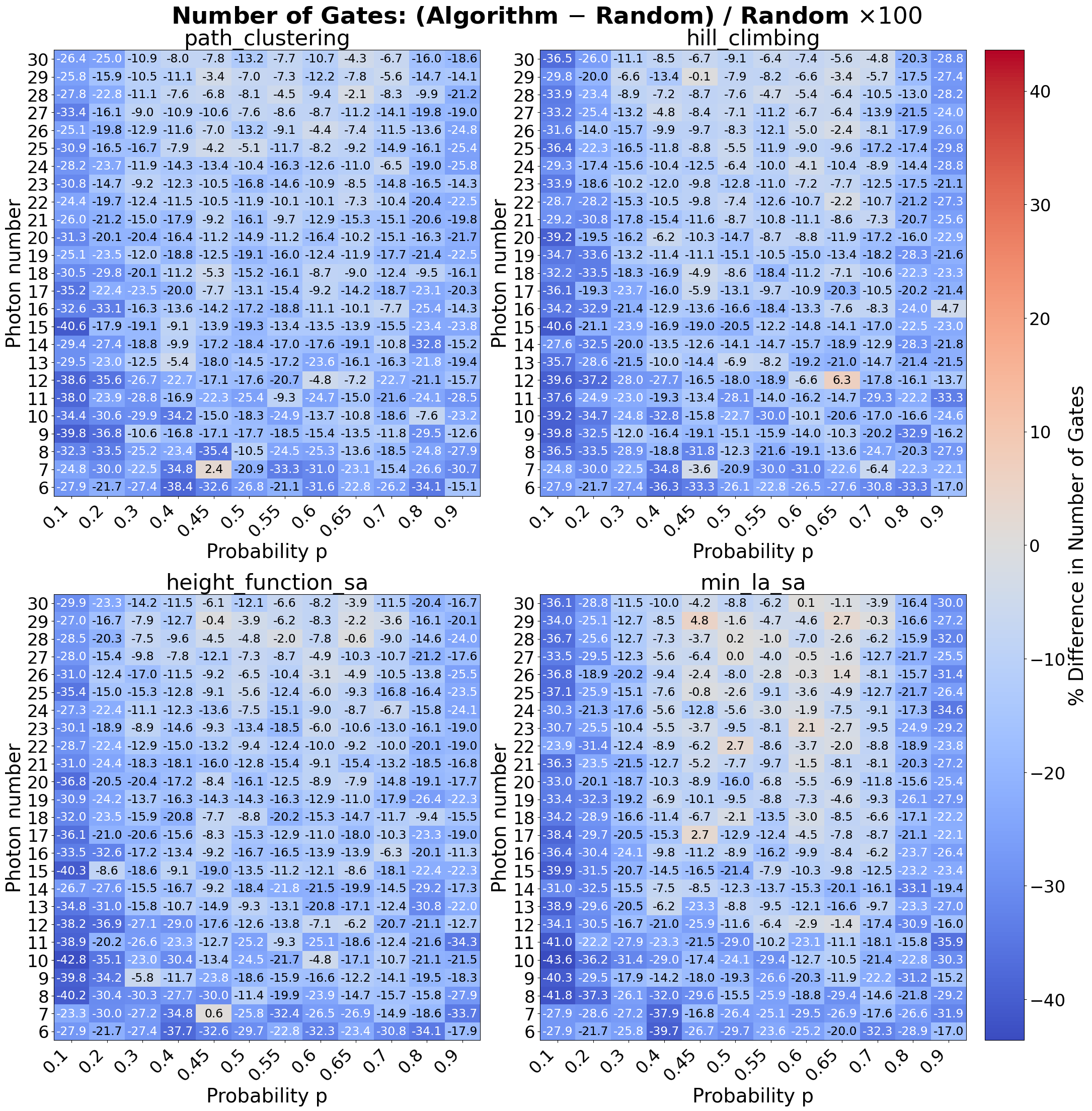}
    \caption{The comparison between the emission ordering given by our algorithms and a random emission, for the gate count of the emission circuit when they are tested on the dataset of random Erd\H{o}s-R\'enyi graphs.}
    \label{fig:best_case_comp_gates}
\end{figure*}

We proceed with the residual percentage comparison for the emitter CNOTs presented in Figure \ref{fig:best_case_comp_cnot}. In this case, we also observe that we can improve the required CNOTs more accurately for the low-density states. However, this improvement is not as great as in the case of the emitters, which is anticipated as we optimize with respect to the number of emitters.
The last metric, for which the results for the complete parameter space will be presented, is the total gate count. In Figure \ref{fig:best_case_comp_gates}, we present these results. In contrast with the previous cases, now there are a few instances where the total gate count becomes larger. However, the property that the optimization generally positively affects in the low-density graphs is preserved. Due to the above, it becomes clear that emitter optimization endeavors can also yield considerable improvements for the rest of the metrics, even if this was not the main objective, which is, of course, desirable.

\begin{figure}
    \centering
    \includegraphics[width=1.0\linewidth]{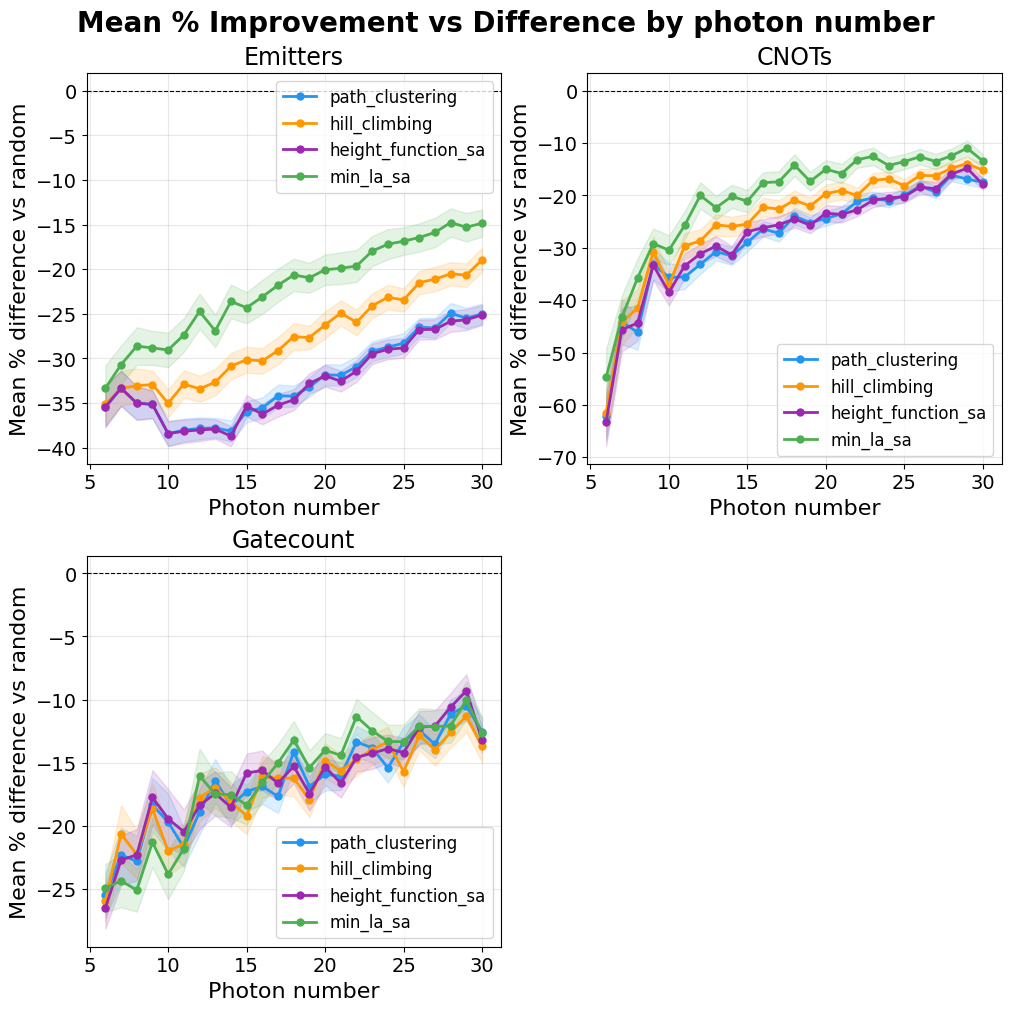}
    \caption{The figure displays the mean difference for each of the metrics. All algorithms, as expected, outperform random labeling, but the percentage improvement decreases as the photon number increases.}
    \label{fig:best_case_comp_percent_comp}
\end{figure}

Let us turn our focus to comparing the performance of the developed algorithms to each other. Despite already having an indication regarding the algorithm advantage for each metric, we conduct a rigorous analysis to determine which algorithm is best for each of the three metrics we consider. In Figure \ref{fig:best_case_comp_percent_comp}, we depict the mean percentage difference between random photon emission order and the one obtained from our algorithms. As it is expected, all algorithms outperform random ordering. However, this improvement decreases as the number of photons required for the state increases. In general, it is expected that as the number of photons increases, the improvement will not be as notable since the problem scales $|V|!$ and thus the probability of being trapped in a local minimum increases. However, as the number of vertices increases, it is more probable that the total resource requirement increases, and thus even a small mean percentage decrease can have a drastic impact on the required resources. 

\begin{figure}
    \centering
    \includegraphics[width=1.0\linewidth]{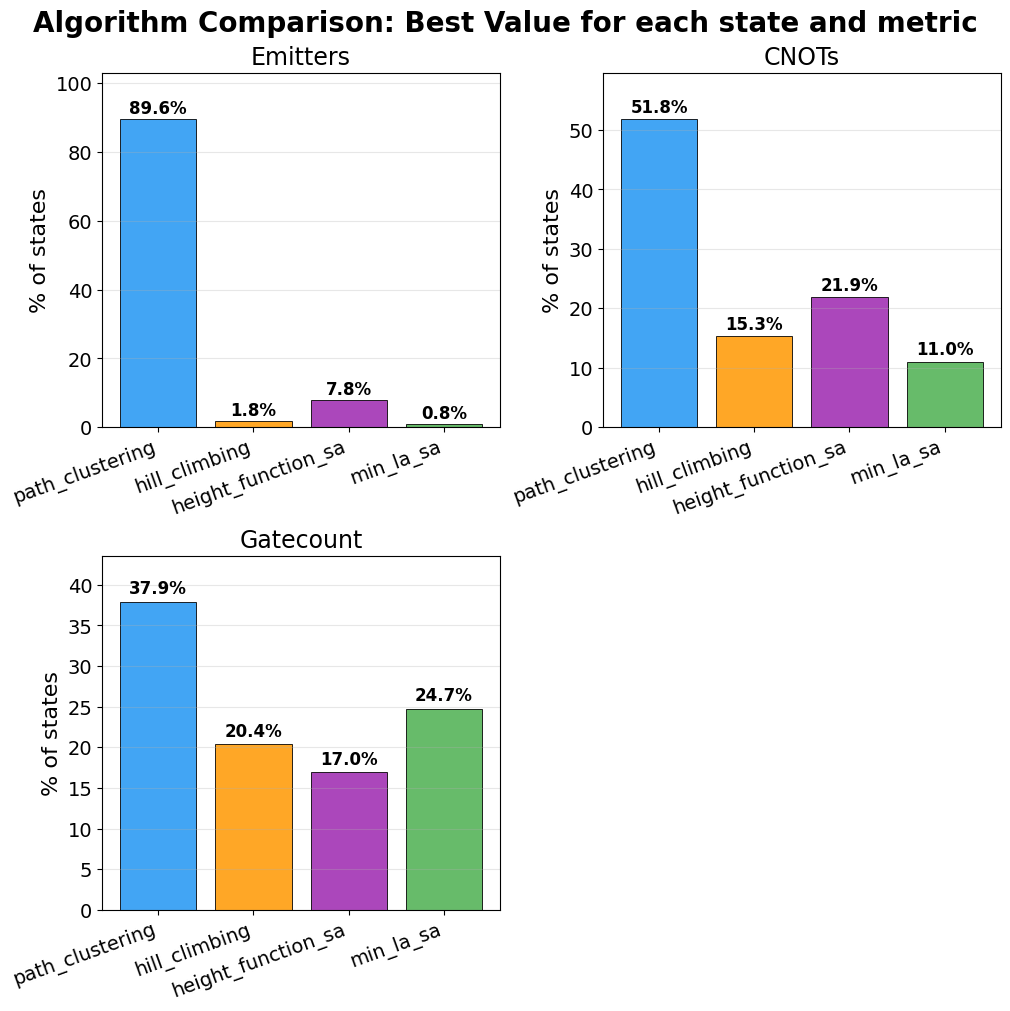}
    \caption{This figure presents a head-to-head comparison across graph instances, identifying for each metric the algorithm that achieves the best result. It offers a first clear indication of which algorithm performs best for each individual metric.}
    \label{fig:best_case_comp_hist}
\end{figure}

Another complementary piece of information relevant for the performance comparison is a head-to-head comparison for each state and the number of times an algorithm beats the rest, without including potential cases that yield equally optimized results. We illustrate this comparison in Figure \ref{fig:best_case_comp_hist}. In this setup for the number of emitters, the \texttt{path\_clustering} is clearly advantageous since it outperforms the rest in the $89.6\%$ of the cases, where a tie does not exist. This clear advantage is also preserved for the number of emitter CNOTs. However, not as strongly, since this is the case, this is true for almost half $(51.8\%)$ of the cases. For both the aforementioned metrics, the second-best algorithm is the \texttt{height\_function\_sa}. For the total gate count, the path clustering approach has a slight advantage, but no clear winner is found. 

\begin{figure}
    \centering
    \includegraphics[width=0.8\linewidth]{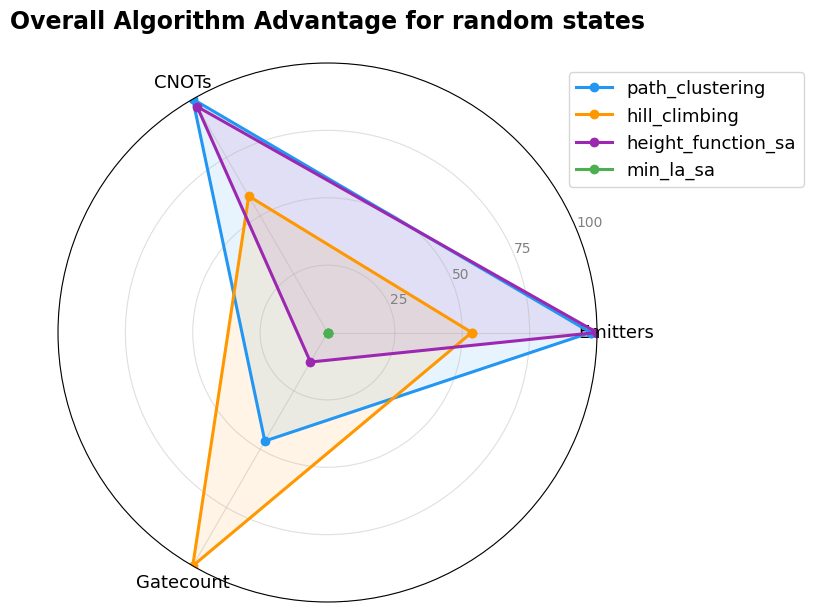}
    \caption{In the figure, we present a clear answer to which algorithm performs best for each of the three metrics we are using. We are displaying the normalized score ranging from 0 (lower) to 100 (higher) per metric.}
    \label{fig:best_case_comp_radar}
\end{figure}

To further elaborate on the algorithm comparison, we must account for the possibility that an algorithm outperforms the other in the majority of cases but by a negligible margin, while performing considerably worse in the remaining cases. To address this, we compute, for each algorithm and metric, the mean difference relative to random emission across all states. We then negate this value, so that an algorithm requiring fewer resources yields a positive improvement score. Within each metric, these scores are then rescaled across the four algorithms via min-max normalization to the range $[0,100]$, so that the algorithm with the greatest improvement is assigned the score $100$ and the one with the smallest the score $0$, we refer to this as the \emph{normalized score} and we are going to use it for every graph state family examined in this work. The resulting normalizing scores are displayed in Figure \ref{fig:best_case_comp_radar}. 

The first metric we discuss is the number of emitters. In this respect, the \texttt{path\_clustering} and \texttt{height\_function\_sa} perform best, with closely comparable results, and a similar pattern holds for the number of CNOTs. However, the \texttt{path\_clustering} is the second best for the total gate count. For this metric, the \texttt{hill\_climbing} algorithm is best, and the \texttt{height\_function\_sa} is the third. Therefore, Figure \ref{fig:best_case_comp_radar} raises a further point worth discussing. For a given family of states, an algorithm that performs better on one metric may not be the best choice for another metric. This tradeoff must be taken into consideration when selecting an algorithm for a practical implementation, since a small increase in the number of emitters may be acceptable if it yields a substantial reduction in emitter CNOT overhead. Analogous tradeoffs between metrics have also been reported elsewhere in the literature \cite{ghanbari_framework}.

\subsection{Emission ordering optimization as a preprocessing step}

\begin{figure}
    \centering
    \includegraphics[width=0.7\linewidth]{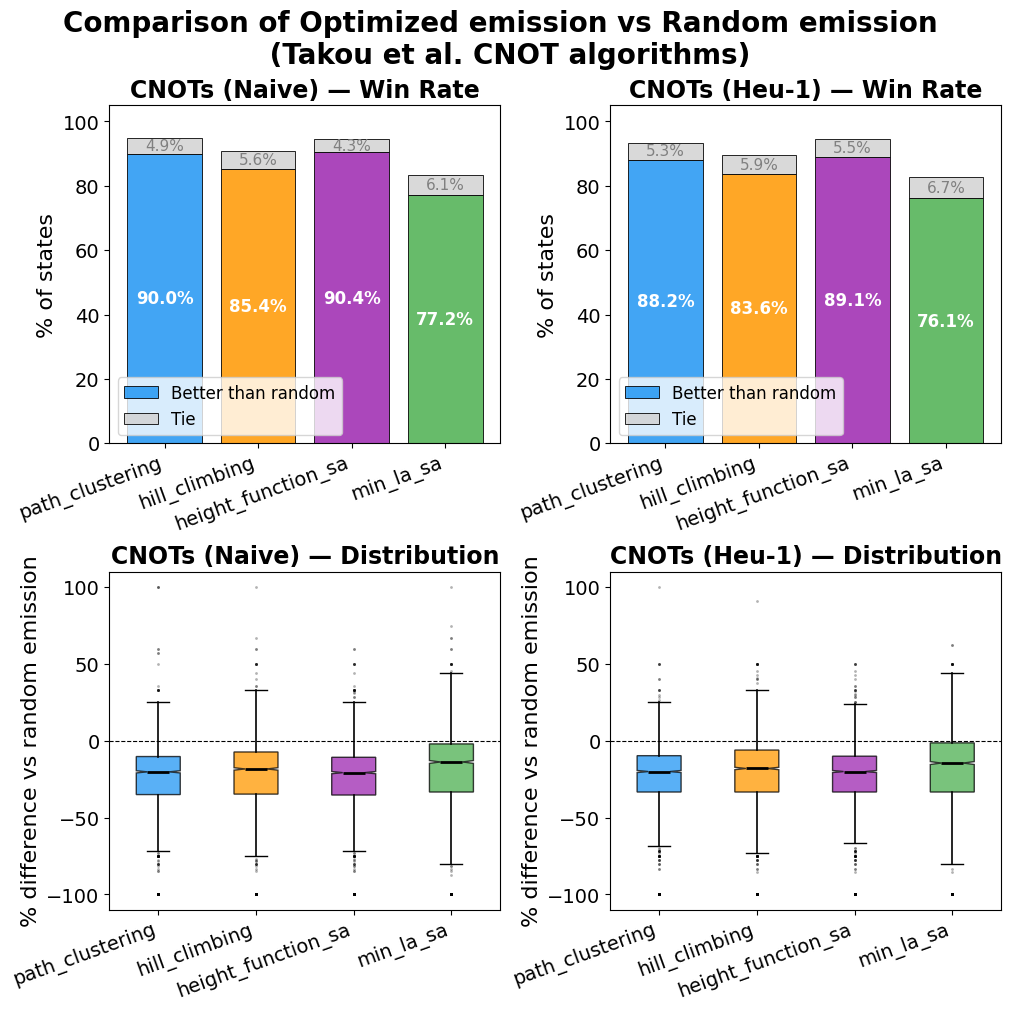}
    \caption{The figure provides a comparison between the CNOT count of the algorithms in \cite{eva_opt}, in case the ordering found by our algorithms is imposed, compared to the random ordering. In the top panel, we have the win rates, while in the bottom one, we have the percentage difference distributions. The top panel validates that, indeed, for the vast majority of cases, an advantage is obtained. The bottom panel validates that this is the case on average, while some outliers do exist.}
    \label{fig:comp_takou_hist}
\end{figure}

Having established these tradeoffs between the emission ordering algorithms, we now turn to a complementary question: whether the choice of vertex ordering also affects the performance of the algorithms aiming to reduce the emitter CNOTs. Such algorithms were presented by Takou et al. in \cite{eva_opt}, where they introduced 3 heuristic algorithms to reduce the emitter CNOT overhead in the emission circuit, admitting substantial reduction when they tested them on random states. It is therefore natural to examine whether using the emission ordering obtained from our algorithms affects the emitter CNOT count. For this reason, we use the so-called Naive and Heuristic-1 (Heu-1) algorithms presented in \cite{eva_opt} and compare their results for a random emission order and the one each algorithm we presented yields. In Figure \ref{fig:comp_takou_hist}, we present the percentage of states where we find an advantage compared to the random emission. In the top figure, we realize that for the vast majority of cases, our algorithms yield a reduced CNOT count. We quantify this difference in the bottom panel of the figure, where we present the probability distribution of the percentage average reduction. The exact numbers are given in table \ref{tab:takou-further-reduction}, making clear that in general our algorithms are an appropriate preprocessing step that can enhance the impact of the algorithms presented in \cite{eva_opt}.

\begin{table}[h]
\centering
\caption{In this table, we present the mean percentage change compared to the random emission order using the emission algorithm of \cite{li_algo}. We are considering the two algorithms (Naive and Heu-1) developed in \cite{eva_opt} when combined with our algorithms, compared to the case where a random labeling is used.}
\label{tab:takou-further-reduction}
\begin{tabular}{lcccc}
\toprule
\textbf{Algorithm}  & \textbf{CNOTs (Naive)} & \textbf{CNOTs (Heu-1)} \\ \midrule
\texttt{path\_clustering}& -25.84\%                  & -24.88\%               \\
\texttt{hill\_climbing}& -24.09\%                  & -23.08\%               \\
\texttt{height\_function\_sa}               & -26.21\%                  & -25.22\%               \\
\texttt{min\_la\_sa}                   & -20.46\%                  & -20.07\%               \\ \bottomrule
\end{tabular}
\end{table}

\begin{figure}
    \centering
    \includegraphics[width=1.0\linewidth]{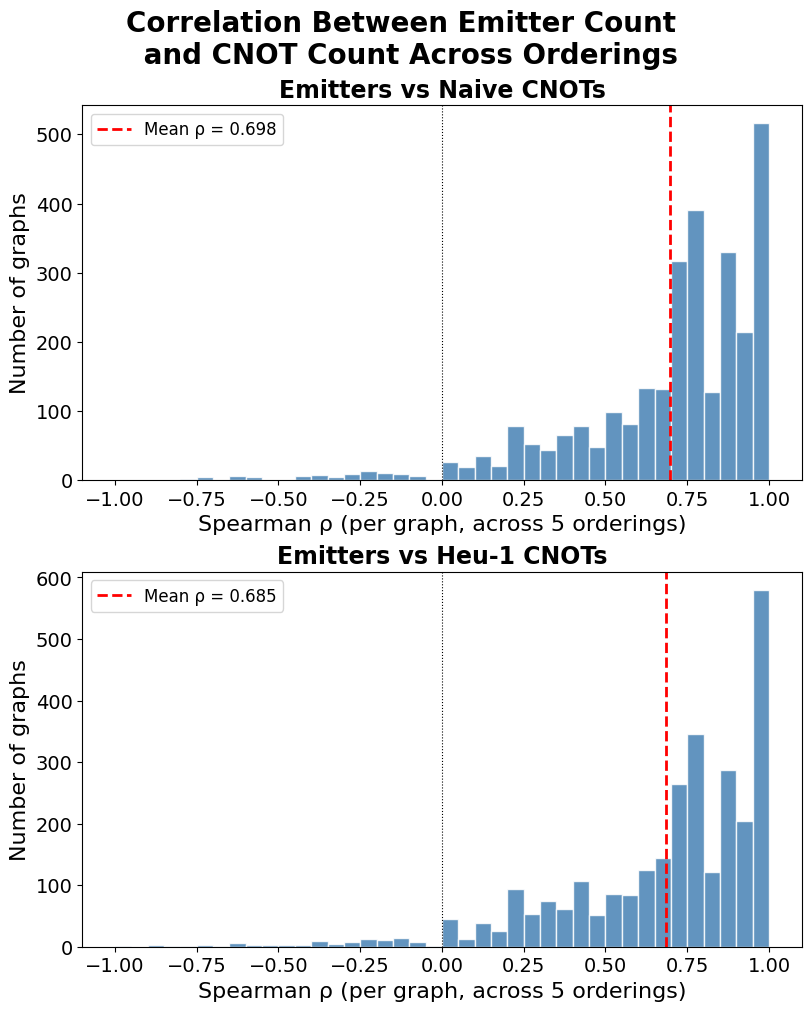}
    \caption{Spearman rank correlation between emitter and CNOT count across orderings. On the left, we have the Naive algorithm, while on the right, the Heu-1. The closer the value is to 1, the stronger the indication that reducing the emitters reduces the CNOT count. The red, dashed line marks the mean $\rho$ across all the graphs.}
    \label{fig:comp_spearman_hist}
\end{figure}

A crucial remaining question is whether the algorithms designed to reduce the number of emitters can have a positive effect on reducing the emitter CNOTs once Takou's et al. algorithms are considered. To address this, we turn to a statistical analysis of our results. Specifically, for each state, we compute the Spearman rank correlation coefficient, denoted $\rho$, between the emitter count and the CNOT count across the $m=5$ orderings (the four from our algorithms and the random one), yielding one value of $\rho$ for the Naive algorithm, and one for the Heu-1 algorithm of \cite{eva_opt}. The Spearman correlation indicates the strength and direction of a monotonic relationship between two ranked variables, defined as
\begin{equation}
    \rho = 1- \frac{6\sum_{i}d_i^2}{m(m^2-1)},
\end{equation}
where $d_i$ is the difference between the ranks assigned to the $i$-th ordering under the two variables being compared, and $m$ is the number of orderings being ranked, here $m=5$. A value close to $1$ indicates that the orderings associated with fewer emitters also tend to yield lower CNOT counts, while a value close to $-1$ indicates the opposite trend. If $\rho\approx 0$, this is an indication of no monotonic relationship. The result distributions of $\rho$ across all states are presented in Figure \ref{fig:comp_spearman_hist}. 

\begin{figure}
    \centering
    \includegraphics[width=1.0\linewidth]{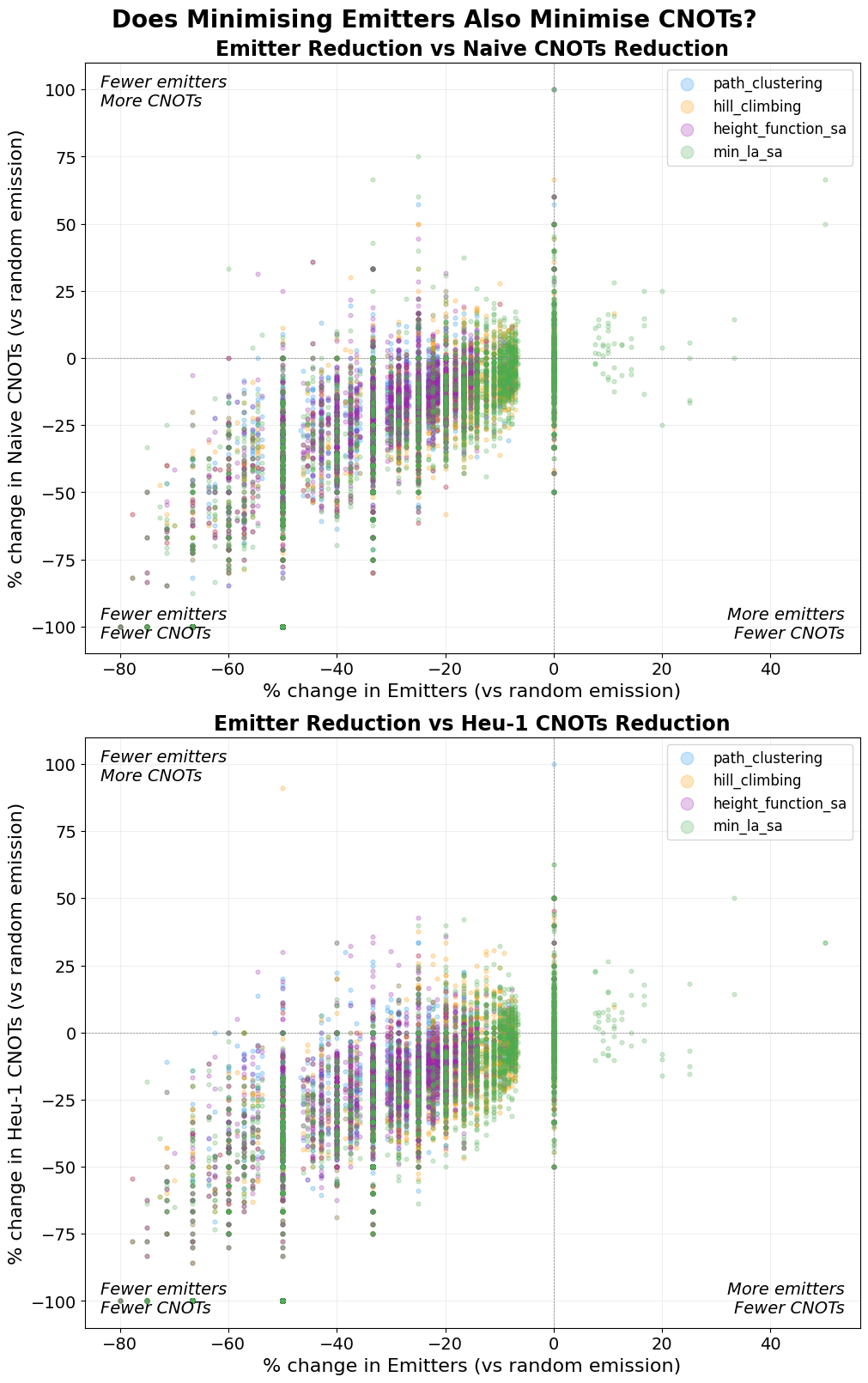}
    \caption{In this plot, we have the percentage change in the number of emitters along with the percentage difference on the projected number of CNOTs for the Naive (left) and the Heu1 (right) algorithms. Here, there is an indication that an algorithm designed to reduce the number of emitters can have a positive impact on the effects of the Naive and Hue1 algorithms.}
    \label{fig:comp_takou_scatter}
\end{figure}

To strengthen our intuition and be able to differentiate the performance on reducing the CNOT count each of our algorithms has when combined with either the Naive or Heu1 algorithms, we present Figure \ref{fig:comp_takou_scatter}. In this case, there is an indication that an algorithm designed to reduce the number of emitters can have a positive impact on the effects of the Naive and Heu-1 algorithms. It is a matter of future work to address this connection in a more concrete way, as well as find the elements that assist in optimizing both the number of emitters and emitter CNOT simultaneously.

At this point, we conclude this subsection, where we calculated the three metrics we are interested in for the algorithms we presented in this work. In general, in terms of reduction across all metrics, the \texttt{path\_clustering} algorithm has a clear advantage. We also addressed the impact of our emitter optimization algorithms on other CNOT optimization algorithms presented in \cite{eva_opt} by showing that the performance of the latter can be further improved using our emission ordering optimizers.

\section{Application on Prominent Graph State Families}\label{sec:applications}
In this section, we test the performance of our algorithms on various graph state families. More specifically, we include resource states for the implementation of Shor's algorithm in the MBQC framework, as well as on  Raussendorf-Harrington-Goyal (RHG) lattices, and finally, we examine various Quantum Error Correcting Codes (QECC). In this section, we aim to further uncover the performance of our algorithms and test them on states that are much more challenging to create, as in some cases, we require more than 400 photonic qubits, but they are crucial for various aspects of quantum technologies.

\subsection{Resource states for a MBQC version of Shor's algorithm}
Shor's algorithm \cite{shorsalgo}, which factors a composite integer $N$ in polynomial time by reducing the factoring problem to the order-finding problem, is known for revolutionizing quantum information and computation. In the MBQC framework, every unitary circuit is compiled into a measurement pattern that acts on a large resource state. The size and topology of the graph state depend on the algorithm intended for implementation \cite{mbqc_flow}. In this section, we are focusing on Shor's algorithm, but this idea can be expanded to every algorithm one may need to utilize. In the main text, we only present the analysis of the data we obtained after evaluating the three metrics we are interested in. For details on how we implemented the algorithm, extracted and optimized the required resource state using \textsc{graphix} \cite{graphix1,graphix2}, and finally, how the corresponding graph database was created, we refer the reader to the Appendix \ref{app:shor}. 

\begin{figure*}
    \centering
    \includegraphics[width=1.0\linewidth]{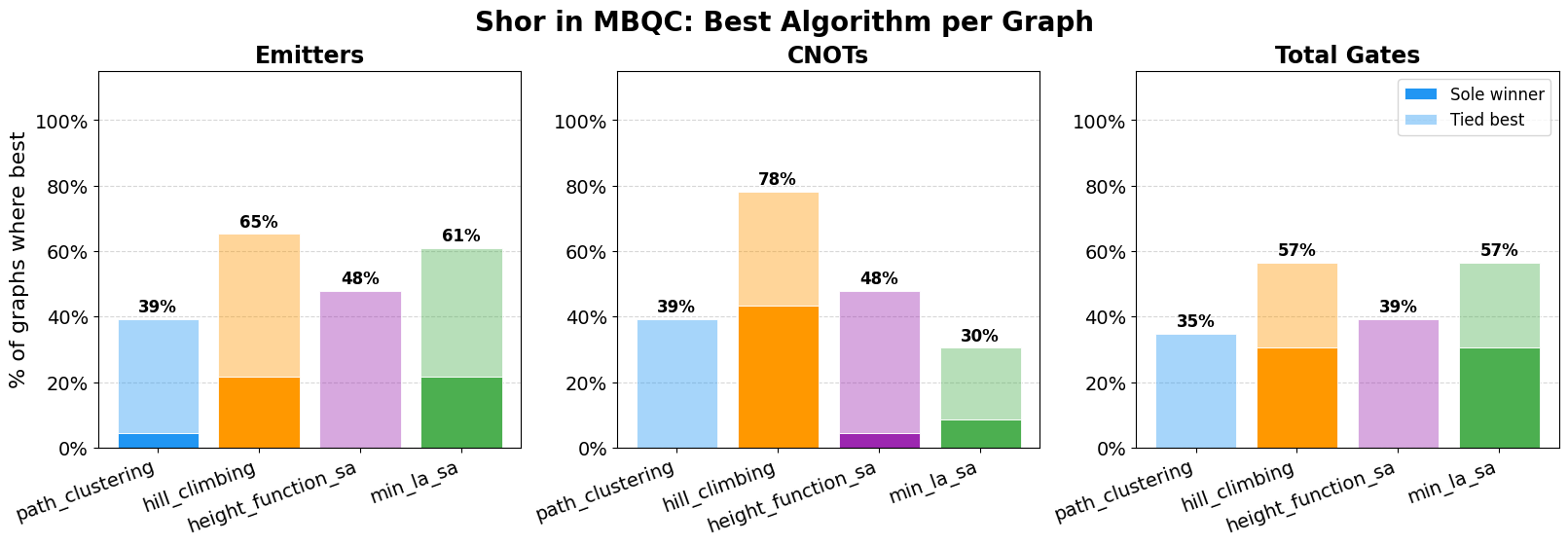}
    \caption{Histogram of the number of cases where each algorithm is the best one or a tie exists.}
    \label{fig:app-shor_hist}
\end{figure*}

\begin{figure}
    \centering
    \includegraphics[width=0.8\linewidth]{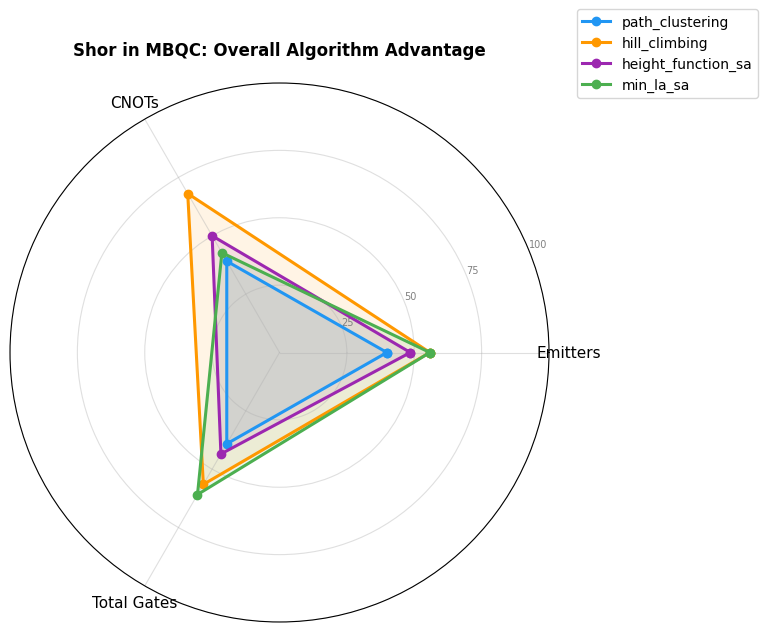}
    \caption{In the figure, we present a clear answer to which algorithm performs best for each of the three metrics we are using. We are displaying the normalized score ranging from 0 (lower) to 100 (higher) per metric.}
    \label{fig:app-shor_radar}
\end{figure}

In Figures \ref{fig:app-shor_hist} and \ref{fig:app-shor_radar}, we present the simulation results. Figure \ref{fig:app-shor_hist} shows a histogram of the number of cases where each algorithm is the best one or a tie exists, while Figure \ref{fig:app-shor_radar} shows a clear hierarchy of the performance each algorithm has for each of the metrics. For the number of emitters and the CNOT count, there is no clear winner. However, \texttt{min\_la\_sa} is the best in terms of the number of emitters, while \texttt{hill\_climbing} is the best in terms of the CNOT count. This behavior can be explained by the fact that the resource graphs for this implementation of Shor's algorithm are sparse and tend to be trees or close to them, a structure for which the minimum linear arrangement proxy used by \texttt{min\_la\_sa} is particularly well suited. We stress, however, that this advantage is specific to this graph family.

It is noteworthy that in this case we examined states requiring from $7$ to $365$ photonic qubits. Apart from their interesting nature from the quantum computing perspective, here, but also in general in this Section, we want to test our algorithms on states requiring a few hundred photons, since we are getting a better understanding of how the metrics evolve in this regime. It is rather interesting that the \texttt{min\_la\_sa} appears to perform better as the number of photons increases. This result validates our choice of using other graph theory properties as proxies to reduce the number of required emitters.

\subsection{RHG Lattice}
The RHG lattice \cite{rhg_1,rhg_2} is the first instance of a graph state designed for fault-tolerant MBQC. It can be seen as a foliation of the surface code \cite{foliation1}, and it consists of cubic unit cells with nodes placed at the centers of the faces and edges of each cube. We consider here $(m_x,m_y,m_z)$-RHG lattice as defined in \cite{rhg_3}, where each parameter $m_{x,y,z}$ determines the number of unit cells stacked along the corresponding axis. For this graph state family, we calculated every possible lattice until $(3,4,4)$, obtaining the number of emitters, emitter CNOTs, and total gate count. The complete table is presented in the Table \ref{tab:rhg_results-total}. However, it would be interesting to compare our results with the emitter and CNOT predicted in \cite{ghanbari_framework}. For this reason, we are concentrating only on the number of emitters and CNOTs. In the Table \ref{tab:rhg_results-ghan} we report the best result our algorithms yield, prioritizing the number of emitters, then the CNOT, and finally the gate count. Regarding the number of emitters, we predict the same number apart from the case of $(2,2,2)$, where we require $11$ while in \cite{ghanbari_framework} they require $12$.

Turning our focus on the emitter CNOTs, we typically require fewer except in the case $(1,1,2)$, where we have $30$, and they need $28$. One crucial remark is that we used the emission circuit obtained by the algorithm presented in \cite{li_algo}, and since in Section \ref{subsec:best-case-and-cnot-red} we have provided clear evidence that the CNOT optimizers from Ref. \cite{eva_opt} combined with our emitter optimizers provided even more improved results, it is therefore highly probable that the aforementioned combination will yield fewer emitter CNOTs than the one presented in \cite{ghanbari_framework}, for the RHG graph states. To test this claim, we used the ordering of \texttt{hill\_climbing} and then the CNOT reduction algorithm of \cite{eva_opt}, and the new number of CNOTs required to generate the (1,1,2) RHG lattice was 26, which is lower than the 28 CNOTs presented in \cite{ghanbari_framework}.

\begin{table}[]
    \centering
    \caption{Comparison of our algorithms with the results presented in \cite{ghanbari_framework}. Here, HC denotes \texttt{hill\_climbing}, HFSA denotes \texttt{height\_function\_sa}, PC denotes \texttt{path\_clustering}, and MLSA denotes \texttt{min\_la\_sa}. We report the number of emitters, emitter CNOTs. The CNOT count is calculated using the emission algorithm presented by \cite{li_algo}.}
\label{tab:rhg_results-ghan}
    
    \begin{tabular}{@{}lccccccc@{}}
\toprule
Dims & $|V|$ & $|E|$ & Alg. & Emit. & CNOTs \\
\midrule

$(1,1,1)$ & 18 & 24 & HC & 4 & 14 \\
$(1,1,1)$ & 18 & 24 & \cite{ghanbari_framework} & 4 & 14 \\

\hline
$(1,1,2)$ & 31 & 44 & HC & 4 & 30 \\
$(1,1,2)$ & 31 & 44 & \cite{ghanbari_framework} & 4 & 28 \\

\hline
$(1,1,3)$ & 44 & 64 & SA & 4 & 39 \\
$(1,1,3)$ & 44 & 64 & \cite{ghanbari_framework} & 4 & 42 \\

\hline
$(1,2,2)$ & 53 & 80 & HC & 7 & 50 \\
$(1,2,2)$ & 53 & 80 & \cite{ghanbari_framework} & 7 & 56 \\
\hline
$(1,2,3)$ & 75 & 116 & HC & 7 & 79 \\
$(1,2,3)$ & 75 & 116 & \cite{ghanbari_framework} & 7 & 84 \\

\hline
$(2,2,2)$ & 90 & 144 & PC & 11 & 101 \\
$(2,2,2)$ & 90 & 144 & \cite{ghanbari_framework} & 12 & 108 \\

\hline
$(1,3,3)$ & 106 & 168 & PC & 10 & 110 \\
$(1,3,3)$ & 106 & 168 & \cite{ghanbari_framework} & 10 & 126 \\

\hline
$(2,3,3)$ & 179 & 300 & HFSA & 17 & 226 \\
$(2,3,3)$ & 179 & 300 & \cite{ghanbari_framework} & 17 & 240 \\

\hline
$(3,3,3)$ & 252 & 432 & HFSA & 21 & 303 \\
$(3,3,3)$ & 252 & 432 & \cite{ghanbari_framework} & 21 & 354 \\

\bottomrule
\end{tabular}

\end{table}

\begin{figure*}
    \centering
    \includegraphics[width=1.0\linewidth]{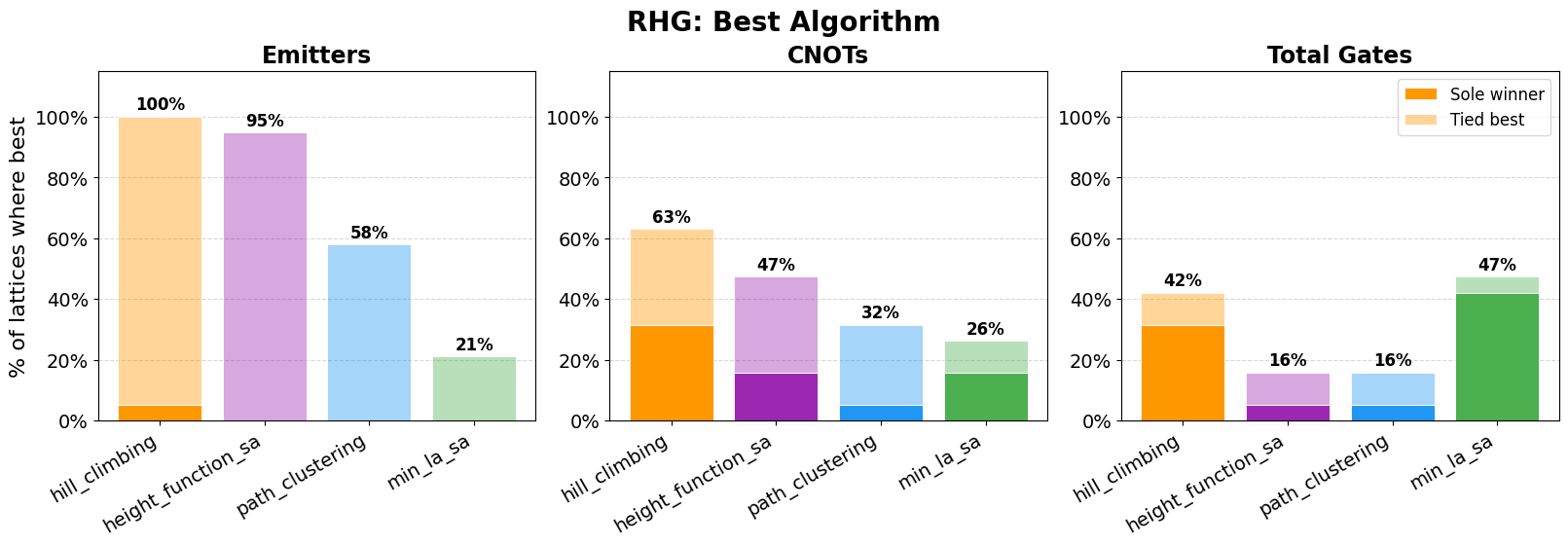}
    \caption{Histogram of the number of cases where each algorithm is the best one or a tie exists.}
    \label{fig:app-rhg_hist}
\end{figure*}

\begin{figure}
    \centering
    \includegraphics[width=0.8\linewidth]{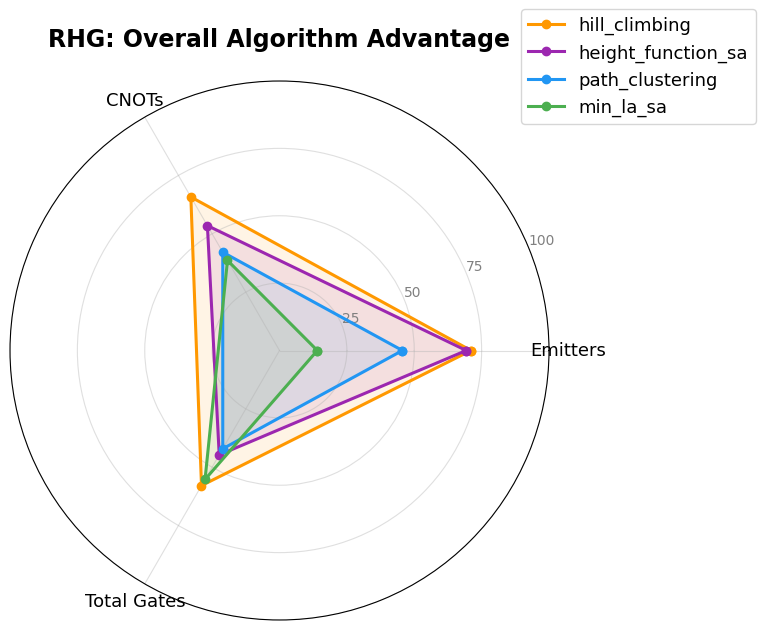}
    \caption{In the figure, we present a clear answer to which algorithm performs best for each of the three metrics we are using. We are displaying the normalized score ranging from 0 (lower) to 100 (higher) per metric.}
    \label{fig:app-rhg_radar}
\end{figure}

Let us turn our focus to comparing our algorithms with respect to the 4 metrics we are widely using, for every lattice until $(3,4,4)$. In Figure \ref{fig:app-rhg_hist}, we have a histogram where for each algorithm we compare them for each metric, denoting the times where there is a tie or one of them achieves a better result. In Figure \ref{fig:app-rhg_radar}, we have the comparison with the normalized score introduced in Section \ref{subsec:best-case-and-cnot-red}, where we are also taking into consideration how much better an algorithm is compared with the others for a specific metric. From both figures, we have that, for the number of emitters, \texttt{hill\_climbing} and \texttt{height\_function\_sa} are clearly the best, with the \texttt{path\_clustering} following and \texttt{min\_la\_sa} being the last. Similar is the situation for the emitter CNOT, but in this case, the \texttt{hill\_climbing} and \texttt{height\_function\_sa} do not have as clear an advantage as they had before. However, the situation is different for the total gate count, where the \texttt{min\_la\_sa} has a slight advantage, while the rest of the algorithms have almost the same performance. 


\subsection{Quantum Error Correcting Codes}
In this subsection, we turn our focus to graph states corresponding to quantum error correcting codes (QECC).
We investigate several interesting families of interesting Calderbank-Shor-Steane (CSS) codes.
We first consider small example codes such as the $\llbracket 7, 1, 3\rrbracket$ Steane code \cite{stean1,stean2,stean3}, the smallest QECC which is CSS, and the $\llbracket 9, 1, 3 \rrbracket$ Shor code \cite{shor1}.

We also consider quantum low-density-parity-check families of codes, namely hypergraph product (HGP) codes \cite{hgp_code1,hgp_code2} and bivariate bicyclic (BB) codes \cite{bb_code}. These codes have a larger encoding rate, i.e., the ratio between the number of logical qubits and the physical qubits, than the surface code and could reduce the resource overhead of fault-tolerant quantum information processing.

% We start with the Steane code \cite{stean1,stean2,stean3} which is a $[[7,1,3]]$ self-dual code Calderbank–Shor–Steane (CSS) code, and noteworthy is the smallest CSS code to correct a single error. It is constructed using the classical binary $[7,1,3]$ Hamming code for protecting against both $X$ and $Z$ errors. We proceed with the first quantum error correcting code, the Shor code presented in \cite{shor1}. 

% Next, we have a family of codes, the Hypergraph product (HGP) code \cite{hgp_code1,hgp_code2}, which is a member of the CSS codes whose stabilizer generator matrix is obtained from a hypergraph product of two classical linear binary codes. Finally, we have the Bivariate bicycle (BB) code \cite{bb_code}, which is among the several codes that admit time-optimal syndrome measurement circuits implementable in a two-layer architecture, which generalizes the square-lattice architecture known to be optimal for surface codes.

\paragraph{CSS code foliation}
We turn our focus on discussing various ways known in the literature that we used to extend the used codes and thus the corresponding graphs. The first one is the framework of foliation for QECC \cite{foliation1,foliation2}. The RHG-lattice construction can be generalized to arbitrary CSS codes through foliation, where the resulting graph state has fault-tolerant properties coming from the CSS code. We construct the fault-tolerant graph by connecting each alternating layer of its $X$-Tanner graph and $Z$-Tanner graph, the resulting graphs obtained by taking its parity check matrices as bi-adjacency matrices.

% By connecting repeated layers of the code's stabilizer graph, it produces an MBQC resource state that preserves the error-correction properties of the original code, while enabling fault-tolerance via local measurements and entangling operations between the layers.
\paragraph{Graph codes from stabilizer codes}

Ref. \cite{shor_and_gen_graph_rep}, presented a method for creating an equivalent graph code from any stabilizer QECC. In the encode-respecting form, each qubit stabilizer code \cite{shor_and_gen_graph_rep,gen_graph_rep2} admits a representation as a bipartite graph with $k$ input and $n$ output nodes, where the first share no mutual connections. Converting the stabilizer tableau to graphs requires $\mathcal{O}(n^2)$ time, while the inverse map from graphs to stabilizer codes necessitates $\mathcal{O}(n^3)$ both via ZX calculus. The properties of the underlying graph directly reflect properties of the code. For example, bipartite encoder-respecting graphs yield CSS codes, and graph degree controls the bounds on the code distance, stabilizer weight, and encoding-circuit depth. In the following, we present the graph families we included as well as the different ways we combined and expanded the codes.

\paragraph{Logical $\ket{0}^k$ states}
From this graph code construction, we obtain a graph with $k+n$ nodes representing the stabilizer code. The induced graph state corresponding to the subgraph  on the $n$-physical nodes corresponds to a local-Clifford equivalent graph code initialized in the $\ket{0}^k$ state.

\paragraph{Logical Bell pairs}
Using this equivalent code with graph properties, we proceed with the construction that prepares a graph-state representation of an encoded Bell-like logical state between two copies of the same code. Starting from two identical encoded graph states, the logical nodes are paired together, and we create edges between any neighbor of one logical node and any neighbor of the other logical node. We then remove the logical nodes from the final graph. The resulting graph state corresponds to a logical entangled state of the form $(\ket{0,+}_L+\ket{1,-}_L)^{\otimes}$, where the logical information is encoded nonlocally across the remaining graph. Finally, for each QECC we discussed, we have the simple isolated version. These resulting graph states have the same error correction properties as an encoded state using the same initial stabilizer code and is useful for applications in quantum communications, such as all-photonic quantum repeaters \cite{ewert2017ultrafast, murphy2026simplified}.

\begin{figure*}
    \centering
    \includegraphics[width=1.0\linewidth]{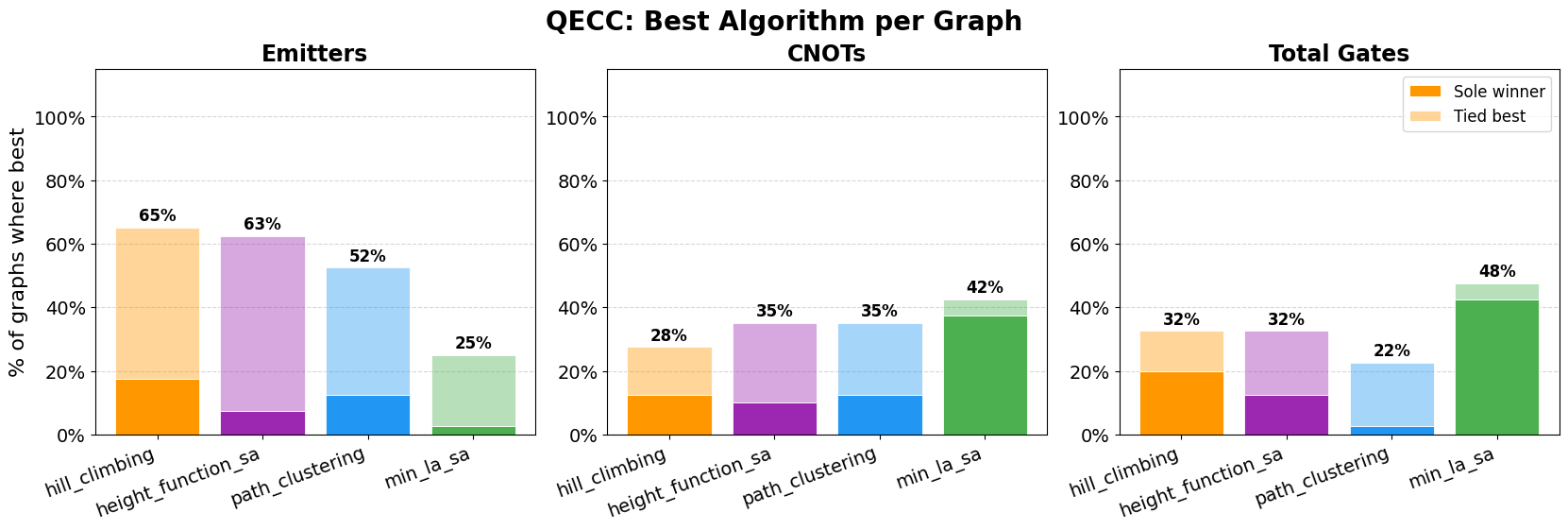}
    \caption{Histogram of the number of cases for which each algorithm is the best one or a tie exists.}
    \label{fig:qecc-hist}
\end{figure*}

\begin{figure}
    \centering
    \includegraphics[width=0.8\linewidth]{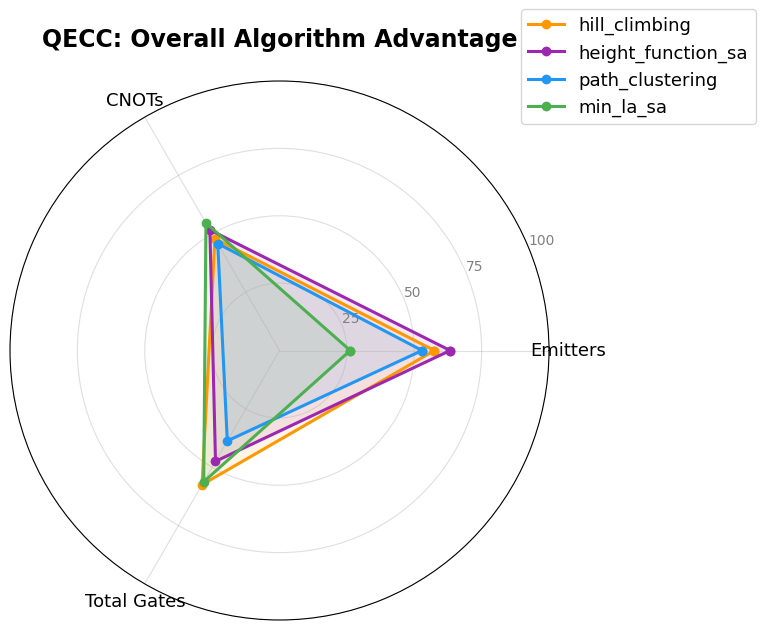}
    \caption{In the figure, we present a clear answer to which algorithm performs best for each of the three metrics we are using. We are displaying the normalized score ranging from 0 (lower) to 100 (higher) per metric.}
    \label{fig:qecc-radar}
\end{figure}

\paragraph{Set of fault-tolerant graph states}
The complete list of graphs, as well as the corresponding table with the exact simulation results, is given in the Supplementary material. In this case, we concentrated on the number of emitters, the CNOT count, and the total gate count. In this dataset, the smallest graph is the one corresponding to the Steane code with $7$ qubits, and the largest one is the HGP graph with $34$ qubits after imposing $5$ foliation layers, resulting in a graph with $490$ vertices. In Figure \ref{fig:qecc-hist}, we have a histogram with the best comparison of each algorithm for the three metrics we are considering, including cases where a tie exists. In Figure \ref{fig:qecc-radar}, we have a fair comparison of each algorithm based on the normalized score as we have already described. For the number of emitters, it seems that apart from the \texttt{min\_la\_sa}, the rest of the algorithms have similar performance. However, for the number of CNOTs and total gate count, the \texttt{min\_la\_sa} is slightly better.

With this, we conclude this subsection, where we examined various graph state families with relevance for quantum computation and, in general, quantum technologies. It became apparent that each one of the presented algorithms provided at least one advantage over one of the three metrics we used in this work. This makes it clear that all of them should be used when one is aiming to prepare a photonic graph state with quantum emitters, and then based on the corresponding results, opt for the choice that best fits any requirement. 
\section{Conclusion}\label{sec:conclusion}

In this work, we study the problem of reducing the number of quantum emitters required for deterministic photonic graph state generation. While previous works provide a general algorithm for addressing the problem, solving the problem for specific graph state families, or reducing the two-qubit gates between the emitters, the underlying assumption is that the emitter ordering is fixed. However, this step is crucial since it determines the number of emitters the generation recipe requires. Optimizing the ordering is an NP-hard problem since it is equivalent to solving the linear rank-width problem. This makes the exact optimization intractable for large graph states and motivates the development of efficient heuristic algorithms.

We introduced four heuristic algorithms for finding improved emission orderings. The first two, \texttt{hill\_climbing} and \texttt{height\_function\_sa}, directly optimize the width of the ordering, i.e., the maximum of the height function. At the beginning, they are the same since they start with the same non-random orderings based on the Fiedler vector, the reverse Cuthill-McKee algorithm, and the minimum degree. \texttt{hill\_climbing} is an algorithm based on local hill-climbing approaches, while \texttt{height\_function\_sa} uses simulated annealing as its final refinement step. Then we have the \texttt{path\_clustering} algorithm, which exploits the fact that caterpillar graphs (and equivalent graphs under local complementation) are the only ones with linear rank-width one, and then we use a divide and conquer approach for the local and global refinement. Finally, with the \texttt{min\_la\_sa} we are using a simulated annealing approach, but we optimize based not on the linear rank-width but on the minimum linear arrangement problem, showing that using such proxies can be beneficial in certain cases.

Our numerical results show that emission order optimization can substantially reduce the number of emitters compared to the random ordering case. On Erd\H{o}s-R\'enyi random graphs, the best performing algorithm achieves up to $30\%$ reduction. We also examined the number of emitter CNOTs and the total gate count. Among the tested methods, the \texttt{path\_clustering} shows the strongest overall performance across the main resource metrics on random graphs. The \texttt{hill\_climbing} and \texttt{height\_function\_sa} methods were also particularly competitive for emitter reduction and, depending on the graph family, can outperform other approaches.

Another important remark is that optimizing the emission ordering can further enhance the performance of algorithms that aim to reduce the number of two-qubit gates, an endeavor that is rather important for quantum technologies. In particular, by having our algorithms as a preprocessing step, we can have at least $20\%$ further improvement to the performance of the algorithms presented in \cite{eva_opt}. Numerical evidence presented here suggests a positive correlation between reducing the number of emitters and the emitter CNOTs, but a more detailed theoretical and algorithm analysis could lead to joint optimization methods that target both resources simultaneously.

We also tested our algorithms on several cases of graph states, ranging from rather small ones to states with more than 400 photonic qubits. More specifically, we analyzed resource states for the measurement-based quantum computation versions of Shor's original algorithm, on Raussendorf-Harrington-Goyal lattices, and various states emerging from a plethora of quantum error correction codes constructions. The results indicate that there is not always a clearly advantageous algorithm, and within the same family, one algorithm can be advantageous for one metric, but it is never advantageous for everyone.

Several research directions remain open. The first one is a detailed and probably analytical examination of the tradeoffs our metrics have. As it was also addressed in the discussion and conclusion section of \cite{ghanbari_framework}, there are cases where allowing a slight increase in the number of emitters can not only lead to a substantial decrease in the rest of the required resources but also could be more feasible in practice. Additionally, since our work addresses the literature gap of optimization algorithms for reducing the number of emitters, the next natural step is a combined framework where the algorithms that focus on one aspect are combined in a coherent theoretical and computational framework. Further, our work is a key element to connect the deterministic approach we discussed with the probabilistic fusion-based ones.

In conclusion, this work provides algorithms that assist in systematically reducing the resources required for the photonic graph state generation with quantum emitters. Furthermore, we showed that systematic methods can reduce the rest of the computational overhead. Therefore, we provide a key ingredient in the realization of photonic graph states, which have a plethora of applications for quantum computation and quantum technologies in general.

\section{Acknowledgments}
We thank Félix Berthier, Axel Daboust, and Eliott Flechtner who wrote the script for  code to graph representation, reproducing existing results as part of a student project.
PH is supported by France 2030 under the French National Research Agency award number “ANR-22-PNCQ-0002”.
KRR thanks the Quriosity team at the LTCI laboratory of Télécom Paris, and the IDIA department of Institut Polytechnique de Paris for hosting him from early February to early March of 2026, during which the core development of the presented algorithms took place (funded by ANR-22-PNCQ-0002).

\bibliography{bibliography}

@article{repeaters,
  title = {Quantum repeaters: From quantum networks to the quantum internet},
  author = {Azuma, Koji and Economou, Sophia E. and Elkouss, David and Hilaire, Paul and Jiang, Liang and Lo, Hoi-Kwong and Tzitrin, Ilan},
  journal = {Rev. Mod. Phys.},
  volume = {95},
  issue = {4},
  pages = {045006},
  numpages = {66},
  year = {2023},
  month = {Dec},
  publisher = {American Physical Society},
  doi = {10.1103/RevModPhys.95.045006},
  url = {https://link.aps.org/doi/10.1103/RevModPhys.95.045006}
}

@article{repeaters2,
  title = {Rate-distance tradeoff and resource costs for all-optical quantum repeaters},
  author = {Pant, Mihir and Krovi, Hari and Englund, Dirk and Guha, Saikat},
  journal = {Phys. Rev. A},
  volume = {95},
  issue = {1},
  pages = {012304},
  numpages = {14},
  year = {2017},
  month = {Jan},
  publisher = {American Physical Society},
  doi = {10.1103/PhysRevA.95.012304},
  url = {https://link.aps.org/doi/10.1103/PhysRevA.95.012304}
}

@article{repeaters3,
  title = {One-Way Quantum Repeater Based on Near-Deterministic Photon-Emitter Interfaces},
  author = {Borregaard, Johannes and Pichler, Hannes and Schr\"oder, Tim and Lukin, Mikhail D. and Lodahl, Peter and S\o{}rensen, Anders S.},
  journal = {Phys. Rev. X},
  volume = {10},
  issue = {2},
  pages = {021071},
  numpages = {13},
  year = {2020},
  month = {Jun},
  publisher = {American Physical Society},
  doi = {10.1103/PhysRevX.10.021071},
  url = {https://link.aps.org/doi/10.1103/PhysRevX.10.021071}
}

@article{repeaters4,
   title={Resource requirements for efficient quantum communication using all-photonic graph states generated from a few matter qubits},
   volume={5},
   ISSN={2521-327X},
   url={http://dx.doi.org/10.22331/q-2021-02-15-397},
   DOI={10.22331/q-2021-02-15-397},
   journal={Quantum},
   publisher={Verein zur Forderung des Open Access Publizierens in den Quantenwissenschaften},
   author={Hilaire, Paul and Barnes, Edwin and Economou, Sophia E.},
   year={2021},
   month=Feb, pages={397} }

@article{repeaters5,
  title = {Error-correcting entanglement swapping using a practical logical photon encoding},
  author = {Hilaire, Paul and Barnes, Edwin and Economou, Sophia E. and Grosshans, Fr\'ed\'eric},
  journal = {Phys. Rev. A},
  volume = {104},
  issue = {5},
  pages = {052623},
  numpages = {12},
  year = {2021},
  month = {Nov},
  publisher = {American Physical Society},
  doi = {10.1103/PhysRevA.104.052623},
  url = {https://link.aps.org/doi/10.1103/PhysRevA.104.052623}
}

@article{repeaters6,
   title={Performance analysis of quantum repeaters enabled by deterministically generated photonic graph states},
   volume={7},
   ISSN={2521-327X},
   url={http://dx.doi.org/10.22331/q-2023-02-16-924},
   DOI={10.22331/q-2023-02-16-924},
   journal={Quantum},
   publisher={Verein zur Forderung des Open Access Publizierens in den Quantenwissenschaften},
   author={Zhan, Yuan and Hilaire, Paul and Barnes, Edwin and Economou, Sophia E. and Sun, Shuo},
   year={2023},
   month=Feb, pages={924} }

@article{repeaters7,
  title = {Linear Optical Logical Bell State Measurements with Optimal Loss-Tolerance Threshold},
  author = {Hilaire, Paul and Castor, Yaron and Barnes, Edwin and Economou, Sophia E. and Grosshans, Fr\'ed\'eric},
  journal = {PRX Quantum},
  volume = {4},
  issue = {4},
  pages = {040322},
  numpages = {18},
  year = {2023},
  month = {Nov},
  publisher = {American Physical Society},
  doi = {10.1103/PRXQuantum.4.040322},
  url = {https://link.aps.org/doi/10.1103/PRXQuantum.4.040322}
}

@article{repeaters8,
  author  = {Niu, Daoheng and Zhang, Yuxuan and Shabani, Alireza and Shapourian, Hassan},
  title   = {All-photonic one-way quantum repeaters with measurement-based error correction},
  journal = {npj Quantum Information},
  year    = {2023},
  volume  = {9},
  number  = {1},
  pages   = {106},
  doi     = {10.1038/s41534-023-00775-9},
  url     = {https://doi.org/10.1038/s41534-023-00775-9},
  issn    = {2056-6387}
}

@article{repeaters10,
  title = {Measurement-based quantum repeaters},
  author = {Zwerger, M. and D\"ur, W. and Briegel, H. J.},
  journal = {Phys. Rev. A},
  volume = {85},
  issue = {6},
  pages = {062326},
  numpages = {11},
  year = {2012},
  month = {Jun},
  publisher = {American Physical Society},
  doi = {10.1103/PhysRevA.85.062326},
  url = {https://link.aps.org/doi/10.1103/PhysRevA.85.062326}
}

@article{qec_gen1,
  title = {Quantum-error-correcting codes using qudit graph states},
  author = {Looi, Shiang Yong and Yu, Li and Gheorghiu, Vlad and Griffiths, Robert B.},
  journal = {Phys. Rev. A},
  volume = {78},
  issue = {4},
  pages = {042303},
  numpages = {11},
  year = {2008},
  month = {Oct},
  publisher = {American Physical Society},
  doi = {10.1103/PhysRevA.78.042303},
  url = {https://link.aps.org/doi/10.1103/PhysRevA.78.042303}
}

@article{qec_gen3,
  title = {Topological graph states and quantum error-correction codes},
  author = {Liao, Pengcheng and Sanders, Barry C. and Feder, David L.},
  journal = {Phys. Rev. A},
  volume = {105},
  issue = {4},
  pages = {042418},
  numpages = {18},
  year = {2022},
  month = {Apr},
  publisher = {American Physical Society},
  doi = {10.1103/PhysRevA.105.042418},
  url = {https://link.aps.org/doi/10.1103/PhysRevA.105.042418}
}

@article{shettell_graph_metrology,
  title = {Graph states as a resource for quantum metrology},
  author = {Shettell, Nathan and Markham, Damian},
  journal = {Phys. Rev. Lett.},
  volume = {124},
  pages = {110502},
  year = {2020},
  month = {Mar},
  publisher = {American Physical Society},
  doi = {10.1103/PhysRevLett.124.110502},
  url = {https://link.aps.org/doi/10.1103/PhysRevLett.124.110502}
}

@article{sense2,
  title = {Continuous-variable graph states for quantum metrology},
  author = {Wang, Yunkai and Fang, Kejie},
  journal = {Phys. Rev. A},
  volume = {102},
  issue = {5},
  pages = {052601},
  numpages = {7},
  year = {2020},
  month = {Nov},
  publisher = {American Physical Society},
  doi = {10.1103/PhysRevA.102.052601},
  url = {https://link.aps.org/doi/10.1103/PhysRevA.102.052601}
}

@article{notwoqubit1,
  author  = {Knill, E. and Laflamme, R. and Milburn, G. J.},
  title   = {A scheme for efficient quantum computation with linear optics},
  journal = {Nature},
  year    = {2001},
  volume  = {409},
  number  = {6816},
  pages   = {46--52},
  doi     = {10.1038/35051009},
  url     = {https://doi.org/10.1038/35051009},
  issn    = {1476-4687}
}

@article{raussendorf_oneway,
  title = {A one-way quantum computer},
  author = {Raussendorf, Robert and Briegel, Hans J.},
  journal = {Phys. Rev. Lett.},
  volume = {86},
  pages = {5188--5191},
  year = {2001},
  month = {May},
  publisher = {American Physical Society},
  doi = {10.1103/PhysRevLett.86.5188},
  url = {https://link.aps.org/doi/10.1103/PhysRevLett.86.5188}
}

@article{briegel_persistent,
  title = {Persistent Entanglement in Arrays of Interacting Particles},
  author = {Briegel, Hans J. and Raussendorf, Robert},
  journal = {Phys. Rev. Lett.},
  volume = {86},
  pages = {910--913},
  year = {2001},
  month = {Jan},
  publisher = {American Physical Society},
  doi = {10.1103/PhysRevLett.86.910},
  url = {https://link.aps.org/doi/10.1103/PhysRevLett.86.910}
}

@article{hein_multiparty,
  title = {Multiparty entanglement in graph states},
  author = {Hein, M. and Eisert, J. and Briegel, H. J.},
  journal = {Phys. Rev. A},
  volume = {69},
  pages = {062311},
  year = {2004},
  month = {Jun},
  publisher = {American Physical Society},
  doi = {10.1103/PhysRevA.69.062311},
  url = {https://link.aps.org/doi/10.1103/PhysRevA.69.062311}
}

@misc{hein_graphstates_review,
  title = {Entanglement in Graph States and its Applications},
  author = {Hein, Marc and D{\"u}r, Wolfgang and Eisert, Jens and Raussendorf, Robert and Van den Nest, Maarten and Briegel, Hans J.},
  year = {2006},
  eprint = {quant-ph/0602096},
  archivePrefix = {arXiv},
  primaryClass = {quant-ph},
  url = {https://arxiv.org/abs/quant-ph/0602096}
}

@article{van_den_nest_lc,
  title = {Graphical description of the action of local Clifford transformations on graph states},
  author = {Van den Nest, Maarten and Dehaene, Jeroen and De Moor, Bart},
  journal = {Phys. Rev. A},
  volume = {69},
  pages = {022316},
  year = {2004},
  month = {Feb},
  publisher = {American Physical Society},
  doi = {10.1103/PhysRevA.69.022316},
  url = {https://link.aps.org/doi/10.1103/PhysRevA.69.022316}
}

@article{qubit_orbits,
  title = {Optimal preparation of graph states},
  author = {Cabello, Ad\'an and Danielsen, Lars Eirik and L\'opez-Tarrida, Antonio J. and Portillo, Jos\'e R.},
  journal = {Phys. Rev. A},
  volume = {83},
  issue = {4},
  pages = {042314},
  numpages = {7},
  year = {2011},
  month = {Apr},
  publisher = {American Physical Society},
  doi = {10.1103/PhysRevA.83.042314},
  url = {https://link.aps.org/doi/10.1103/PhysRevA.83.042314}
}

@article{mbqc_flow,
  doi = {10.1088/1367-2630/9/8/250},
  url = {https://doi.org/10.1088/1367-2630/9/8/250},
  year = {2007},
  month = {aug},
  volume = {9},
  number = {8},
  pages = {250},
  author = {Browne, Daniel E. and Kashefi, Elham and Mhalla, Mehdi and Perdrix, Simon},
  title = {Generalized flow and determinism in measurement-based quantum computation},
  journal = {New Journal of Physics}
}

@article{oum_rankwidth,
  title = {Rank-width and vertex-minors},
  author = {Oum, Sang-il and Seymour, Paul},
  journal = {Journal of Combinatorial Theory, Series B},
  volume = {95},
  number = {1},
  pages = {79--100},
  year = {2005},
  doi = {10.1016/j.jctb.2005.03.003},
  url = {https://doi.org/10.1016/j.jctb.2005.03.003}
}

@article{Oum_rank_width_rev,
   title={Rank-width: Algorithmic and structural results},
   volume={231},
   ISSN={0166-218X},
   url={http://dx.doi.org/10.1016/j.dam.2016.08.006},
   DOI={10.1016/j.dam.2016.08.006},
   journal={Discrete Applied Mathematics},
   publisher={Elsevier BV},
   author={Oum, Sang-il},
   year={2017},
   month=Nov, pages={15–24} }

@article{oum_linear_rankwidth,
  title = {Excluded vertex-minors for graphs of linear rank-width at most k},
  author = {Jeong, Jisu and Kwon, O-joung and Oum, Sang-il},
  journal = {European Journal of Combinatorics},
  volume = {41},
  pages = {242--257},
  year = {2014},
  doi = {10.1016/j.ejc.2014.05.003},
  url = {https://doi.org/10.1016/j.ejc.2014.05.003}
}

@article{rank_width_nphard,
author = {Oum, Sang-Il},
title = {Approximating rank-width and clique-width quickly},
year = {2008},
issue_date = {November 2008},
publisher = {Association for Computing Machinery},
address = {New York, NY, USA},
volume = {5},
number = {1},
issn = {1549-6325},
url = {https://doi.org/10.1145/1435375.1435385},
doi = {10.1145/1435375.1435385},
journal = {ACM Trans. Algorithms},
month = dec,
articleno = {10},
numpages = {20}
}

@article{fiedler_algebraic_connectivity,
  title = {Algebraic connectivity of graphs},
  author = {Fiedler, Miroslav},
  journal = {Czechoslovak Mathematical Journal},
  volume = {23},
  number = {2},
  pages = {298--305},
  year = {1973},
  doi = {10.21136/CMJ.1973.101168},
  url = {https://doi.org/10.21136/CMJ.1973.101168}
}

@inproceedings{cuthill_mckee,
  title = {Reducing the bandwidth of sparse symmetric matrices},
  author = {Cuthill, Elizabeth and McKee, James},
  booktitle = {Proceedings of the 1969 24th National Conference},
  pages = {157--172},
  year = {1969},
  doi = {10.1145/800195.805928},
  url = {https://doi.org/10.1145/800195.805928}
}

@article{kirkpatrick_sa,
  title = {Optimization by simulated annealing},
  author = {Kirkpatrick, S. and Gelatt, C. D. and Vecchi, M. P.},
  journal = {Science},
  volume = {220},
  number = {4598},
  pages = {671--680},
  year = {1983},
  doi = {10.1126/science.220.4598.671},
  url = {https://doi.org/10.1126/science.220.4598.671}
}

@book{van1987simulated,
  title={Simulated Annealing: Theory and Applications},
  author={van Laarhoven, P.J. and Aarts, E.H.},
  isbn={9789027725134},
  lccn={lc87009666},
  series={Mathematics and Its Applications},
  url={https://books.google.de/books?id=-IgUab6Dp_IC},
  year={1987},
  publisher={Springer Netherlands}
}

@phdthesis{george1971computer,
  author       = {George, Alan},
  title        = {Computer Implementation of the Finite Element Method},
  school       = {Stanford University},
  department   = {Computer Science Department},
  year         = {1971},
  type         = {Ph.D. dissertation},
  number       = {STAN-CS-71-208}
}

@article{cater_are_1,
  title={A Note on Graphs of Linear Rank-Width 1},
  author={Binh-Minh Bui-Xuan and Mamadou Moustapha Kant{\'e} and Vincent Limouzy},
  journal={ArXiv},
  year={2013},
  volume={abs/1306.1345},
  url={https://api.semanticscholar.org/CorpusID:12655025}
}

@article{caters_def,
title = {The number of caterpillars},
journal = {Discrete Mathematics},
volume = {6},
number = {4},
pages = {359-365},
year = {1973},
issn = {0012-365X},
doi = {https://doi.org/10.1016/0012-365X(73)90067-8},
url = {https://www.sciencedirect.com/science/article/pii/0012365X73900678},
author = {Frank Harary and Allen J. Schwenk}
}

@article{Ghanbari_opt,
  title = {Optimization of deterministic photonic-graph-state generation via local operations},
  author = {Ghanbari, Sobhan and Lin, Jie and MacLellan, Benjamin and Robichaud, Luc and Roztocki, Piotr and Lo, Hoi-Kwong},
  journal = {Phys. Rev. A},
  volume = {110},
  issue = {5},
  pages = {052605},
  numpages = {11},
  year = {2024},
  month = {Nov},
  publisher = {American Physical Society},
  doi = {10.1103/PhysRevA.110.052605},
  url = {https://link.aps.org/doi/10.1103/PhysRevA.110.052605}
}

@misc{ghanbari_framework,
  title = {Cost-aware Photonic Graph State Generation: A Graphical Framework},
  author = {Ghanbari, Sobhan and Lo, Hoi-Kwong},
  year = {2025},
  eprint = {2509.22777},
  archivePrefix = {arXiv},
  primaryClass = {quant-ph},
  url = {https://arxiv.org/abs/2509.22777}
}

@article{eva_opt,
  title = {Optimization complexity and resource minimization of emitter-based photonic graph state generation protocols},
  author = {Takou, Evangelia and Barnes, Edwin and Economou, Sophia E.},
  journal = {npj Quantum Information},
  volume = {11},
  number = {1},
  pages = {108},
  year = {2025},
  doi = {10.1038/s41534-025-01056-3},
  url = {https://doi.org/10.1038/s41534-025-01056-3}
}

@article{li_algo,
  title = {Photonic resource state generation from a minimal number of quantum emitters},
  author = {Li, Bikun and Economou, Sophia E. and Barnes, Edwin},
  journal = {npj Quantum Information},
  volume = {8},
  number = {1},
  pages = {11},
  year = {2022},
  doi = {10.1038/s41534-022-00522-6},
  url = {https://doi.org/10.1038/s41534-022-00522-6}
}

@article{lindner_cluster_strings,
  title = {Proposal for pulsed on-demand sources of photonic cluster state strings},
  author = {Lindner, Netanel H. and Rudolph, Terry},
  journal = {Phys. Rev. Lett.},
  volume = {103},
  pages = {113602},
  year = {2009},
  month = {Sep},
  publisher = {American Physical Society},
  doi = {10.1103/PhysRevLett.103.113602},
  url = {https://link.aps.org/doi/10.1103/PhysRevLett.103.113602}
}

@article{gimeno_large_cluster,
  title = {Deterministic generation of large-scale entangled photonic cluster state from interacting solid state emitters},
  author = {Gimeno-Segovia, Mercedes and Rudolph, Terry and Economou, Sophia E.},
  journal = {Phys. Rev. Lett.},
  volume = {123},
  pages = {070501},
  year = {2019},
  month = {Aug},
  publisher = {American Physical Society},
  doi = {10.1103/PhysRevLett.123.070501},
  url = {https://link.aps.org/doi/10.1103/PhysRevLett.123.070501}
}

@article{russo_graph_generation,
  title = {Photonic graph state generation from quantum dots and color centers for quantum communications},
  author = {Russo, Antonio and Barnes, Edwin and Economou, Sophia E.},
  journal = {Phys. Rev. B},
  volume = {98},
  pages = {085303},
  year = {2018},
  month = {Aug},
  publisher = {American Physical Society},
  doi = {10.1103/PhysRevB.98.085303},
  url = {https://link.aps.org/doi/10.1103/PhysRevB.98.085303}
}

@article{schwartz_cluster,
  title = {Deterministic generation of a cluster state of entangled photons},
  author = {Schwartz, I. and Cogan, D. and Schmidgall, E. R. and Don, Y. and Gantz, L. and Kenneth, O. and Lindner, N. H. and Gershoni, D.},
  journal = {Science},
  volume = {354},
  number = {6311},
  pages = {434--437},
  year = {2016},
  doi = {10.1126/science.aah4758},
  url = {https://doi.org/10.1126/science.aah4758}
}

@article{huet_reconfigurable,
  title = {Deterministic and reconfigurable graph state generation with a single solid-state quantum emitter},
  author = {Huet, H. and Vitale, N. and Caillet, X. and Krebs, O. and Lema\^itre, A. and Somaschi, N. and Senellart, P. and Ollivier, H.},
  journal = {Nature Communications},
  volume = {16},
  pages = {4270},
  year = {2025},
  doi = {10.1038/s41467-025-59693-3},
  url = {https://doi.org/10.1038/s41467-025-59693-3}
}

@article{kaur_loss_aware,
  title = {Resource-efficient loss-aware photonic-graph-state preparation using atomic emitters},
  author = {Kaur, Eneet and Patil, Ashlesha and Guha, Saikat},
  journal = {Phys. Rev. A},
  volume = {112},
  issue = {6},
  pages = {062608},
  numpages = {22},
  year = {2025},
  month = {Dec},
  publisher = {American Physical Society},
  doi = {10.1103/2cbn-448l},
  url = {https://link.aps.org/doi/10.1103/2cbn-448l}
}

@article{lin_graphiq,
  title = {GraphiQ: Quantum circuit design for photonic graph states},
  author = {Lin, Jie and MacLellan, Benjamin and Ghanbari, Sobhan and Belleville, Julie and Tran, Khuong and Robichaud, Luc and Melko, Roger G. and Lo, Hoi-Kwong and Roztocki, Piotr},
  journal = {Quantum},
  volume = {8},
  pages = {1453},
  year = {2024},
  doi = {10.22331/q-2024-08-28-1453},
  url = {https://doi.org/10.22331/q-2024-08-28-1453}
}

@article{cashman2025finding,
  title = {Finding trail covers: near-optimal decompositions of graph states as linear fusion networks},
  author = {Cashman, William and de Felice, Giovanni and Kissinger, Aleks},
  journal = {arXiv preprint arXiv:2508.18375},
  year = {2025},
  eprint = {2508.18375},
  archivePrefix = {arXiv},
  primaryClass = {quant-ph},
  url = {https://arxiv.org/abs/2508.18375}
}

@article{shorsalgo,
  author = {Shor, Peter W.},
  title = {Polynomial-Time Algorithms for Prime Factorization and Discrete Logarithms on a Quantum Computer},
  journal = {SIAM Journal on Computing},
  volume = {26},
  number = {5},
  pages = {1484--1509},
  year = {1997},
  doi = {10.1137/S0097539795293172},
  url = {https://doi.org/10.1137/S0097539795293172}
}

@software{graphix1,
  author = {Uldemolins, Mateo and Fukushima, Masato and Graham, Emlyn and Nair, Pranav and Sasaki, Daichi and Shiratani, Sora and Watanabe, Yuki and Martinez, Thierry and Garnier, Maxime and Sunami, Shinichi},
  title = {Graphix},
  month = feb,
  year = {2026},
  publisher = {Zenodo},
  version = {v0.3.4},
  doi = {10.5281/zenodo.18503266},
  url = {https://doi.org/10.5281/zenodo.18503266}
}

@misc{graphix2,
  title = {Graphix: optimizing and simulating measurement-based quantum computation on local-Clifford decorated graph},
  author = {Sunami, Shinichi and Fukushima, Masato},
  year = {2022},
  eprint = {2212.11975},
  archivePrefix = {arXiv},
  primaryClass = {quant-ph},
  url = {https://arxiv.org/abs/2212.11975}
}

@article{rhg_1,
  title = {A fault-tolerant one-way quantum computer},
  journal = {Annals of Physics},
  volume = {321},
  number = {9},
  pages = {2242--2270},
  year = {2006},
  issn = {0003-4916},
  doi = {10.1016/j.aop.2006.01.012},
  url = {https://doi.org/10.1016/j.aop.2006.01.012},
  author = {Raussendorf, R. and Harrington, J. and Goyal, K.}
}

@article{rhg_2,
  doi = {10.1088/1367-2630/9/6/199},
  url = {https://doi.org/10.1088/1367-2630/9/6/199},
  year = {2007},
  month = {jun},
  volume = {9},
  number = {6},
  pages = {199},
  author = {Raussendorf, R. and Harrington, J. and Goyal, K.},
  title = {Topological fault-tolerance in cluster state quantum computation},
  journal = {New Journal of Physics}
}

@article{rhg_3,
  doi = {10.22331/q-2023-12-20-1212},
  url = {https://doi.org/10.22331/q-2023-12-20-1212},
  title = {Graph-theoretical optimization of fusion-based graph state generation},
  author = {Lee, Seok-Hyung and Jeong, Hyunseok},
  journal = {Quantum},
  issn = {2521-327X},
  publisher = {Verein zur F{"o}rderung des Open Access Publizierens in den Quantenwissenschaften},
  volume = {7},
  pages = {1212},
  month = dec,
  year = {2023}
}

@article{fbqc,
  author = {Bartolucci, Sara and Birchall, Patrick and Bomb{\'i}n, Hector and Cable, Hugo and Dawson, Chris and Gimeno-Segovia, Mercedes and Johnston, Eric and Kieling, Konrad and Nickerson, Naomi and Pant, Mihir and Pastawski, Fernando and Rudolph, Terry and Sparrow, Chris},
  title = {Fusion-based quantum computation},
  journal = {Nature Communications},
  year = {2023},
  volume = {14},
  number = {1},
  pages = {912},
  doi = {10.1038/s41467-023-36493-1},
  url = {https://doi.org/10.1038/s41467-023-36493-1},
  issn = {2041-1723}
}

@article{Hilaire_Near-deterministic,
   title={Near-deterministic hybrid generation of arbitrary photonic graph states using a single quantum emitter and linear optics},
   volume={7},
   ISSN={2521-327X},
   url={http://dx.doi.org/10.22331/q-2023-04-27-992},
   DOI={10.22331/q-2023-04-27-992},
   journal={Quantum},
   publisher={Verein zur Forderung des Open Access Publizierens in den Quantenwissenschaften},
   author={Hilaire, Paul and Vidro, Leonid and Eisenberg, Hagai S. and Economou, Sophia E.},
   year={2023},
   month=apr, pages={992} }

@article{azuma_all_photonic,
  title = {All-photonic quantum repeaters},
  author = {Azuma, Koji and Tamaki, Kiyoshi and Lo, Hoi-Kwong},
  journal = {Nature Communications},
  volume = {6},
  pages = {6787},
  year = {2015},
  doi = {10.1038/ncomms7787},
  url = {https://doi.org/10.1038/ncomms7787}
}

@article{buterakos,
  title = {Deterministic Generation of All-Photonic Quantum Repeaters from Solid-State Emitters},
  author = {Buterakos, Donovan and Barnes, Edwin and Economou, Sophia E.},
  journal = {Phys. Rev. X},
  volume = {7},
  issue = {4},
  publisher = {American Physical Society},
  pages = {041023},
  numpages = {10},
  year = {2017},
  month = {Oct},
  doi = {10.1103/PhysRevX.7.041023},
  url = {https://link.aps.org/doi/10.1103/PhysRevX.7.041023}
}

@article{bell_graph_qec,
  title = {Experimental demonstration of a graph state quantum error-correction code},
  author = {Bell, B. A. and Herrera-Mart\'i, D. A. and Tame, M. S. and Markham, D. and Wadsworth, W. J. and Rarity, J. G.},
  journal = {Nature Communications},
  volume = {5},
  pages = {3658},
  year = {2014},
  doi = {10.1038/ncomms4658},
  url = {https://doi.org/10.1038/ncomms4658}
}

@article{browne_rudolph,
  title = {Resource-efficient linear optical quantum computation},
  author = {Browne, Daniel E. and Rudolph, Terry},
  journal = {Phys. Rev. Lett.},
  volume = {95},
  pages = {010501},
  year = {2005},
  month = {Jun},
  publisher = {American Physical Society},
  doi = {10.1103/PhysRevLett.95.010501},
  url = {https://link.aps.org/doi/10.1103/PhysRevLett.95.010501}
}

@article{dots1,
  title = {Optically Generated 2-Dimensional Photonic Cluster State from Coupled Quantum Dots},
  author = {Economou, Sophia E. and Lindner, Netanel and Rudolph, Terry},
  journal = {Phys. Rev. Lett.},
  volume = {105},
  issue = {9},
  pages = {093601},
  numpages = {4},
  year = {2010},
  month = {Aug},
  publisher = {American Physical Society},
  doi = {10.1103/PhysRevLett.105.093601},
  url = {https://link.aps.org/doi/10.1103/PhysRevLett.105.093601}
}

@article{dots3,
  title = {Deterministic Generation of Entangled Photonic Cluster States from Quantum Dot Molecules},
  author = {Vezvaee, Arian and Hilaire, Paul and Doty, Matthew F. and Economou, Sophia E.},
  journal = {Phys. Rev. Appl.},
  volume = {18},
  issue = {6},
  pages = {L061003},
  numpages = {6},
  year = {2022},
  month = {Dec},
  publisher = {American Physical Society},
  doi = {10.1103/PhysRevApplied.18.L061003},
  url = {https://link.aps.org/doi/10.1103/PhysRevApplied.18.L061003}
}

@article{color1,
  title = {Generation of entangled photon strings using NV centers in diamond},
  author = {Rao, D. D. Bhaktavatsala and Yang, Sen and Wrachtrup, J\"org},
  journal = {Phys. Rev. B},
  volume = {92},
  issue = {8},
  pages = {081301(R)},
  numpages = {4},
  year = {2015},
  month = {Aug},
  publisher = {American Physical Society},
  doi = {10.1103/PhysRevB.92.081301},
  url = {https://link.aps.org/doi/10.1103/PhysRevB.92.081301}
}

@article{color2,
  author  = {Vasconcelos, Rui and Reisenbauer, Sarah and Salter, Cameron and Wachter, Georg and Wirtitsch, Daniel and Schmiedmayer, J{\"o}rg and Walther, Philip and Trupke, Michael},
  title   = {Scalable spin--photon entanglement by time-to-polarization conversion},
  journal = {npj Quantum Information},
  year    = {2020},
  volume  = {6},
  number  = {1},
  pages   = {9},
  doi     = {10.1038/s41534-019-0236-x},
  url     = {https://doi.org/10.1038/s41534-019-0236-x},
  issn    = {2056-6387}
}

@article{color3,
   title={Multidimensional cluster states using a single spin-photon interface coupled strongly to an intrinsic nuclear register},
   volume={5},
   ISSN={2521-327X},
   url={http://dx.doi.org/10.22331/q-2021-10-19-565},
   DOI={10.22331/q-2021-10-19-565},
   journal={Quantum},
   publisher={Verein zur Forderung des Open Access Publizierens in den Quantenwissenschaften},
   author={Michaels, Cathryn P. and Arjona Martínez, Jesús and Debroux, Romain and Parker, Ryan A. and Stramma, Alexander M. and Huber, Luca I. and Purser, Carola M. and Atatüre, Mete and Gangloff, Dorian A.},
   year={2021},
   month=Oct, pages={565} }

@article{atom1,
   title={Efficient generation of entangled multiphoton graph states from a single atom},
   volume={608},
   ISSN={1476-4687},
   url={http://dx.doi.org/10.1038/s41586-022-04987-5},
   DOI={10.1038/s41586-022-04987-5},
   number={7924},
   journal={Nature},
   publisher={Springer Science and Business Media LLC},
   author={Thomas, Philip and Ruscio, Leonardo and Morin, Olivier and Rempe, Gerhard},
   year={2022},
   month=Aug, pages={677–681} }

@article{atom2,
   title={Fusion of deterministically generated photonic graph states},
   volume={629},
   ISSN={1476-4687},
   url={http://dx.doi.org/10.1038/s41586-024-07357-5},
   DOI={10.1038/s41586-024-07357-5},
   number={8012},
   journal={Nature},
   publisher={Springer Science and Business Media LLC},
   author={Thomas, Philip and Ruscio, Leonardo and Morin, Olivier and Rempe, Gerhard},
   year={2024},
   month=May, pages={567–572} }

@article{stean1,
  title = {Error Correcting Codes in Quantum Theory},
  author = {Steane, A. M.},
  journal = {Phys. Rev. Lett.},
  volume = {77},
  issue = {5},
  pages = {793--797},
  year = {1996},
  month = {Jul},
  publisher = {American Physical Society},
  doi = {10.1103/PhysRevLett.77.793},
  url = {https://link.aps.org/doi/10.1103/PhysRevLett.77.793}
}

@article{stean2,
  title = {Multiple-particle interference and quantum error correction},
  author = {Steane, Andrew M.},
  volume = {452},
  number = {1954},
  journal = {Proceedings of the Royal Society of London. Series A: Mathematical, Physical and Engineering Sciences},
  publisher = {The Royal Society},
  year = {1996},
  month = {Nov},
  pages = {2551--2577},
  doi = {10.1098/rspa.1996.0136},
  url = {https://doi.org/10.1098/rspa.1996.0136}
}

@article{stean3,
  title = {Encoding one logical qubit into six physical qubits},
  volume = {78},
  url = {https://doi.org/10.1103/PhysRevA.78.012337},
  doi = {10.1103/PhysRevA.78.012337},
  number = {1},
  pages = {012337},
  journal = {Phys. Rev. A},
  publisher = {American Physical Society},
  author = {Shaw, Bilal and Wilde, Mark M. and Oreshkov, Ognyan and Kremsky, Isaac and Lidar, Daniel A.},
  year = {2008},
  month = {Jul}
}

@article{shor1,
  author = {Shor, Peter W.},
  title = {Scheme for reducing decoherence in quantum computer memory},
  doi = {10.1103/PhysRevA.52.R2493},
  journal = {Phys. Rev. A},
  volume = {52},
  number = {4},
  pages = {R2493--R2496},
  year = {1995},
  url = {https://link.aps.org/doi/10.1103/PhysRevA.52.R2493}
}

@article{shor_and_gen_graph_rep,
  title = {Universal Graph Representation of Stabilizer Codes},
  volume = {6},
  url = {https://doi.org/10.1103/1gjs-2rhx},
  doi = {10.1103/1gjs-2rhx},
  number = {4},
  journal = {PRX Quantum},
  publisher = {American Physical Society},
  author = {Khesin, Andrey Boris and Lu, Jonathan Z. and Shor, Peter W.},
  year = {2025},
  month = {Nov}
}

@article{gen_graph_rep2,
  title = {Improved graph formalism for quantum circuit simulation},
  volume = {105},
  url = {https://doi.org/10.1103/PhysRevA.105.022432},
  doi = {10.1103/PhysRevA.105.022432},
  number = {2},
  pages = {022432},
  journal = {Phys. Rev. A},
  publisher = {American Physical Society},
  author = {Hu, Alexander Tianlin and Khesin, Andrey Boris},
  year = {2022},
  month = {Feb}
}

@article{foliation1,
  title = {Foliated Quantum Error-Correcting Codes},
  volume = {117},
  url = {https://doi.org/10.1103/PhysRevLett.117.070501},
  doi = {10.1103/PhysRevLett.117.070501},
  number = {7},
  pages = {070501},
  journal = {Phys. Rev. Lett.},
  publisher = {American Physical Society},
  author = {Bolt, A. and Duclos-Cianci, G. and Poulin, D. and Stace, T. M.},
  year = {2016},
  month = {Aug}
}

@article{foliation2,
  title = {Universal fault-tolerant measurement-based quantum computation},
  volume = {2},
  url = {https://doi.org/10.1103/PhysRevResearch.2.033305},
  doi = {10.1103/PhysRevResearch.2.033305},
  number = {3},
  pages = {033305},
  journal = {Physical Review Research},
  publisher = {American Physical Society},
  author = {Brown, Benjamin J. and Roberts, Sam},
  year = {2020},
  month = {Aug}
}

@article{hgp_code1,
  author = {Tillich, Jean-Pierre and Z\'emor, Gilles},
  journal = {IEEE Transactions on Information Theory},
  title = {Quantum LDPC Codes With Positive Rate and Minimum Distance Proportional to the Square Root of the Blocklength},
  year = {2014},
  volume = {60},
  number = {2},
  pages = {1193--1202},
  doi = {10.1109/TIT.2013.2292061}
}

@inproceedings{hgp_code2,
  author = {Kovalev, Alexey A. and Pryadko, Leonid P.},
  booktitle = {2012 IEEE International Symposium on Information Theory Proceedings},
  title = {Improved quantum hypergraph-product LDPC codes},
  year = {2012},
  pages = {348--352},
  doi = {10.1109/ISIT.2012.6284206}
}

@article{bb_code,
  author = {Bravyi, Sergey and Cross, Andrew W. and Gambetta, Jay M. and Maslov, Dmitri and Rall, Patrick and Yoder, Theodore J.},
  title = {High-threshold and low-overhead fault-tolerant quantum memory},
  journal = {Nature},
  year = {2024},
  volume = {627},
  number = {8005},
  pages = {778--782},
  doi = {10.1038/s41586-024-07107-7},
  url = {https://doi.org/10.1038/s41586-024-07107-7},
  issn = {1476-4687}
}

@inproceedings{Cohen_Optimal_LA,
author={Cohen, Johanne and Fomin, Fedor and Heggernes, Pinar and Kratsch, Dieter and Kucherov, Gregory},
title={Optimal Linear Arrangement of Interval Graphs},
booktitle={Mathematical Foundations of Computer Science},
year={2006},
publisher={Springer Berlin Heidelberg},
doi = {https://doi.org/10.1007/11821069_24},
pages={267--279}
}

@inproceedings{Koren_scale_algo_LA, 
author = {Koren, Yehuda and Harel, David}, 
title = {A Multi-scale Algorithm for the Linear Arrangement Problem}, 
year = {2002}, 
isbn = {3540003312}, 
publisher = {Springer-Verlag}, 
address = {Berlin, Heidelberg}, 
booktitle = {Revised Papers from the 28th International Workshop on Graph-Theoretic Concepts in Computer Science}, 
pages = {296–309},
doi = {https://doi.org/10.1007/3-540-36379-3_26},
numpages = {14}, 
series = {WG '02} }

@inproceedings{Garey_minLA_NP-complete,
author = {Garey, M. R. and Johnson, D. S. and Stockmeyer, L.},
title = {Some simplified NP-complete problems}, 
year = {1974},
isbn = {9781450374231}, 
publisher = {Association for Computing Machinery}, 
url = {https://doi.org/10.1145/800119.803884}, 
doi = {10.1145/800119.803884}, booktitle = {Proceedings of the Sixth Annual ACM Symposium on Theory of Computing}, 
pages = {47–63}, 
numpages = {17} }

@misc{github,
  author       = {Revis, Konstantinos-Rafail},
  title        = {photonic\_state\_generation},
  year         = {2026},
  howpublished = {\url{https://github.com/reviskostis/photonic_state_generation}},
  note         = {GitHub repository}
}

@misc{graph_theory,
      title={An introduction to graph theory}, 
      author={Darij Grinberg},
      year={2025},
      eprint={2308.04512},
      archivePrefix={arXiv},
      primaryClass={math.HO},
      url={https://arxiv.org/abs/2308.04512}, 
}

@article{fpt_rank_width,
  author={Jeong, Jisu and Kim, Eun Jung and Oum, Sang-il},
  journal={IEEE Transactions on Information Theory}, 
  title={The “Art of Trellis Decoding” Is Fixed-Parameter Tractable}, 
  year={2017},
  volume={63},
  number={11},
  pages={7178-7205},
  doi={10.1109/TIT.2017.2740283}}

@article{schmidt_measure,
  title = {Schmidt measure as a tool for quantifying multiparticle entanglement},
  author = {Eisert, Jens and Briegel, Hans J.},
  journal = {Phys. Rev. A},
  volume = {64},
  issue = {2},
  pages = {022306},
  numpages = {4},
  year = {2001},
  month = {Jul},
  publisher = {American Physical Society},
  doi = {10.1103/PhysRevA.64.022306},
  url = {https://link.aps.org/doi/10.1103/PhysRevA.64.022306}
}

@article{BOUCHET199375,
title = {Recognizing locally equivalent graphs},
journal = {Discrete Mathematics},
volume = {114},
number = {1},
pages = {75-86},
year = {1993},
issn = {0012-365X},
doi = {https://doi.org/10.1016/0012-365X(93)90357-Y},
url = {https://www.sciencedirect.com/science/article/pii/0012365X9390357Y},
author = {André Bouchet}
}

@article{lc_minimization,
  doi = {10.22331/q-2026-02-09-2001},
  url = {https://doi.org/10.22331/q-2026-02-09-2001},
  title = {Minimising the number of edges in {LC}-equivalent graph states},
  author = {Sharma, Hemant and Goodenough, Kenneth and Borregaard, Johannes and Rozp{\k{e}}dek, Filip and Helsen, Jonas},
  journal = {{Quantum}},
  issn = {2521-327X},
  publisher = {{Verein zur F{\"{o}}rderung des Open Access Publizierens in den Quantenwissenschaften}},
  volume = {10},
  pages = {2001},
  year = {2026}
}

@article{Adcock_2020,
   title={Mapping graph state orbits under local complementation},
   volume={4},
   ISSN={2521-327X},
   url={http://dx.doi.org/10.22331/q-2020-08-07-305},
   DOI={10.22331/q-2020-08-07-305},
   journal={Quantum},
   publisher={Verein zur Forderung des Open Access Publizierens in den Quantenwissenschaften},
   author={Adcock, Jeremy C. and Morley-Short, Sam and Dahlberg, Axel and Silverstone, Joshua W.},
   year={2020},
   month=Aug, pages={305} }

@misc{qutrit_orbits,
      title={Orbit classification and analysis of qutrit graph states under local complementation and local scaling}, 
      author={Konstantinos-Rafail Revis and Hrachya Zakaryan and Zahra Raissi},
      year={2025},
      eprint={2506.05478},
      archivePrefix={arXiv},
      primaryClass={quant-ph},
      url={https://arxiv.org/abs/2506.05478}, 
}

@misc{metropolis,
      title={The Metropolis-Hastings algorithm}, 
      author={Christian P. Robert},
      year={2016},
      eprint={1504.01896},
      archivePrefix={arXiv},
      primaryClass={stat.CO},
      url={https://arxiv.org/abs/1504.01896}, 
}

@book{path_npcomplete,
  title={Combinatorial Optimization: Polyhedra and Efficiency},
  author={Schrijver, A.},
  number={Bd. 1},
  isbn={9783540443896},
  lccn={2002036693},
  series={Algorithms and Combinatorics},
  url={https://books.google.de/books?id=mqGeSQ6dJycC},
  year={2003},
  publisher={Springer}
}

@article{tsp_nphard,
title = {A survey on the Traveling Salesman Problem and its variants in a warehousing context},
journal = {European Journal of Operational Research},
volume = {322},
number = {1},
pages = {1-14},
year = {2025},
issn = {0377-2217},
doi = {https://doi.org/10.1016/j.ejor.2024.04.014},
url = {https://www.sciencedirect.com/science/article/pii/S0377221724002959},
author = {Stefan Bock and Stefan Bomsdorf and Nils Boysen and Michael Schneider}
}

@Inbook{linear_layout,
author="Pardo, Eduardo G.
and Mart{\'i}, Rafael
and Duarte, Abraham",
editor="Mart{\'i}, Rafael
and Pardalos, Panos M. 
and Resende, Mauricio G. C.",
title="Linear Layout Problems",
bookTitle="Handbook of Heuristics",
year="2018",
publisher="Springer International Publishing",
address="Cham",
pages="1025--1049",
isbn="978-3-319-07124-4",
doi="10.1007/978-3-319-07124-4_45",
url="https://doi.org/10.1007/978-3-319-07124-4_45"
}

@article{bell2023optimizing,
  title = {Optimizing Graph Codes for Measurement-Based Loss Tolerance},
  author = {Bell, Thomas J. and Pettersson, Love A. and Paesani, Stefano},
  journal = {PRX Quantum},
  volume = {4},
  issue = {2},
  pages = {020328},
  numpages = {20},
  year = {2023},
  month = {May},
  publisher = {American Physical Society},
  doi = {10.1103/PRXQuantum.4.020328},
  url = {https://link.aps.org/doi/10.1103/PRXQuantum.4.020328}
}

@article{varnava2006loss,
  title = {Loss Tolerance in One-Way Quantum Computation via Counterfactual Error Correction},
  author = {Varnava, Michael and Browne, Daniel E. and Rudolph, Terry},
  journal = {Phys. Rev. Lett.},
  volume = {97},
  issue = {12},
  pages = {120501},
  numpages = {4},
  year = {2006},
  month = {Sep},
  publisher = {American Physical Society},
  doi = {10.1103/PhysRevLett.97.120501},
  url = {https://link.aps.org/doi/10.1103/PhysRevLett.97.120501}
}

@article{coste2023high,
  author = {Coste, N. and Fioretto, D. A. and Belabas, N. and Wein, S. C. and Hilaire, P. and Frantzeskakis, R. and Gundin, M. and Goes, B. and Somaschi, N. and Morassi, M. and Lemaître, A. and Sagnes, I. and Harouri, A. and Economou, S. E. and Auffeves, A. and Krebs, O. and Lanco, L. and Senellart, P.},
  title = {High-rate entanglement between a semiconductor spin and indistinguishable photons},
  journal = {Nature Photonics},
  year = {2023},
  volume = {17},
  number = {7},
  pages = {582--587},
  doi = {10.1038/s41566-023-01186-0},
  issn = {1749-4893},
}

@article{cogan2023deterministic,
  author = {Cogan, Dan and Su, Zu-En and Kenneth, Oded and Gershoni, David},
  title = {Deterministic generation of indistinguishable photons in a cluster state},
  journal = {Nature Photonics},
  year = {2023},
  volume = {17},
  number = {4},
  pages = {324--329},
  doi = {10.1038/s41566-022-01152-2},
  issn = {1749-4893},
}

@article{meng2024deterministic,
  author = {Meng, Yijian and Chan, Ming Lai and Nielsen, Rasmus B. and Appel, Martin H. and Liu, Zhe and Wang, Ying and Bart, Nikolai and Wieck, Andreas D. and Ludwig, Arne and Midolo, Leonardo and Tiranov, Alexey and Sørensen, Anders S. and Lodahl, Peter},
  title = {Deterministic photon source of genuine three-qubit entanglement},
  journal = {Nature Communications},
  year = {2024},
  month = {sep},
  volume = {15},
  number = {1},
  pages = {7774},
  doi = {10.1038/s41467-024-52086-y},
  issn = {2041-1723},
}

@article{thomas2022efficient,
  title={Efficient generation of entangled multiphoton graph states from a single atom},
  author={Thomas, Philip and Ruscio, Leonardo and Morin, Olivier and Rempe, Gerhard},
  journal={Nature},
  volume={608},
  number={7924},
  pages={677--681},
  year={2022},
  doi = {10.1038/s41586-022-04987-5},
  publisher={Nature Publishing Group UK London}
}

@article{yang2022sequential,
  title={Sequential generation of multiphoton entanglement with a Rydberg superatom},
  author={Yang, Chao-Wei and Yu, Yong and Li, Jun and Jing, Bo and Bao, Xiao-Hui and Pan, Jian-Wei},
  journal={Nature Photonics},
  volume={16},
  number={9},
  pages={658--661},
  year={2022},
  doi = {10.1038/s41566-022-01054-3},
  publisher={Nature Publishing Group UK London}
}

@article{besse2020realizing,
  title={Realizing a deterministic source of multipartite-entangled photonic qubits},
  author={Besse, Jean-Claude and Reuer, Kevin and Collodo, Michele C and Wulff, Arne and Wernli, Lucien and Copetudo, Adrian and Malz, Daniel and Magnard, Paul and Akin, Abdulkadir and Gabureac, Mihai and others},
  journal={Nature communications},
  volume={11},
  number={1},
  pages={4877},
  year={2020},
  doi = {10.1038/s41467-020-18635-x},
  publisher={Nature Publishing Group UK London}
}

@article{ewert2017ultrafast,
  title={Ultrafast fault-tolerant long-distance quantum communication with static linear optics},
  author={Ewert, Fabian and Van Loock, Peter},
  journal={Physical Review A},
  volume={95},
  number={1},
  pages={012327},
  year={2017},
  publisher={APS}
}

@article{murphy2026simplified,
  title={Simplified circuit-level decoding using Knill error correction},
  author={Murphy, Ewan and Sahu, Subhayan and Vasmer, Michael},
  journal={arXiv preprint arXiv:2603.05320},
  year={2026}
}

\appendix
\section{Implementation benchmarking and further tests on graph state families}\label{app:algo-bench}
A technical detail is the validation that the implementation of the presented algorithms is correct. For this, we have two main ways to test our code base. The first is that, given a fixed ordering, the number of emitters is independently calculated from our codebase and the one given by \cite{eva_opt}. The dataset we used to test was the random graph used in Section \ref{subsec:best-case-and-cnot-red}, and for every graph state and a fixed ordering, we got exactly the same number of emitters. 

The second way is to test various graph state families for which the minimum number of emitters is known. So the goal here is not only the validation of our implementation, but also to provide further insights regarding the performance of the algorithms, but this time focusing only on the number of emitters, as this was our primary focus. 

\begin{figure}
    \centering
    \includegraphics[width=1.0\linewidth]{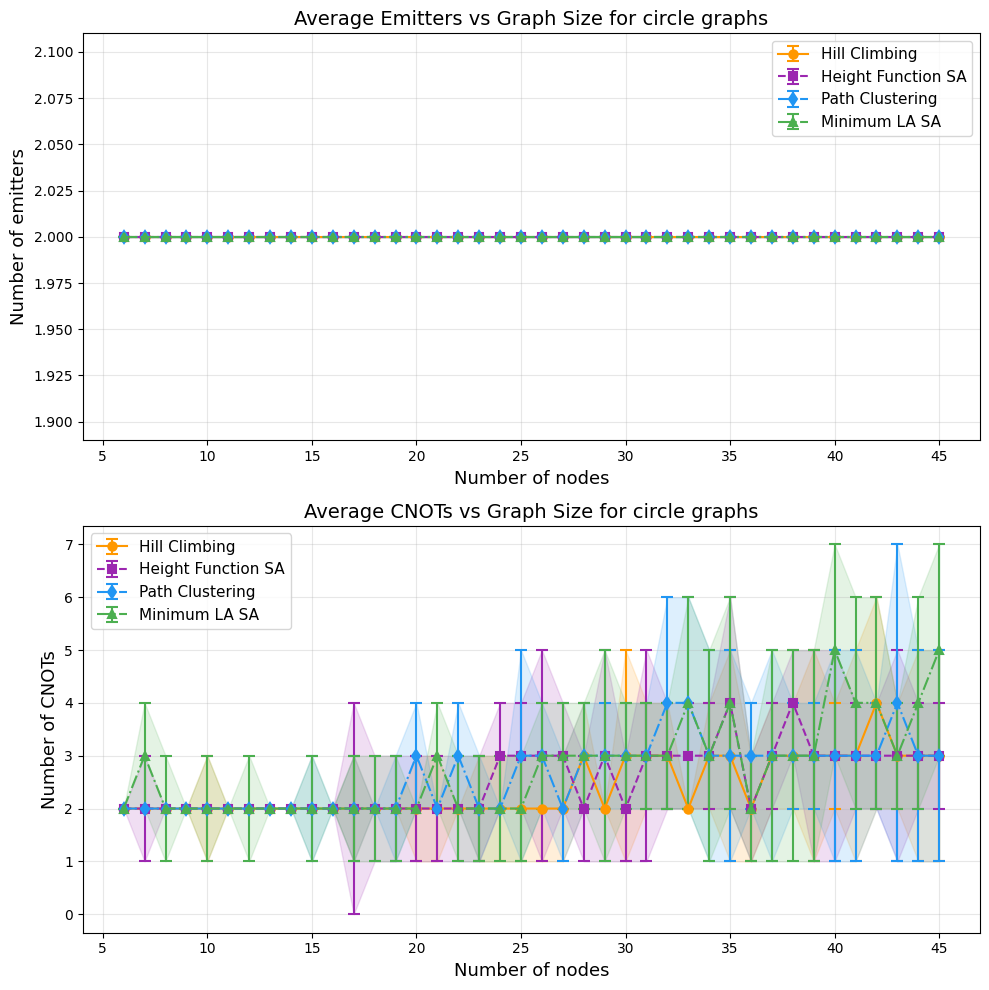}
    \caption{Number of emitters and CNOT for graphs that are LC equivalent to the cycle graphs, which have a known linear rank-width of $2$.}
    \label{fig:bench-circ}
\end{figure}

Let us start with the family of cycle graphs. This family is of great relevance since it is a basic resource state of the fusion-based quantum computation \cite{fbqc}. In our case, states with photon number ranging from $5$ to $45$, we created $30$ different states by applying a sequence of random local complementations. Then we obtained the number of emitters. It is a well-known result that the cycle graph and the LC equivalent to them have linear rank-width of $2$. This is exactly what the result we obtained and we present in Figure \ref{fig:bench-circ}.

\begin{figure}
    \centering
    \includegraphics[width=1.0\linewidth]{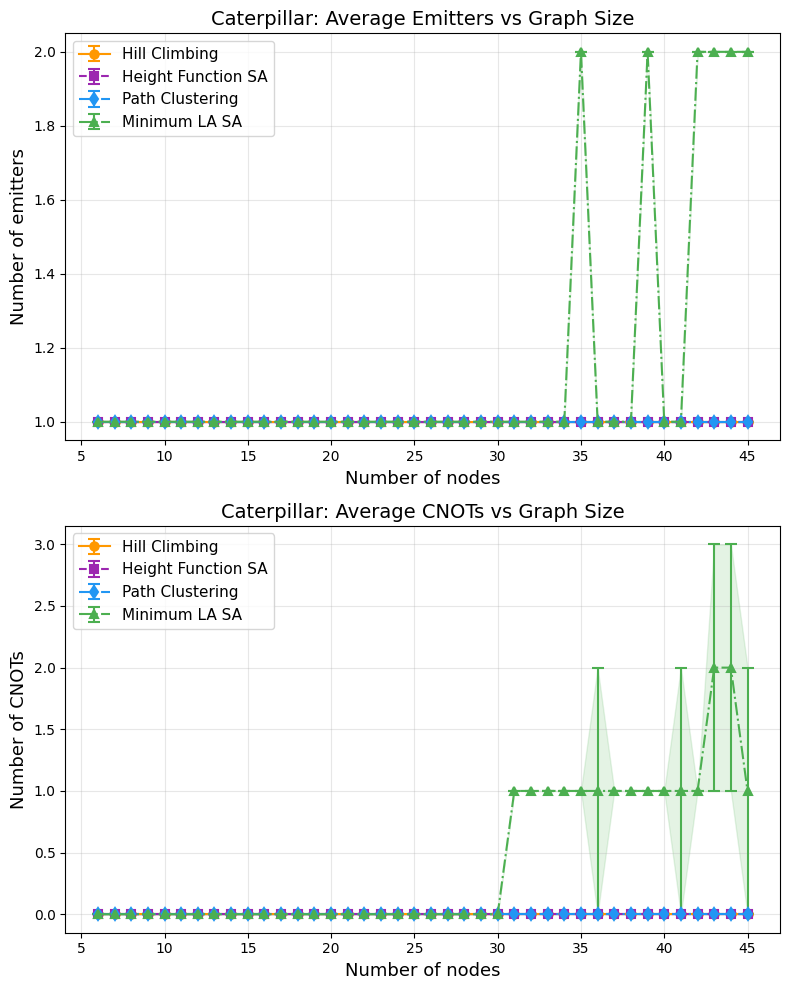}
    \caption{Number of emitters and CNOT for graphs that are LC equivalent to the caterpillar graphs, which have a known linear rank-width of $1$.}
    \label{fig:bench-cater}
\end{figure}

We proceed with the graph states that are LC equivalent to the caterpillar. Once again, we proceed by starting from a caterpillar state, and for each photon number, we derive $30$ distinct graphs for which we calculate the number of emitters and CNOTs. In Figure \ref{fig:bench-cater} we present our findings. Even if it is known that such graphs have linear rank-width $1$, we must keep in mind that our algorithms are probabilistic and thus there are some cases, for instance here for some of the results of the \texttt{min\_la\_sa} that we are not obtaining the well-known minimum.

\begin{figure}
    \centering
    \includegraphics[width=1.0\linewidth]{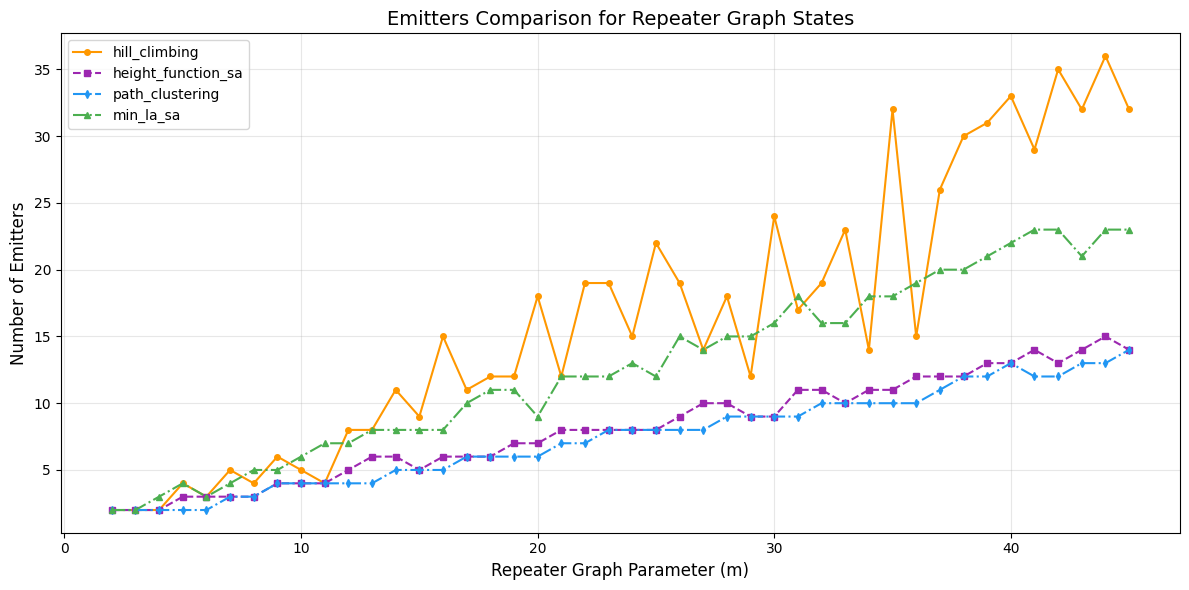}
    \caption{Number of emitters RGS, for which a protocol to generate them with one emitter couples to a single qubit is known.}
    \label{fig:bench-rgs}
\end{figure}

Photonic repeater graph states (RGS) are promising candidates for eneabling long distace quantum communication, without the need for matter-based memory qubits \cite{repeaters}. We are working with the $4m$-vertex repeater that has $2m$ vertices that create a fully connected graph. Then on this graph, each vertex is connected with another vertex, resulting in the total $4m$ vertex graph. However, it must be noted that in \cite{buterakos}, they showed that arbitrarily large repeater states can be generated using only one emitter coupled to a single qubit. In this case, we present the number of emitters in Figure \ref{fig:bench-rgs}. In this case, the \texttt{path\_clustering} algorithm is the best, which is expected since the first step, which is finding the largest path of the graph, appears to serve well as an initial condition. However, this result here indicates something very interesting. This is that if one is interested in a very specific graph state family, using our algorithms can be a great first step, but tailored protocols could potentially yield even better results.

\section{Resource states for MBQC implementation of Shor's and Deutsch-Jozsa's algorithms}\label{app:shor}
\subsection{Implementation details}
In this Appendix, we describe the details necessary to construct the database of resource states required to implement the MBQC version of Shor's algorithm. Let us start with the circuit-level implementation as presented in the original work in Ref. \cite{shorsalgo}. Given a register $N$ to factor the following, we required three types of qubit registers:
\begin{itemize}
    \item Counting register, $n_c = 2\lceil\log_2 N\rceil$ qubits. This register stores the superposition that is then analyzed by the inverse Quantum Fourier Transform (QFT). Using $2n$ counting qubits, with $n=\lceil\log_2 N\rceil$, ensures a sufficiently high success probability for resolving the period via continued factor-expansion.
    \item Work register, $n_w = \lceil\log_2 N\rceil$ qubits. The register encodes the result of the modular exponentiation $a^x\mod{N}$ in the computational basis.
    \item Ancilla register, $n_a = \max(n_w-1,0)$ qubits. These ancillae are used by the Toffoli-cascade decomposition of multi-controlled NOT gates that appear in the general modular exponentiation oracle.
\end{itemize}
Let $a$ be an integer coprime to $N$, chosen uniformly at random from $\{2,\cdots,N-1\}$ with $\gcd(a,N) = 1$. We firstly initialize all qubits at $\ket{0}$. We then apply a Hadamard gate in the counting register
\begin{equation}
    \frac{1}{\sqrt{2^{n_c}}} \sum_{x=0}^{2^{n_c}-1} \ket{x}\ket{0}.
\end{equation}
Then the least-significant qubit is flipped to $\ket{1}$ yielding
\begin{equation}
    \frac{1}{\sqrt{2^{n_c}}} \sum_{x=0}^{2^{n_c}-1} \ket{x}\ket{1}.
\end{equation}
The core step is applying the transformation
\begin{equation}
    \ket{x}\ket{1} \longmapsto \ket{x}\ket{a^x \bmod N}.
\end{equation}
This is decomposed into a product of controlled modular multiplications $U_{a^{2^j}}$ for each one of the counting qubit $j$ ($j = 0,\cdots,n_c-1$), such that

\begin{equation}
    U_a = \prod_{j=0}^{n_c - 1} \bigl(U_{a^{2^j}}\bigr)^{c_j},
\end{equation}
where the exponent $c_j$ denotes the control on counting qubit $j$. Each $U_{a^{2^j}}$ maps $\ket{y}\mapsto\ket{y \cdot a^{2^j}\bmod N}$ on the work register. Then, we apply the inverse QFT to the counting register. Finally, we measure the counting register in the computational basis. The outcome is processed with the continued-fraction algorithm to extract the order $r$ of $a\mod{N}$, from which the factors of $N$ are obtained via $\gcd(a^{r/2}\pm1,N)$.

The Python package \textsc{Graphix} \cite{graphix1,graphix2} supports the following gateset
\begin{align*}
    &\{H, X, Y, Z, S, R_x, R_y, R_z,\\
    &\mathrm{CNOT}, \mathrm{CCX},\mathrm{SWAP}, R_{zz}\}
\end{align*}
and thus we need to implement some of the gates required by Shor's algorithm for it. The first gate is the control phase gate that applies that $\ket{11}\mapsto
e^{i\theta}\ket{11}$ with $\theta = 2\pi / 2^k$. Therefore, we decompose as
\begin{align*}
    CR_k(\theta) = & R_z^{(t)}\left(\tfrac{\theta}{2}\right)\cdot\mathrm{CNOT}(c,t)\cdot R_z^{(t)}\left(-\tfrac{\theta}{2}\right)\\
    &\cdot \mathrm{CNOT}(c,t)\cdot R_z^{(c)}\left(\tfrac{\theta}{2}\right),
\end{align*}
where the subscripts $c$ and $t$ correspond to the control and target qubits, respectively. Then we have the implementation of the inverse QFT. For qubit indices $q_0,\cdots,q_{n-1}$ the procedure iterates from $i = n-1,\cdots , 0$ and for each $i$, it applies $CR^{-1}_k$, with $k=j-i+1$ and $\theta = -2\pi/2^k$ between the control qubit $q_j$ and the target qubit $q_i$ for every $j>i$. Then, there is a Hadamard gate on $q_i$. Finally, the SWAP gates reverse the qubit ordering:
\begin{align*}
     \mathrm{QFT}^{-1} =&
  \left(\prod_{i=0}^{\lfloor n/2\rfloor - 1} \mathrm{SWAP}(q_i, q_{n-1-i})\right)\\
  &\cdot \prod_{i=0}^{n-1} \left[H(q_i) \prod_{j=i+1}^{n-1} CR_{j-i+1}^{-1}(q_j, q_i) \right].
\end{align*}
Then the control SWAP gate is decomposed as
\begin{align*}
      \mathrm{CSWAP}(c, t_1, t_2) = &\mathrm{CNOT}(t_2, t_1)\cdot\\
  &\mathrm{CCX}(c, t_1, t_2)\cdot \mathrm{CNOT}(t_2, t_1).
\end{align*}
Finally, we have to consider the $n$-controlled NOT gate with $n\geq3$. We decompose it into a cascade of Toffoli gates using $n-2$ ancilla qubits, all initialized to $\ket{0}$:
\begin{enumerate}
    \item Compute the $CCX(c_0,c_1,a_0)$ and then for $i = 2,\cdots,n-1$ compute $CCX(c_i,a_{i-2},a_{i-1})$.
    \item Apply the Toffoli gates in reverse order to restore all ancillae to $\ket{0}$.
\end{enumerate}
For an arbitrary $N$ and any base $a$ coprime to $N$ the map $\pi_a\colon y \mapsto y \cdot a \bmod N$ is a permutation on $\{0, 1, \dots, N-1\}$. We implement the controlled version of this permutation through the following three-step procedure.
\begin{itemize}
    \item \textbf{Cycle Decomposition.} The permutation $\pi_a$ is decomposed into disjoint cycles. Each cycle $c_0,\cdots,c_{\ell}$ with $\pi_a(c_i) = c_{i+1 \bmod (\ell+1)}$ is then further decomposed into a product of transpositions:
\begin{equation}
    (c_0, c_1, \dots, c_\ell) = (c_0, c_\ell)(c_0, c_{\ell-1})\cdots(c_0, c_1).
\end{equation}
\item \textbf{Controlled basis-state transportation.} Each transportation $(v_i,v_j)$ swaps the computational basis $\ket{v_i}$ and $\ket{v_j}$ in the $n_w$ work qubit register, controlled on the Shor counting qubit $c_j$.
\item The full controlled multiplication $U_{a^{2^j}}$ is the sequential application of all the controlled transpositions arising from the cycle decomposition. The complete modular exponentiation is then
\begin{equation}
  U_a = \prod_{j=0}^{n_c-1} U_{a^{2^j}}^{(c_j)},
\end{equation}
where $a^{2^j}$ is computed classically as $\texttt{pow}(a, 2^j, N)$.  If $a^{2^j} \equiv 1 \pmod{N}$, the corresponding controlled multiplication is the identity and is skipped.
\end{itemize}

The main reason for finding the exact circuit to implement Shor's algorithm is to transpile it into an MBQC pattern using \textsc{graphix}, which yields the number of nodes, the entangling operations, the measurements, and the Pauli corrections that will act on the graph state. However, the raw MBQC pattern is typically large, and thus further optimization is required. For this, we use the following three \textsc{graphix} optimization schemes sequentially.
\begin{enumerate}
    \item \textbf{Standardization}, which reorders the commands so that all node preparations precede all entangling operations, which precede all measurements and corrections, and brings the pattern into a standard form.
    \item \textbf{Signal shifting} modifies the measurement angles to absorb Pauli operators that arise from earlier adaptive measurements, reducing the depth of the classical feedforward.
    \item \textbf{Pauli measurement preprocessing} identifies the measurements whose angles are multiples of $\pi/2$ and performs them analytically. Each such measurement eliminates one node from the graph state and may also remove or introduce edges among its neighbors via local complementation rules. A final standardization pass is applied afterwards.
\end{enumerate}

After the aforementioned steps, the resource graph state is extracted from the optimized patterns and is typically much smaller than the one before the pattern optimization.

At this junction, let us briefly discuss how the corresponding database is created. For each integer from 3 to 17, we enumerate all the coprime numbers $a\in\{2,\cdots,N-1\}$. For each $(a,N)$ pair, the multiplicative order $r\mod{N}$ (so the smallest positive integer such that $a^r=1\mod{N}$) is computed. The total number of $(N,a)$ configurations is determined by the Euler totient function
\begin{equation}
\phi(N) = N\prod_{p\mid N}\left(1-\frac{1}{p}\right).
\end{equation}
We then follow the transpilation to MBQC and extract the graphs we have to prepare their measurement patterns. However, for the vast majority of them, they are not connected, and thus, we pick the largest graph since it will probably be the one that requires more emitters, CNOTs, and gates to be created. Finally, we relabel the nodes to range from $\{0,\cdots,|V|-1\}$ and they are randomly permuted. An important remark is that the transpilation procedure can require substantial time, and thus, we limited ourselves to 10 seconds on our machine. From this point, we have the dataset, and we proceed with the well-known optimization pipeline for the number of emitters. We present every metric for each one of the corresponding resource graphs in the Table \ref{tab:shor_results}.

\subsection{Deutsch-Jozsa's Algorithm}
\begin{figure}
    \centering
    \includegraphics[width=1.0\linewidth]{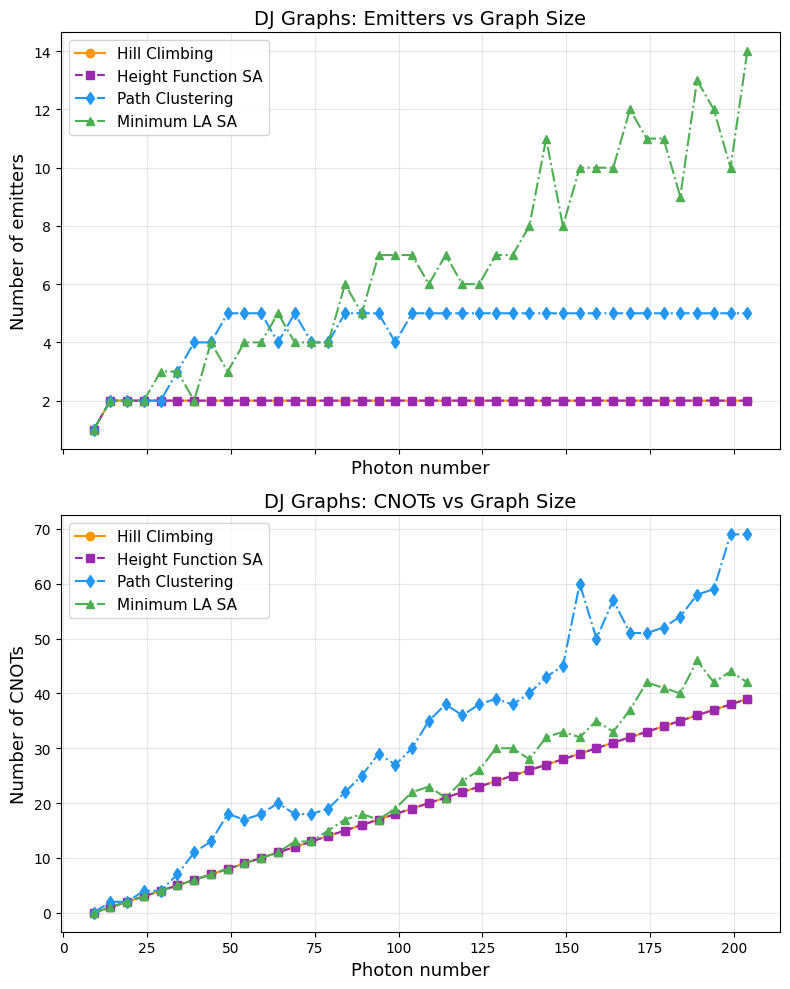}
    \caption{Number of emitters and CNOT for tree graphs emerging from the non-transpiled MBQC version of the Deutsch-Jozsa algorithm.}
    \label{fig:dj-tree}
\end{figure}

Here, we turn our focus on the MBQC implementation of the Deutsch-Jozsa algorithm. In the documentation of the \textsc{graphix}\cite{graphix1,graphix2}, they explicitly present how to implement the algorithm and transpile it. However, if we do not impose the transpilation step we only obtain tree graphs. Therefore, we choose not to test our algorithms on tree graphs. We present our result in Figure \ref{fig:dj-tree}. It is noteworthy that the \texttt{hill\_climbing} and \texttt{height\_function\_sa} algorithms are the best in both the emitter and CNOT metrics, while the \texttt{path\_clustering} comes third on the emitters and last on the emitter CNOT.

\section{Algorithms details and pseudocodes}\label{app:pseudocodes}
Here, we provide the detailed algorithms discussed in the main text. 
\subsection{\texttt{hill\_climbing}}\label{app-RW}
\begin{algorithm}[H]
\caption{General Pipeline of \texttt{hill\_climbing}}
\label{alg:pipeline-rank-width}
\begin{algorithmic}[1]
  \renewcommand{\algorithmicrequire}{\textbf{Require:}}
  \renewcommand{\algorithmicensure}{\textbf{Ensure:}}
  \Require Graph $G = (V, E)$, parameters \texttt{refine}, \texttt{max\_iter}, \texttt{window}
  \Ensure Approximation $\widehat{lrw}(G)$ and the witness ordering $\pi_{\text{opt}}$
  
  \State Relabel $V \to \{0, \dots, n-1\}$; compute $A \in \mathbb{F}^{n \times n}$
  
  \State \textbf{Phase A:} $\mathcal{C} \leftarrow \{\pi_{\text{spec}}, \pi_{\text{rcm}}, \pi_{\text{deg}}\}$ 
  \hfill \Comment{Three initial orderings}
  
  \State \textbf{Phase B:} $(\pi^*, \widehat{k}) \leftarrow \min_{\pi \in \mathcal{C}} n_e(\pi, G)$ 
  \hfill \Comment{Intermediate best candidate}
  
  \State $\pi_{\text{opt}} \leftarrow \pi^*$
  \hfill \Comment{Default to the Phase B candidate}
  
  \If{\texttt{refine} \textbf{and} $\widehat{lrw} > 0$}
    \State \textbf{Phase C:} $(\pi_{\text{opt}}, \widehat{k}) \leftarrow \text{HillClimb}(G, \pi^*, \widehat{k}, \dots)$
  \EndIf
  
  \State \Return $\widehat{lrw} = n_e(\pi_{\text{opt}},G), k)$, $\pi_{\text{opt}}$
\end{algorithmic}
\end{algorithm}

\begin{algorithm}[H]
\caption{Phase C of \texttt{hill\_climbing}}\label{alg:hillclimb}
\begin{algorithmic}[1]
  \Require $G$, ordering $\pi$, current lrw $w$, bottleneck $k^*$, \texttt{max\_iter}, \texttt{window}
  \Ensure Refined ordering $\pi$, refined lrw $w$, refined bottleneck $k^*$
  
  \State $\text{no\_improve} \leftarrow 0$
  \For{iteration $= 1$ to \texttt{max\_iter}}
    \State $\mathcal{N} \leftarrow \{k^* - \text{window}, \dots, k^* + \text{window}\} \cap \{0, \dots, n-1\}$
    \State \textbf{improved} $\leftarrow$ \textbf{false}
    
    \ForAll{$(i, j) \in \binom{\mathcal{N}}{2}$} \Comment{Try swaps within window}
      \State $\pi[i] \leftrightarrow \pi[j]$
      \State $(w', k') \leftarrow \textsc{EvalOrdering}(A, \pi)$
      \If{$w' < w$}
        \State $w \leftarrow w'$; $k^* \leftarrow k'$; \textbf{improved} $\leftarrow$ \textbf{true}
        \State \textbf{break} \Comment{Accept first improvement}
      \Else
        \State $\pi[i] \leftrightarrow \pi[j]$ \Comment{Undo swap}
      \EndIf
    \EndFor

    \If{\textbf{not} improved}
      \State $(i, j) \leftarrow$ random distinct indices in $\{0, \dots, n-1\}$
      \State $\pi[i] \leftrightarrow \pi[j]$; recompute $(w', k')$
      \If{$w' < w$}
        \State $w \leftarrow w'$; $k^* \leftarrow k'$; $\text{no\_improve} \leftarrow 0$
      \Else
        \State $\pi[i] \leftrightarrow \pi[j]$; $\text{no\_improve} \leftarrow \text{no\_improve} + 1$
      \EndIf
    \Else
      \State $\text{no\_improve} \leftarrow 0$
    \EndIf

    \If{$\text{no\_improve} \ge \lfloor \texttt{max\_iter}/2 \rfloor$}
      \State \textbf{break} \Comment{Early stopping}
    \EndIf
  \EndFor
  \State \Return $\pi, w, k^*$
\end{algorithmic}
\end{algorithm}

\subsection{\texttt{height\_function\_sa}}
As discussed in the main text, the SA procedure may include an optional periodic reheating mechanism designed to mitigate premature convergence to a local optimum. During the geometric cooling, the temperature $T$ decreases monotonically, which progressively reduces the acceptance probability of uphill moves and can cause the search to stagnate if it becomes trapped near a suboptimal ordering. To address this issue, the temperature is periodically increased by a reheat factor $r$ for every reheat interval $R$ as: 
\begin{equation}
    T\leftarrow\min\bigl(T\cdot r, T_{\text{start}}\bigr)
\end{equation}
\begin{algorithm}[H]
\caption{Simulated Annealing for \texttt{height\_function\_sa}}
\label{alg:sa}
\begin{algorithmic}[1]
  \renewcommand{\algorithmicrequire}{\textbf{Require:}}
  \renewcommand{\algorithmicensure}{\textbf{Ensure:}}
  
  \Require Graph $G$ with adjacency matrix $\Gamma$; ordering $\pi$; current cost $w$ and bottleneck $b$; parameters $T_{\mathrm{start}}$, $T_{\mathrm{min}}$, $\alpha$, $S$, \texttt{reheat\_interval} $R$, \texttt{reheat\_factor} $r$
  \Ensure Best ordering $\pi^*$, best cost $w^*$, best bottleneck $b^*$
  
  \State $T \leftarrow T_{\mathrm{start}}$; $\pi^* \leftarrow \pi$; $w^* \leftarrow w$; $b^* \leftarrow b$; $c \leftarrow 0$
  \hfill \Comment{cooling step counter}
  
  \While{$T > T_{\mathrm{min}}$}
    \For{$s = 1 \textbf{ to } S$} 
    \hfill \Comment{steps at this temperature}
      \State Draw move type $\tau \sim \{\mathrm{Swap}: 0.5, \mathrm{Reverse}: 0.3, \mathrm{Relocate}: 0.2\}$
      \State $(\pi, \mathit{undo}) \leftarrow \mathrm{Apply}(\tau, \pi)$
      \hfill \Comment{perturb in-place}
      \State $(w', b') \leftarrow \mathrm{Evaluate}(\Gamma, \pi)$
      \hfill \Comment{$\max(h)$ via tableau}
      \State $\Delta \leftarrow w' - w$
      
      \If{$\Delta \le 0$ \textbf{or} $\exp(-\Delta / T) > U(0,1)$}
      \hfill \Comment{Metropolis acceptance}
        \State $w \leftarrow w'$; $b \leftarrow b'$
        \hfill \Comment{accept move}
        \If{$w < w^*$}
          \State $\pi^* \leftarrow \mathrm{copy}(\pi)$; $w^* \leftarrow w$; $b^* \leftarrow b$
          \If{$w^* = 0$}
            \State \Return $\pi^*, w^*, b^*$
            \hfill \Comment{early exit: optimal}
          \EndIf
        \EndIf
      \Else
        \State $\pi \leftarrow \mathrm{Undo}(\pi, \mathit{undo})$
        \hfill \Comment{reject move}
      \EndIf
    \EndFor
    
    \State $T \leftarrow \alpha \cdot T$; $c \leftarrow c + 1$
    \hfill \Comment{cool down}
    
    \If{$R > 0$ \textbf{and} $c \bmod R = 0$}
      \State $T \leftarrow \min(T \cdot r, T_{\mathrm{start}})$
      \hfill \Comment{reheat}
    \EndIf
  \EndWhile
  
  \State \Return $\pi^*, w^*, b^*$
\end{algorithmic}
\end{algorithm}

\subsection{\texttt{path\_clustering}}
\begin{algorithm}[H]
\caption{BFS-Farthest: Approximate Diameter Endpoint}
\label{alg:path-bfs-farthest}
\begin{algorithmic}[1]
  \renewcommand{\algorithmicrequire}{\textbf{Require:}}
  \renewcommand{\algorithmicensure}{\textbf{Ensure:}}
  
  \Require Connected graph $G = (V, E)$; source vertex $s$
  \Ensure Vertex $t$ at maximum BFS distance from $s$
  
  \State $\textit{visited} \leftarrow \{s\}$; $\textit{queue} \leftarrow [s]$; $t \leftarrow s$
  \hfill \Comment{initialize BFS from $s$}
  
  \While{$\textit{queue} \neq \emptyset$}
    \State $\textit{next} \leftarrow []$
    \hfill \Comment{vertices in the next BFS layer}
    
    \For{each $v \in \textit{queue}$}
      \For{each $u \in N(v)$}
        \If{$u \notin \textit{visited}$}
          \State $\textit{visited} \leftarrow \textit{visited} \cup \{u\}$; append $u$ to $\textit{next}$; $t \leftarrow u$
          \hfill \Comment{record the most recently reached vertex}
        \EndIf
      \EndFor
    \EndFor
    
    \State $\textit{queue} \leftarrow \textit{next}$
    \hfill \Comment{advance to the next layer}
  \EndWhile
  
  \State \Return $t$
\end{algorithmic}
\end{algorithm}

\begin{algorithm}[H]
\caption{Greedy-DFS: Greedy Path Extension from a Seed Vertex}
\label{alg:path-greedy-dfs}
\begin{algorithmic}[1]
  \renewcommand{\algorithmicrequire}{\textbf{Require:}}
  \renewcommand{\algorithmicensure}{\textbf{Ensure:}}
  
  \Require Graph $G = (V, E)$; start vertex $s$
  \Ensure Simple path $P = [s, v_2, \ldots, v_\ell]$
  
  \State $P \leftarrow [s]$; $\textit{visited} \leftarrow \{s\}$; $\textit{current} \leftarrow s$
  \hfill \Comment{initialize path from seed vertex}
  
  \Loop
    \State $\textit{nbrs} \leftarrow \{u \in N(\textit{current}) : u \notin \textit{visited}\}$
    \hfill \Comment{unvisited neighbors of current vertex}
    
    \If{$\textit{nbrs} = \emptyset$}
      \State \textbf{break}
      \hfill \Comment{no feasible extension}
    \EndIf
    
    \State $v^* \leftarrow \arg\max_{u \in \textit{nbrs}} |\{w \in N(u) : w \notin \textit{visited}\}|$
    \hfill \Comment{prefer high residual degree}
    
    \State $\textit{visited} \leftarrow \textit{visited} \cup \{v^*\}$; append $v^*$ to $P$; $\textit{current} \leftarrow v^*$
    \hfill \Comment{extend path greedily}
  \EndLoop
  
  \State \Return $P$
\end{algorithmic}
\end{algorithm}

\begin{algorithm}[H]
\caption{Longest Path: Multi-Start Polynomial Long-Path Heuristic}
\label{alg:longest-path}
\begin{algorithmic}[1]
  \renewcommand{\algorithmicrequire}{\textbf{Require:}}
  \renewcommand{\algorithmicensure}{\textbf{Ensure:}}
  
  \Require Graph $G = (V, E)$; number of random starts $k$
  \Ensure Approximately longest simple path $P^*$
  
  \State $P^* \leftarrow []$
  \hfill \Comment{best path found so far}
  
  \State $t_1 \leftarrow \textsc{BFS-Farthest}(G, v_0)$ for arbitrary $v_0$
  \hfill \Comment{first diameter-endpoint estimate}
  \State $t_2 \leftarrow \textsc{BFS-Farthest}(G, t_1)$
  \hfill \Comment{second diameter-endpoint estimate}
  
  \State $\textit{seeds} \leftarrow \{t_1, t_2\} \cup \text{top-3 highest-degree vertices} \cup \text{$k$ random vertices}$
  \hfill \Comment{construct multi-start seed set}
  
  \For{each $s \in \textit{seeds}$}
    \State $P \leftarrow \textsc{GreedyDFS}(G, s)$
    \hfill \Comment{greedily grow path from seed}
    
    \If{$|P| > |P^*|$}
      \State $P^* \leftarrow P$
      \hfill \Comment{update best path}
    \EndIf
    
    \State $P' \leftarrow \textsc{GreedyDFS}(G, \text{last}(P))$
    \hfill \Comment{attempt extension from other end}
    
    \If{$|P'| > |P^*|$}
      \State $P^* \leftarrow P'$
      \hfill \Comment{update best path}
    \EndIf
  \EndFor
  
  \State \Return $P^*$
\end{algorithmic}
\end{algorithm}

\begin{algorithm}[H]
\caption{Extract Clusters: Hierarchical Path-Peeling Decomposition}
\label{alg:path-extract-clusters}
\begin{algorithmic}[1]
  \renewcommand{\algorithmicrequire}{\textbf{Require:}}
  \renewcommand{\algorithmicensure}{\textbf{Ensure:}}
  
  \Require Graph $G = (V, E)$
  \Ensure List of path clusters $\mathcal{C} = [\mathcal{G}_1, \ldots, \mathcal{G}_p]$ covering $V$
  
  \State $\mathcal{C} \leftarrow []$; $\textit{claimed} \leftarrow \emptyset$; $\textit{remaining} \leftarrow V$; $\ell \leftarrow 0$
  \hfill \Comment{initialize cluster list and unclaimed vertices}
  
  \While{$\textit{remaining} \neq \emptyset$}
    \State $H \leftarrow G[\textit{remaining}]$
    \hfill \Comment{subgraph induced by unassigned vertices}
    
    \State $\textit{isolates} \leftarrow$ isolated vertices of $H$
    \hfill \Comment{handle isolated vertices separately}
    
    \For{each connected component $S$ of $H - \textit{isolates}$}
      \If{$|S| = 1$}
        \State Create singleton cluster; append it to $\mathcal{C}$
        \hfill \Comment{degenerate component}
      \Else
        \State $P \leftarrow \textsc{LongestPath}(G[S])$
        \hfill \Comment{extract approximate longest path}
        
        \State $\mathcal{G} \leftarrow \textsc{BuildCluster}(G, P, \textit{claimed})$ with level $\ell$
        \hfill \Comment{build cluster around path}
        
        \State Append $\mathcal{G}$ to $\mathcal{C}$; $\textit{claimed} \leftarrow \textit{claimed} \cup \mathcal{G}.\text{vertices}$
        \hfill \Comment{record claimed vertices}
        
        \State $\textit{remaining} \leftarrow \textit{remaining} \setminus \mathcal{G}.\text{vertices}$
        \hfill \Comment{peel cluster from graph}
      \EndIf
    \EndFor
    
    \For{each $v \in \textit{isolates}$}
      \State Create singleton cluster for $v$; append it to $\mathcal{C}$
      \hfill \Comment{isolated singleton cluster}
      
      \State $\textit{remaining} \leftarrow \textit{remaining} \setminus \{v\}$
      \hfill \Comment{mark isolated vertex as assigned}
    \EndFor
    
    \State $\ell \leftarrow \ell + 1$
    \hfill \Comment{advance hierarchy level}
  \EndWhile
  
  \State \Return $\mathcal{C}$
\end{algorithmic}
\end{algorithm}

\begin{algorithm}[H]
\caption{Brute-Force TSP for Small Instances}
\label{alg:path-tsp-exact}
\begin{algorithmic}[1]
  \renewcommand{\algorithmicrequire}{\textbf{Require:}}
  \renewcommand{\algorithmicensure}{\textbf{Ensure:}}
  
  \Require Distance matrix $D \in \mathbb{R}^{p \times p}$
  \Ensure Optimal Hamiltonian path $\pi^*$ starting from cluster $0$
  
  \State $\pi^* \leftarrow [0, 1, \ldots, p-1]$; $c^* \leftarrow \infty$
  \hfill \Comment{initialize incumbent path and cost}
  
  \For{each permutation $\pi$ of $\{1, \ldots, p-1\}$}
    \State $\pi' \leftarrow [0] \oplus \pi$
    \hfill \Comment{prepend fixed start cluster $0$}
    
    \State $c \leftarrow \sum_{i=0}^{p-2} D[\pi'_i, \pi'_{i+1}]$
    \hfill \Comment{Hamiltonian path cost}
    
    \If{$c < c^*$}
      \State $c^* \leftarrow c$; $\pi^* \leftarrow \pi'$
      \hfill \Comment{update incumbent}
    \EndIf
  \EndFor
  
  \State \Return $\pi^*$
\end{algorithmic}
\end{algorithm}

\begin{algorithm}[H]
\caption{Intra-ClusterSA: Multi-Start SA for a Single Cluster}
\label{alg:path-intra-sa}
\begin{algorithmic}[1]
  \renewcommand{\algorithmicrequire}{\textbf{Require:}}
  \renewcommand{\algorithmicensure}{\textbf{Ensure:}}
  
  \Require Graph $G$; cluster vertices $V_c$; SA parameters
  \Ensure Ordering $\pi^*$ of $V_c$ minimizing $\max(h)$
  
  \State $G_c \leftarrow G[V_c]$
  \hfill \Comment{induced subgraph on cluster vertices}
  
  \State \textbf{Tier 1: Seeding}
  \hfill \Comment{construct initial candidate orderings}
  \State Generate structured orderings using spectral, RCM, and minimum-degree heuristics
  \State Generate $R$ random orderings
  \State Evaluate all candidates and sort them by cost
  
  \State \textbf{Tier 2: SA Refinement}
  \hfill \Comment{refine the best initial candidates}
  \For{$i = 1 \textbf{ to } N_{\textit{sa}}$}
    \State $(\pi_i^*, w_i^*) \leftarrow \textsc{SimulatedAnnealing}(G_c, \pi_i)$
    \hfill \Comment{refine candidate $\pi_i$}
  \EndFor
  
  \State $i^* \leftarrow \arg\min_i w_i^*$
  \hfill \Comment{select best refined candidate}
  
  \State \Return $\pi_{i^*}^*$
\end{algorithmic}
\end{algorithm}

\begin{algorithm}[H]
\caption{Boundary-Biased SA: Global SA Refinement}
\label{alg:path-global-sa-refine}
\begin{algorithmic}[1]
  \renewcommand{\algorithmicrequire}{\textbf{Require:}}
  \renewcommand{\algorithmicensure}{\textbf{Ensure:}}
  
  \Require Graph $G$ with adjacency matrix $\Gamma$; ordering $\pi$; current cost $w$; boundary set $B$; parameters $T_{\mathrm{start}}$, $T_{\min}$, $\alpha$, $S$; boundary bias $p_b$
  \Ensure Refined ordering $\pi^*$ with cost $w^*$
  
  \State $\pi^* \leftarrow \pi$; $w^* \leftarrow w$; $T \leftarrow T_{\mathrm{start}}$
  \hfill \Comment{initialize incumbent ordering}
  
  \While{$T > T_{\min}$}
    \For{$s = 1 \textbf{ to } S$}
      \State Draw move type $\tau$ and apply it to $\pi$ with boundary bias $p_b$
      \hfill \Comment{perturb boundary positions preferentially}
      
      \State $(w', b') \leftarrow \textsc{Evaluate}(\Gamma, \pi)$
      \hfill \Comment{$\max(h)$ via tableau}
      
      \State $\Delta \leftarrow w' - w$
      
      \If{$\Delta \le 0$ \textbf{or} $\exp(-\Delta / T) > U(0,1)$}
        \State $w \leftarrow w'$
        \hfill \Comment{accept move}
        
        \If{$w < w^*$}
          \State $\pi^* \leftarrow \mathrm{copy}(\pi)$; $w^* \leftarrow w$
          \hfill \Comment{update incumbent}
        \EndIf
      \Else
        \State Undo move
        \hfill \Comment{reject move}
      \EndIf
    \EndFor
    
    \State $T \leftarrow \alpha \cdot T$
    \hfill \Comment{cool down}
  \EndWhile
  
  \State \Return $\pi^*, w^*$
\end{algorithmic}
\end{algorithm}

\begin{algorithm}[H]
\caption{\texttt{path\_clustering}: Full Path-Based Clustering Pipeline}
\label{alg:path-com-pipeline}
\begin{algorithmic}[1]
  \renewcommand{\algorithmicrequire}{\textbf{Require:}}
  \renewcommand{\algorithmicensure}{\textbf{Ensure:}}
  
  \Require Graph $G = (V, E)$; SA parameters; flag \texttt{refine}
  \Ensure Approximation $\widehat{\operatorname{lrw}}(G)$ and witness ordering $\pi_{\text{opt}}$
  
  \State Relabel $V \to \{0, \ldots, n-1\}$
  \hfill \Comment{normalize vertex labels}
  
  \State $\mathcal{C} \leftarrow \textsc{ExtractClusters}(G)$
  \hfill \Comment{Phase 1: hierarchical path peeling}
  
  \If{$|\{c \in \mathcal{C} : |c| > 1\}| \le 1$}
    \State Generate initial orderings; evaluate candidates; run SA
    \hfill \Comment{single-cluster fast path}
    \State Map best ordering back to original vertex labels
    \State \Return $\widehat{\operatorname{lrw}}(G), \pi_{\text{opt}}$
  \EndIf
  
  \State Build cluster meta-graph $\mathcal{M}$ with inter-cluster weights
  \hfill \Comment{Phase 3: meta-graph construction}
  
  \State $\pi \leftarrow \textsc{OrderClusters}(\mathcal{C}, \mathcal{M})$
  \hfill \Comment{Phase 4: TSP ordering on meta-graph}
  
  \For{each cluster index $i$ in order $\pi$}
    \State Order vertices within $\mathcal{G}_i$
    \hfill \Comment{Phase 5: intra-cluster ordering}
  \EndFor
  
  \State Assemble global ordering $\pi$ by concatenating the ordered clusters
  \hfill \Comment{construct full vertex ordering}
  
  \State $(w, b) \leftarrow \textsc{Evaluate}(\Gamma, \pi)$
  \hfill \Comment{$\max(h)$ via tableau}
  
  \If{\texttt{refine} \textbf{and} $w > 0$}
    \State $B \leftarrow$ boundary vertices of $\mathcal{C}$
    \hfill \Comment{Phase 6: collect boundary vertices}
    
    \State $(\pi_{\text{opt}}, w^*) \leftarrow \textsc{BoundaryBiasedSA}(G, \pi, w, B)$
    \hfill \Comment{global boundary-biased refinement}
  \Else
    \State $\pi_{\text{opt}} \leftarrow \pi$; $w^* \leftarrow w$
    \hfill \Comment{skip refinement}
  \EndIf
  
  \State Map $\pi_{\text{opt}}$ back to original vertex labels
  \hfill \Comment{restore input labels}
  
  \State \Return $\widehat{\operatorname{lrw}}(G) = w^*, \pi_{\text{opt}}$
\end{algorithmic}
\end{algorithm}

\subsection{Detailed pipeline examination}\label{app:pipeline-exam}
In the main manuscript, we employed the complete pipeline as we described in Figure \ref{fig:pipeline}. However, a complete analysis requires examining how the metrics are affected by each step separately, to assist further future optimization endeavors. For this reason, we created two datasets of random Erd\H{o}s-R\'enyi graph states, one where we fix the photon number and we vary the $p$ parameter, and vice versa for the other one:

\begin{itemize}
    \item Dataset 1: For $n=20$, we create random Erd\H{o}s-R\'enyi graphs states with probabilities $p = [0.1, 0.2, 0.3, 0.4, 0.45, 0.5, 0.55, 0.6, 0.65, 0.7, 0.8, 0.9]$. For each $p$ we create 30 distinct states.
    \item Dataset 2: For $p = 0.65$ we create random Erd\H{o}s-R\'enyi states with nodes ranging from $n = 6,\cdots, 30$. For each instance of $n$ we create a sample of 30 states.
\end{itemize}
The metrics we  calculate and base our comparison upon will be the number of emitters, the emitter CNOTs, and the number of total gates for the emission circuit. The last two are obtained from the emission algorithm of Ref. \cite{li_algo}. Firstly, we aim to identify the effect of edge reduction for each one of the algorithms. For this reason, we fixed $p=0.65$ for dataset 2 such that edge reduction will affect the studied states.  Secondly, we want to identify how well the initial emission orderings $\pi_{ini}$ for phases A and B for the algorithms \texttt{hill\_climbing} and \texttt{height\_function\_sa} perform. Finally, we also include the result for random emission ordering for a complete description and comparison of our results.

\begin{figure*}[t]
    \centering
    % Top row: Dataset 1
    % \noindent\textbf{A)}\par
    \includegraphics[width=0.9\textwidth]{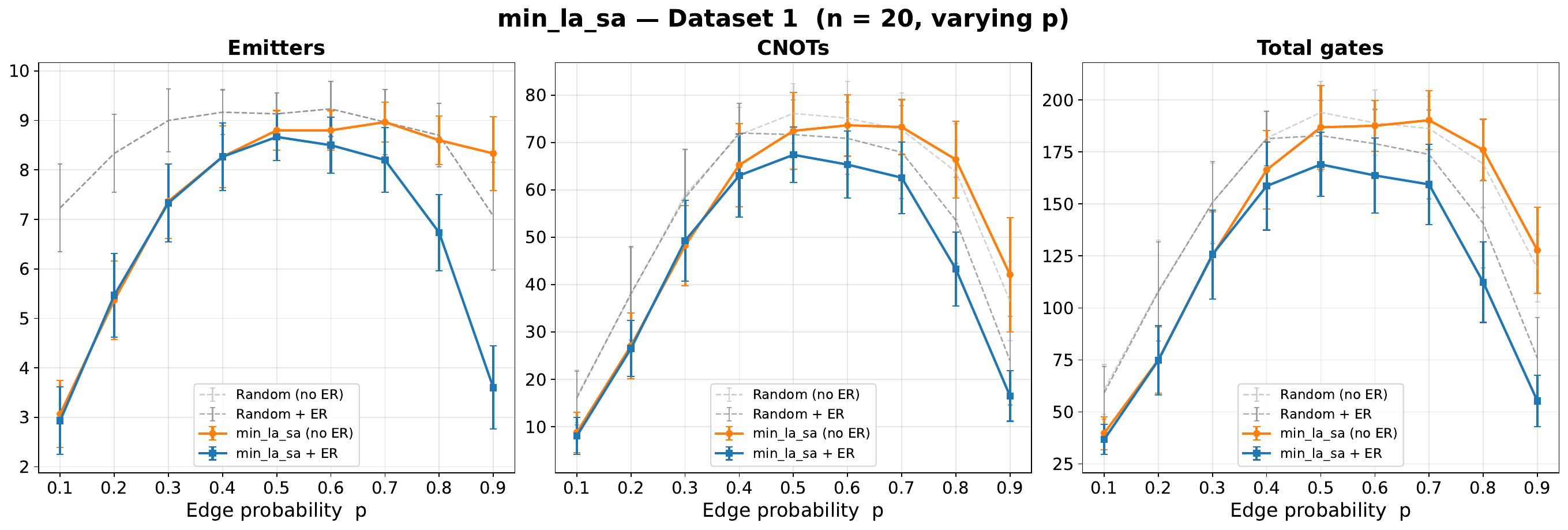}
    
    % \vspace{0.5cm} 
    
    % Bottom row: Dataset 2
    \includegraphics[width=0.9\textwidth]{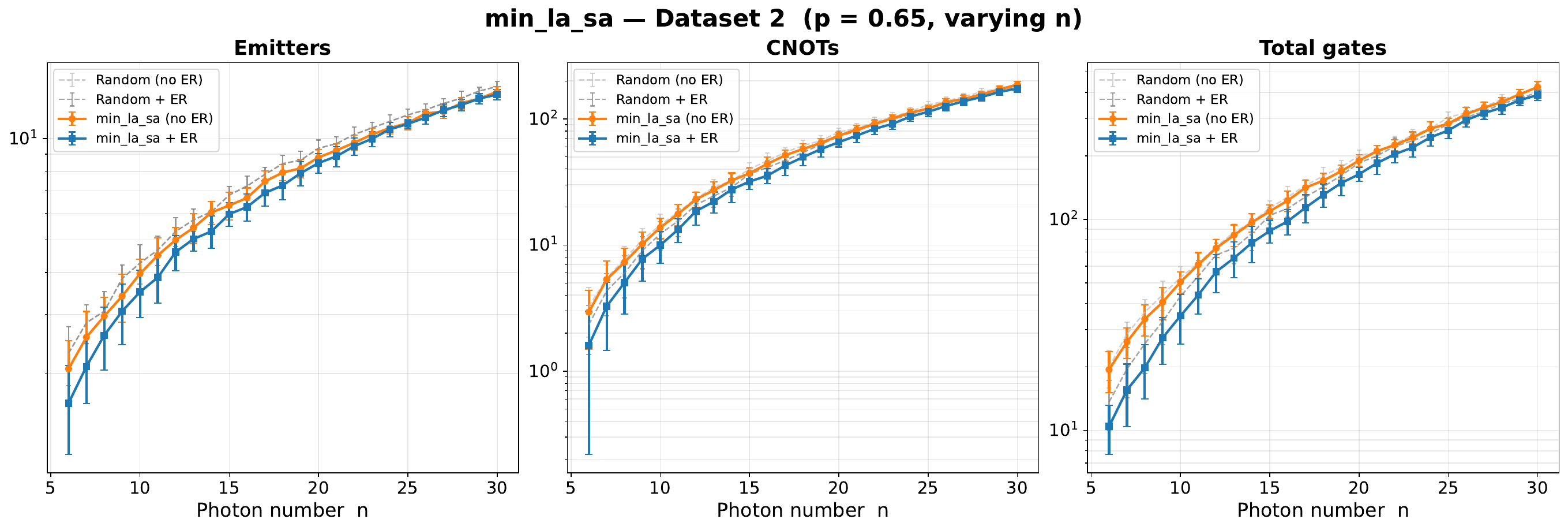}
    
    \caption{\label{fig:pipe_comp_saminla} Resource overhead scaling for photonic graph state generation using \texttt{min\_la\_sa}. The number of Emitters, CNOT gates, and Total gates is plotted for two scenarios. (Top) Performance on Dataset 1 with $n=20$ and varying edge probability $p$. (Bottom) Scaling against the number of nodes $n$ for Dataset 2 at a fixed $p=0.65$.}
\end{figure*}

Let us start with the effect of the edge reduction for a fixed number of $20$ emitted photons.  In Figure \ref{fig:pipe_comp_saminla} and the top panel, we present how the metric evolves as we vary the edge probability $p$ for the \texttt{min\_la\_sa} algorithm. As it is expected for lower probabilities, the results are pretty close, if not exactly the same, since the edge reduction has a minor or no effect. However, as we increase $p$, the edge reduction can have a major impact on the results. In fact, we realize that  having a random photon emission order requires fewer emitters, emitter CNOTs, and total gates than using the \texttt{min\_la\_sa} algorithm without an edge reduction.

\begin{figure*}[t]
    \centering
    % Top row: Dataset 1
    \includegraphics[width=0.9\textwidth]{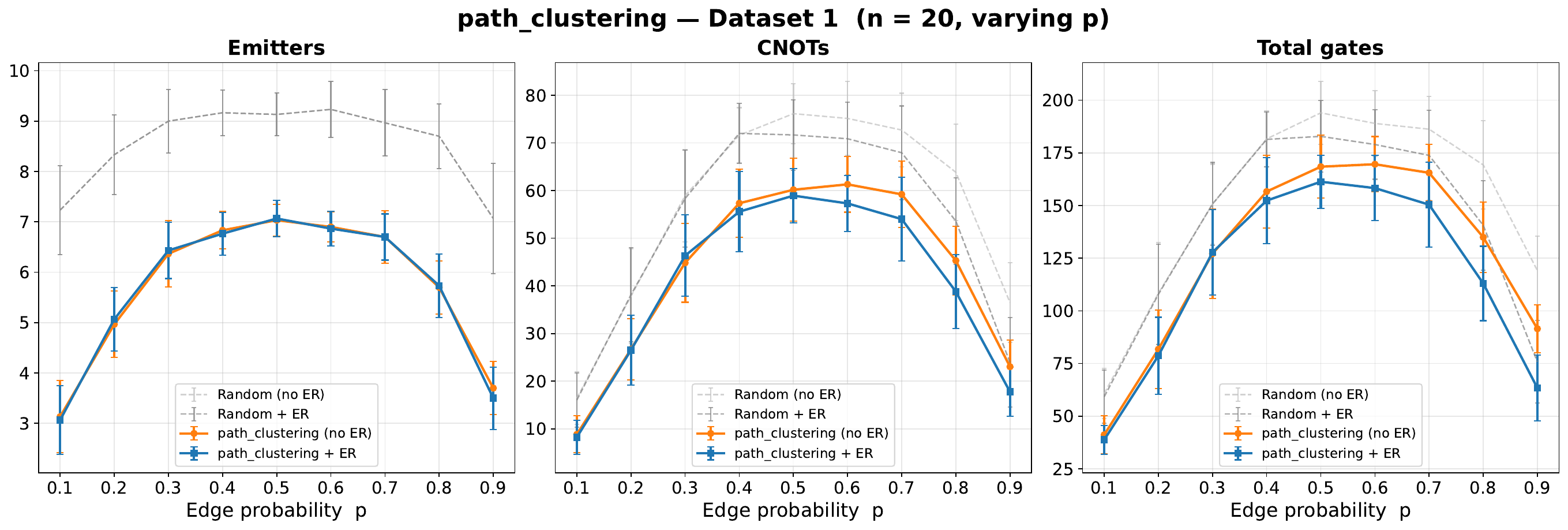}
    
    % \vspace{0.5cm} 
    
    % Bottom row: Dataset 2
    \includegraphics[width=0.9\textwidth]{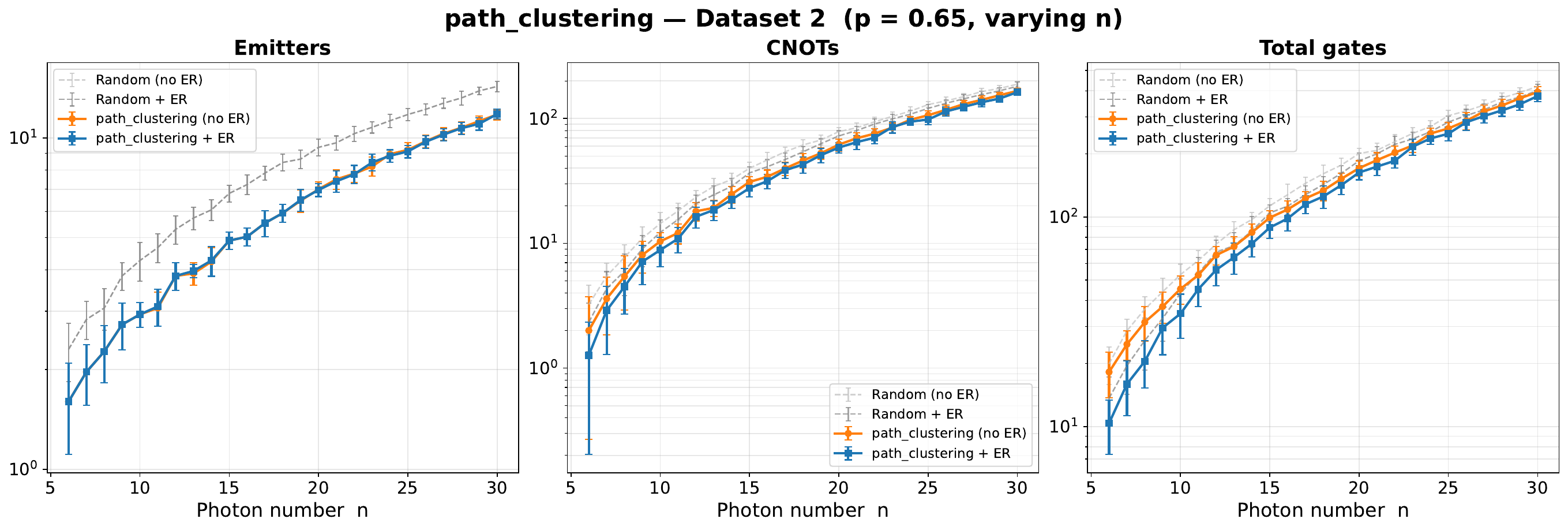}
    
    \caption{\label{fig:pipe_comp_path} Resource overhead scaling for photonic graph state generation using \texttt{path\_clustering}. The number of Emitters, CNOT gates, and Total gates is plotted for two scenarios. (Top) Performance on Dataset 1 with $n=20$ and varying edge probability $p$. (Bottom) Scaling against the number of nodes $n$ for Dataset 2 at a fixed $p=0.65$.}
\end{figure*}

\begin{figure*}[t]
    \centering
    % Top row: Dataset 1
    \includegraphics[width=0.9\textwidth]{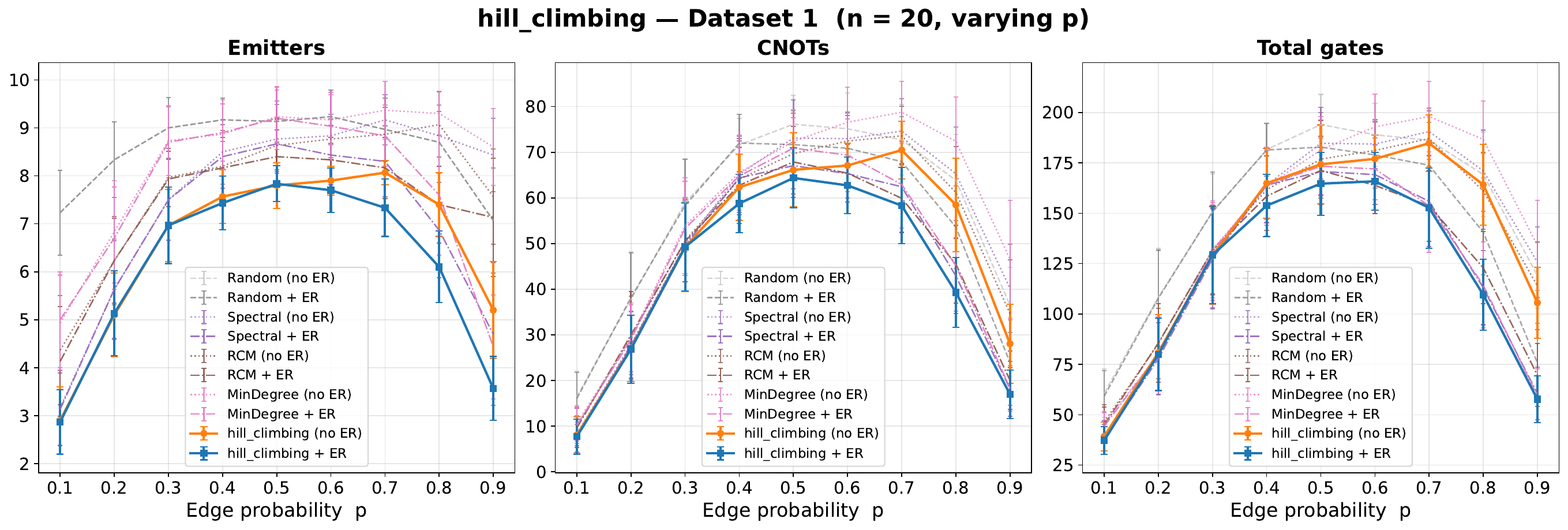}
    
    % \vspace{0.5cm} 
    
    % Bottom row: Dataset 2
    \includegraphics[width=0.9\textwidth]{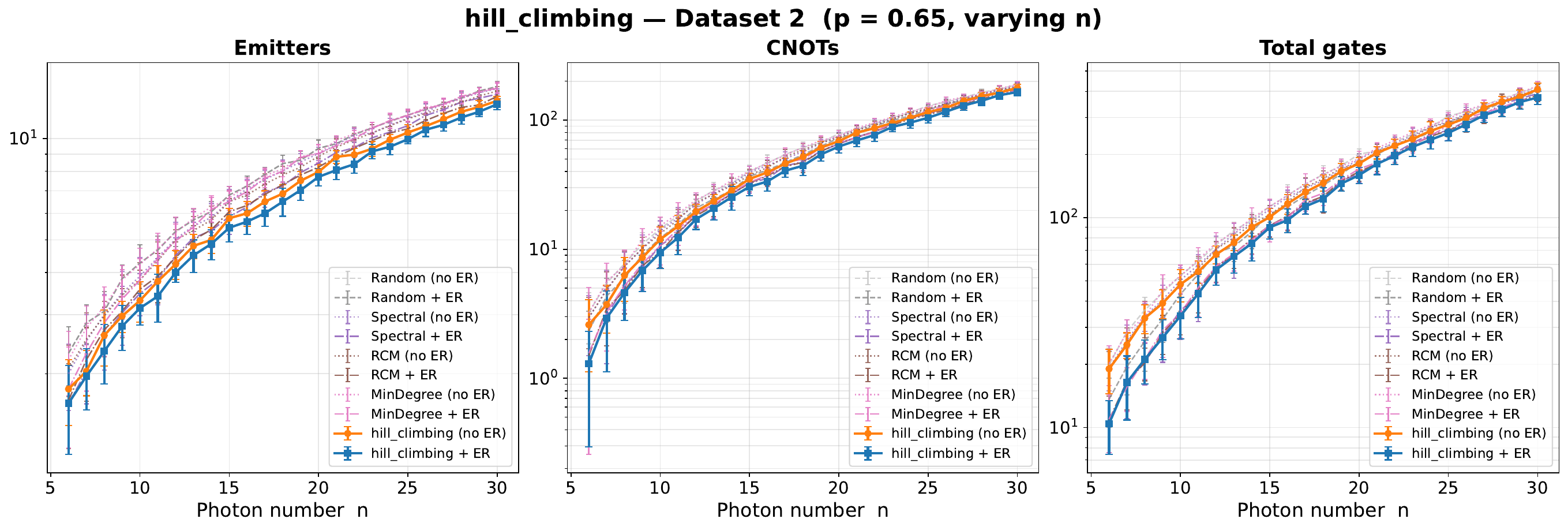}
    
    \caption{\label{fig:pipe_comp_rankwidth} Resource overhead scaling for photonic graph state generation using \texttt{hill\_climbing}. The number of Emitters, CNOT gates, and Total gates is plotted for two scenarios. (Top) Performance on Dataset 1 with $n=20$ and varying edge probability $p$. (Bottom) Scaling against the number of nodes $n$ for Dataset 2 at a fixed $p=0.65$.}
\end{figure*}

However, the edge reduction is not an equally important preprocessing step for each one of our algorithms. As presented in Figure \ref{fig:pipe_comp_path}, for all three metrics, the random emission ordering with edge reduction is inferior to the \texttt{path\_clustering} algorithm without edge reduction. In fact, this algorithm yields the same results for the number of emitters, regardless of edge reduction or not. This is not true for the emitter CNOTs and the total number of gates, where, as expected, for probabilities roughly above $0.5$, the edge reduction assists in further reduction of the resource for the state preparation. The same conclusions regarding our metrics and the effect of edge reduction can be made for \texttt{height\_function\_sa} as they are presented in Figure \ref{fig:pipe_comp_rankwidth}.

Until this point, we have established that for the same choice or algorithm to determine the emission ordering, the edge reduction has either a positive or no impact at all. Surely, it does not negatively affect our goal of reducing the state preparation resources. Bearing this in mind, the question we wish to answer is how the initial orderings $\pi_{ini}$ affect the final result. Since the phases A and B are the same for both \texttt{hill\_climbing} and \texttt{height\_function\_sa}, we present them only for the first in Figure \ref{fig:pipe_comp_rankwidth}. Interestingly, for large densities, so for $p$ close to $1$, the metrics that all $\pi_{ini}$ yield are better than the complete pipeline without the edge reduction step, but worse if the edge reduction is included. This fact highlights that every step included in the pipeline presented in Figure \ref{fig:pipeline} is necessary to achieve the highest reduction. Interestingly, if we have some of the $\pi_{ini}$ orderings for the non-edge reduced case, there is a chance we get worse results than the random emission with also no edge reduction, like for $p>0.7$ and the $\pi_{deg}$.

We now turn our focus on the same analysis but for a fixed $p=0.65$ and states with different photon numbers ranging from $6$ to $30$. As presented in the bottom panel of Figure \ref{fig:pipe_comp_saminla}, the \texttt{min\_la\_sa} algorithm yields comparable results for both cases, with an expected slight advantage of the full pipeline. We also observe that for the total number of gates, the edge-reduced random emission order is superior to the \texttt{min\_la\_sa} with no edge reduction. For the \texttt{path\_clustering}, we present the results in the bottom panel of Figure \ref{fig:pipe_comp_path}. For all metrics, there is a clear hierarchy where the random emission yields always inferior results, and the effect of edge reduction is relatively low, for instance, for the emitter CNOTs and the total gate count, if not negligible, as for the emitters required. We now turn our focus on the initial orderings $\pi_{ini}$ from phases A and B of \texttt{hill\_climbing} and \texttt{height\_function\_sa}. In Figure \ref{fig:pipe_comp_rankwidth}, we present the results, and for all metrics, it becomes clear that the final algorithms yield better results. Second comes the full pipeline without edge reduction, and finally, once more, there is a chance that one of the $\pi_{ini}$ can yield worse results than the random ordering.

\begin{figure*}[t]
    \centering
    % Top row: Dataset 1
    \includegraphics[width=0.9\textwidth]{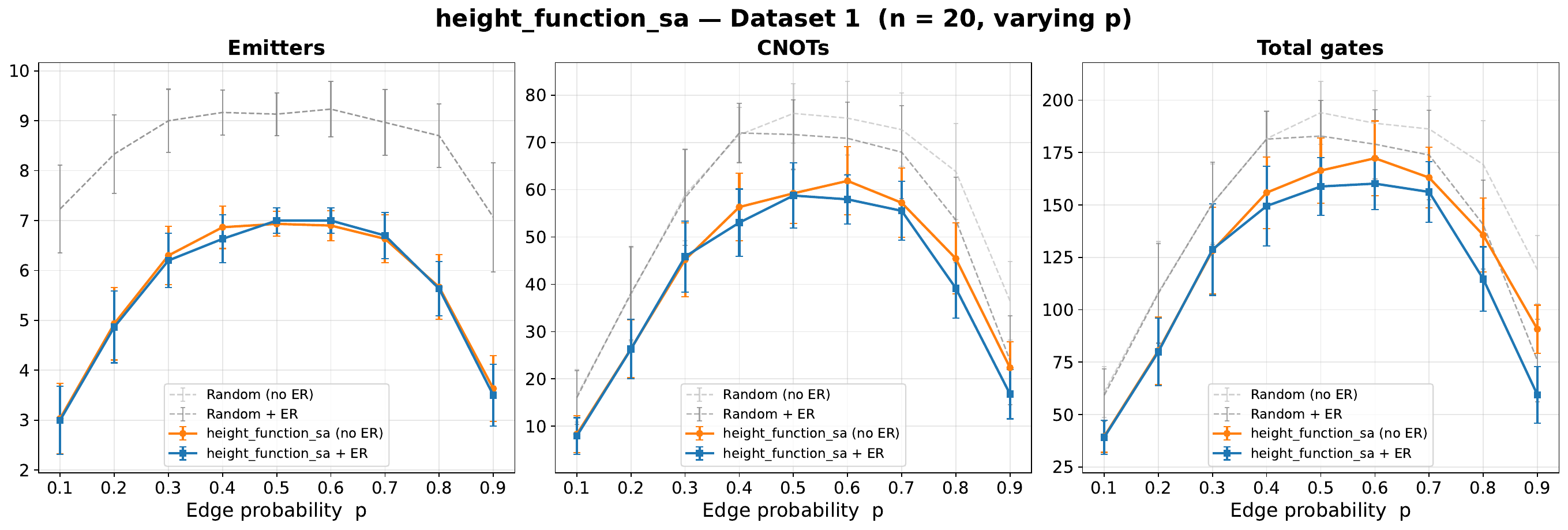}
    
    % \vspace{0.5cm} 
    
    % Bottom row: Dataset 2
    \includegraphics[width=0.9\textwidth]{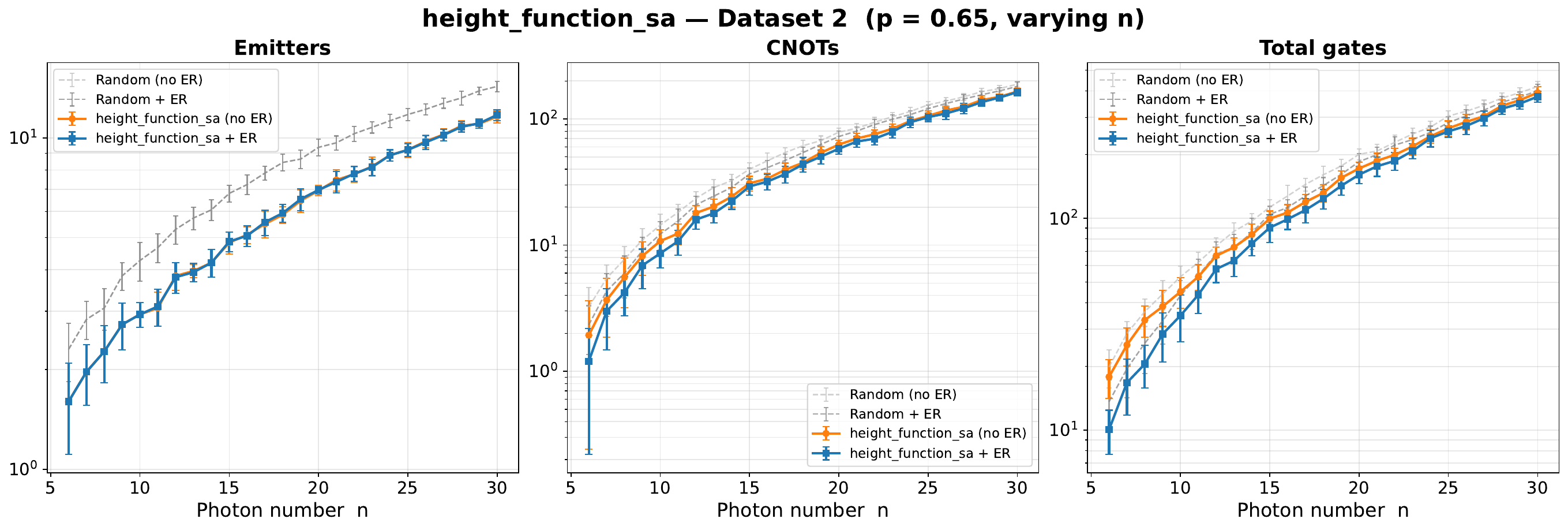}
    
    \caption{\label{fig:pipe_comp_rankwidthSA} Resource overhead scaling for photonic graph state generation using \texttt{height\_function\_sa}. The number of Emitters, CNOT gates, and Total gates is plotted for two scenarios. (Top) Performance on Dataset 1 with $n=20$ and varying edge probability $p$. (Bottom) Scaling against the number of nodes $n$ for Dataset 2 at a fixed $p=0.65$.}
\end{figure*}

With this, we conclude this detailed pipeline examination where we examined how each one of the used steps impacts our endeavor, which is the reduction of the resource overhead for the photonic graph state generation with quantum emitters. However, we tested and analyzed our algorithms on random graph states. Typically, we aim to create an exact one or choose one from a specific family. 

\subsection{Runtime Comparison}\label{app:runtime}
In addition to the comparison of emitter counts, CNOT counts, and total gate counts presented above, we also carried out a dedicated runtime comparison among the four algorithms themselves (\texttt{min\_la\_sa}, \texttt{hill\_climbing}, \texttt{height\_function\_sa}, and \texttt{path\_clustering}), estimating how long the four algorithms take relative to one another.

To ensure a fair comparison, several safeguards were taken into account during the measurement procedure. Each graph is timed three times per algorithm, and the reported runtime is the median over these repetitions rather than a single sample.

\begin{figure}[t]
    \centering
    % Top row: Dataset 1
    \includegraphics[width=0.4\textwidth]{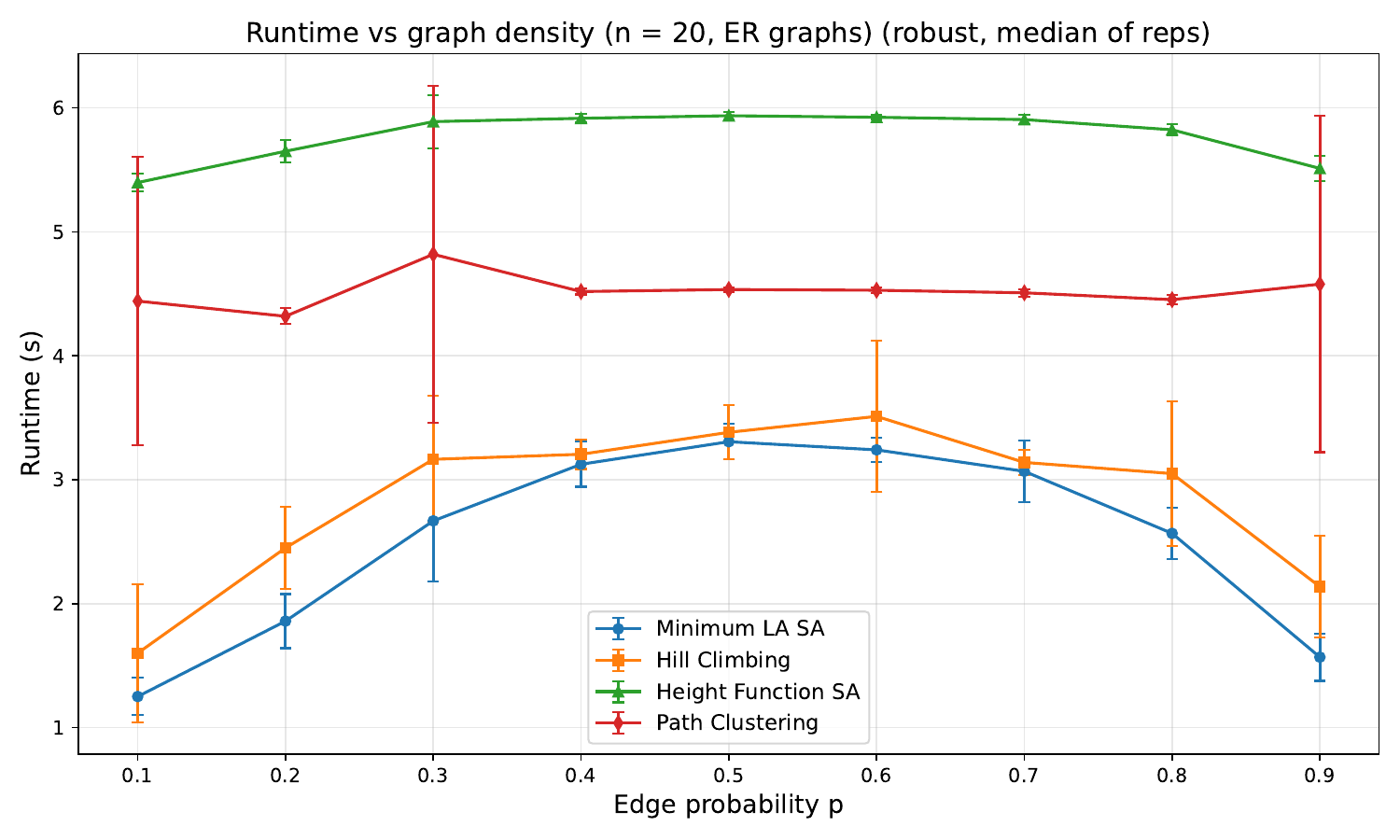}
    
    % \vspace{0.5cm} 
    
    % Bottom row: Dataset 2
    \includegraphics[width=0.4\textwidth]{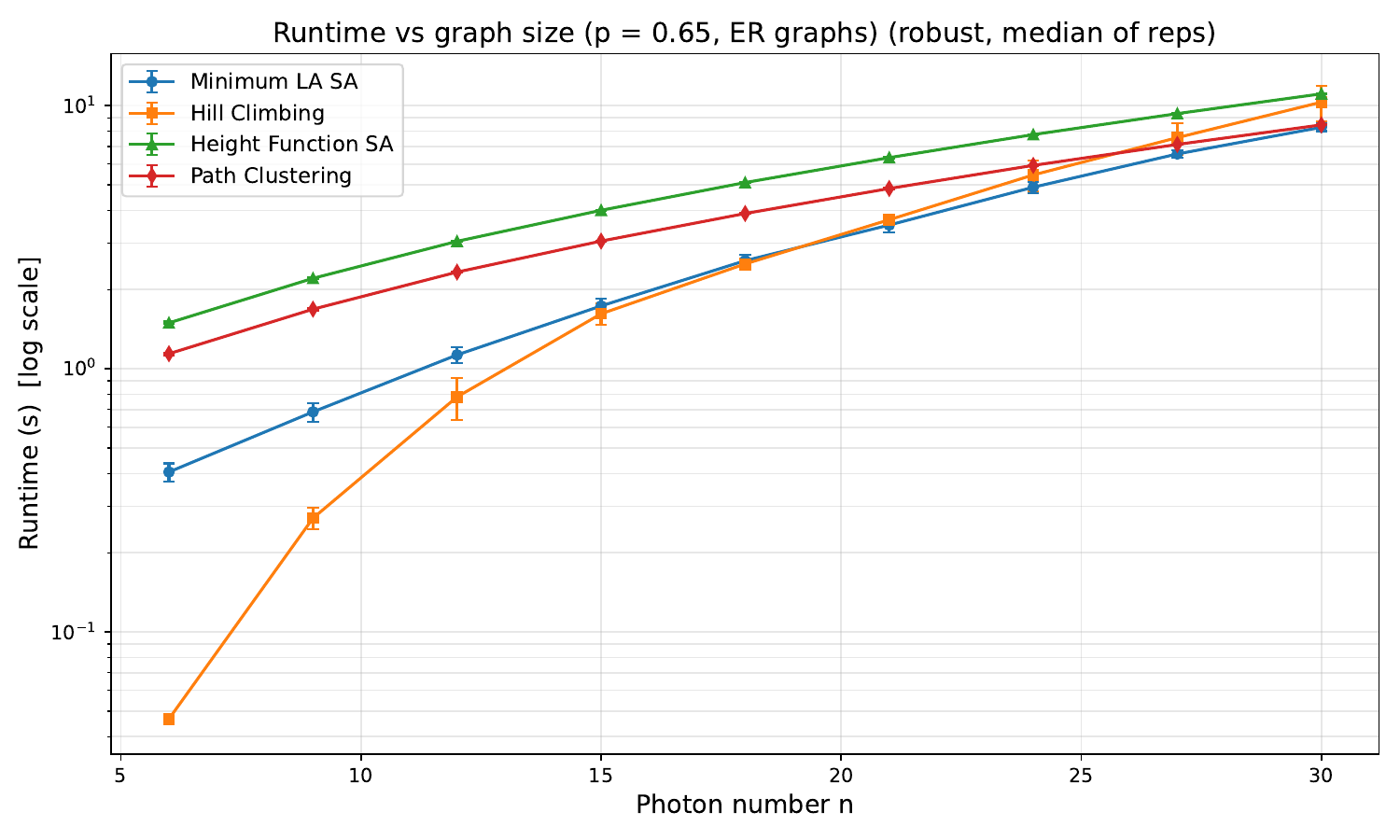}
    
    \caption{Runtime comparison across the four algorithms, reported as the median over repetitions, with error bars indicating variability across repetitions. (Top) Runtime as a function of edge probability $p$ for Erd\H{o}s-R\'enyi graphs with $n=20$ nodes (Dataset 1). (Bottom) Runtime as a function of the number of nodes $n$ for a fixed edge probability $p=0.65$ (Dataset 2), shown on a logarithmic scale.}
    \label{fig:runtime}
\end{figure}

The order in which the four algorithms are timed is independently reshuffled for every repetition, since under a fixed order, any drift in background system load over the course of a run would systematically favor the algorithm that happened to be measured first in every repetition rather than averaging out the noise. We use the same two datasets employed throughout this section: Dataset 1 ($n=20$ fixed, $p$ varying) and Dataset 2 ($p=0.65$ fixed, $n$ varying). The corresponding results are shown in Figure \ref{fig:runtime}.

As a function of edge density (top panel of Figure \ref{fig:runtime}), \texttt{min\_la\_sa} is consistently the fastest of the four algorithms across the entire range of $p$, closely followed by \texttt{hill\_climbing}. Both, however, exhibit a non-monotonic dependence on density: their runtime increases from the sparsest graphs up to a maximum around intermediate density ($p\approx0.5$--$0.6$), before decreasing again toward the densest graphs. In contrast, \texttt{height\_function\_sa} and \texttt{path\_clustering} are markedly slower throughout, with runtimes that depend only weakly on $p$; \texttt{path\_clustering} additionally shows substantially larger run-to-run variability at the sparsest and densest ends of the range, suggesting that its performance is more sensitive to the specific structure of individual graph instances in these regimes.

As a function of graph size at fixed density (bottom panel of Figure \ref{fig:runtime}), all four algorithms show the expected increase in runtime with the number of nodes $n$. For small graphs, \texttt{hill\_climbing} is dramatically faster than the other three algorithms, by roughly two orders of magnitude at $n=6$, with \texttt{min\_la\_sa} the second fastest. This advantage is not preserved as $n$ grows, however: the runtime of \texttt{hill\_climbing} increases more steeply than that of \texttt{min\_la\_sa}, and by $n\approx15$-$18$ the two become comparable, converging together with \texttt{path\_clustering} and \texttt{height\_function\_sa} toward similar runtimes for the largest graphs considered. Across the full range of $n$, \texttt{min\_la\_sa} exhibits the most consistent scaling and remains among the fastest algorithms throughout, while \texttt{height\_function\_sa} remains the slowest, or close to it, at essentially every graph size.

Taken together, these results indicate that \texttt{min\_la\_sa} offers the best overall runtime scaling among the four algorithms, with a particular advantage for larger and sparser graphs, while \texttt{hill\_climbing} is competitive primarily in the small-graph regime. \texttt{height\_function\_sa} and \texttt{path\_clustering}, while advantageous with respect to other metrics in certain regimes, are consistently more computationally expensive across the densities and graph sizes tested here. The non-monotonic density dependence observed for \texttt{min\_la\_sa} and \texttt{hill\_climbing} is consistent with a search landscape whose difficulty peaks at intermediate density, though we leave a more detailed characterization of this effect to future work.

\section{Simulation tables}
In the following, we have tables for the simulation results presented in the main text, for Shor's algorithm, RHG lattices, and quantum error correcting codes.
% \subsection{Shor's Algorithm}
\begin{table*}[t]
% \tiny
\setlength{\tabcolsep}{2.6pt}
\renewcommand{\arraystretch}{1.2}
\caption{Testing our algorithms on 23 examples for implementing Shor's algorithm in MBQC. Here HC denotes \texttt{hill\_climbing}, HFSA denotes \texttt{height\_function\_sa}, PC denotes \texttt{path\_clustering}, and MLSA denotes \texttt{min\_la\_sa}. The graphs used are identified by the parameters $(N,a)$ used to implement Shor's algorithm. The CNOT count is calculated using the emission algorithm presented by \cite{li_algo}.}
\label{tab:shor_results}

\begin{minipage}[t]{0.49\textwidth}
\centering
\begin{tabular}{@{}ccccccc@{}}
\toprule
$(N,a)$ & $|V|$ & $|E|$ & Alg. & Emit. & CNOTs & Gates \\
\midrule

(4,3) & 7 & 6 & HC & 1 & 0 & 8 \\
(4,3) & 7 & 6 & HFSA & 1 & 0 & 8 \\
(4,3) & 7 & 6 & PC & 1 & 0 & 8 \\
(4,3) & 7 & 6 & MLSA & 1 & 0 & 8 \\

\hline
(15,14) & 15 & 14 & HC & 1 & 0 & 16 \\
(15,14) & 15 & 14 & HFSA & 1 & 0 & 16 \\
(15,14) & 15 & 14 & PC & 1 & 0 & 16 \\
(15,14) & 15 & 14 & MLSA & 1 & 0 & 16 \\

\hline
(15,4) & 15 & 14 & HC & 1 & 0 & 16 \\
(15,4) & 15 & 14 & HFSA & 1 & 0 & 16 \\
(15,4) & 15 & 14 & PC & 1 & 0 & 16 \\
(15,4) & 15 & 14 & MLSA & 1 & 0 & 16 \\

\hline
(15,7) & 15 & 14 & HC & 1 & 0 & 16 \\
(15,7) & 15 & 14 & HFSA & 1 & 0 & 16 \\
(15,7) & 15 & 14 & PC & 1 & 0 & 16 \\
(15,7) & 15 & 14 & MLSA & 1 & 0 & 16 \\

\hline
(3,2) & 18 & 44 & HC & 2 & 3 & 25 \\
(3,2) & 18 & 44 & HFSA & 2 & 4 & 33 \\
(3,2) & 18 & 44 & PC & 2 & 4 & 33 \\
(3,2) & 18 & 44 & MLSA & 3 & 5 & 30 \\

\hline
(15,11) & 20 & 72 & HC & 3 & 9 & 52 \\
(15,11) & 20 & 72 & HFSA & 3 & 12 & 51 \\
(15,11) & 20 & 72 & PC & 3 & 12 & 51 \\
(15,11) & 20 & 72 & MLSA & 3 & 11 & 51 \\

\hline
(15,13) & 32 & 152 & HC & 3 & 13 & 64 \\
(15,13) & 32 & 152 & HFSA & 5 & 24 & 72 \\
(15,13) & 32 & 152 & PC & 4 & 19 & 72 \\
(15,13) & 32 & 152 & MLSA & 4 & 18 & 80 \\

\hline
(15,2) & 32 & 152 & HC & 4 & 17 & 83 \\
(15,2) & 32 & 152 & HFSA & 4 & 22 & 87 \\
(15,2) & 32 & 152 & PC & 5 & 22 & 83 \\
(15,2) & 32 & 152 & MLSA & 5 & 21 & 73 \\

\hline
(15,8) & 32 & 152 & HC & 5 & 22 & 83 \\
(15,8) & 32 & 152 & HFSA & 4 & 17 & 78 \\
(15,8) & 32 & 152 & PC & 5 & 19 & 69 \\
(15,8) & 32 & 152 & MLSA & 4 & 15 & 60 \\

\hline
(8,5) & 37 & 96 & HC & 4 & 14 & 72 \\
(8,5) & 37 & 96 & HFSA & 4 & 9 & 63 \\
(8,5) & 37 & 96 & PC & 4 & 9 & 63 \\
(8,5) & 37 & 96 & MLSA & 4 & 11 & 63 \\

\hline
(8,3) & 58 & 224 & HC & 6 & 31 & 130 \\
(8,3) & 58 & 224 & HFSA & 6 & 31 & 130 \\
(8,3) & 58 & 224 & PC & 6 & 31 & 130 \\
(8,3) & 58 & 224 & MLSA & 7 & 31 & 123 \\

\bottomrule
\end{tabular}
\end{minipage}
\hfill
\begin{minipage}[t]{0.49\textwidth}
\centering
\begin{tabular}{@{}ccccccc@{}}
\toprule
$(N,a)$ & $|V|$ & $|E|$ & Alg. & Emit. & CNOTs & Gates \\
\midrule

(6,5) & 76 & 368 & HC & 6 & 35 & 158 \\
(6,5) & 76 & 368 & HFSA & 7 & 35 & 158 \\
(6,5) & 76 & 368 & PC & 7 & 35 & 158 \\
(6,5) & 76 & 368 & MLSA & 7 & 33 & 146 \\

\hline
(5,4) & 77 & 384 & HC & 9 & 45 & 171 \\
(5,4) & 77 & 384 & HFSA & 9 & 45 & 171 \\
(5,4) & 77 & 384 & PC & 8 & 58 & 194 \\
(5,4) & 77 & 384 & MLSA & 9 & 56 & 196 \\

\hline
(7,6) & 127 & 960 & HC & 12 & 69 & 265 \\
(7,6) & 127 & 960 & HFSA & 7 & 68 & 259 \\
(7,6) & 127 & 960 & PC & 15 & 76 & 282 \\
(7,6) & 127 & 960 & MLSA & 7 & 73 & 253 \\

\hline
(16,9) & 130 & 896 & HC & 8 & 67 & 256 \\
(16,9) & 130 & 896 & HFSA & 9 & 82 & 265 \\
(16,9) & 130 & 896 & PC & 12 & 86 & 281 \\
(16,9) & 130 & 896 & MLSA & 11 & 80 & 259 \\

\hline
(8,7) & 134 & 1056 & HC & 14 & 78 & 294 \\
(8,7) & 134 & 1056 & HFSA & 12 & 118 & 342 \\
(8,7) & 134 & 1056 & PC & 13 & 162 & 418 \\
(8,7) & 134 & 1056 & MLSA & 11 & 102 & 320 \\

\hline
(10,9) & 199 & 2192 & HC & 17 & 182 & 522 \\
(10,9) & 199 & 2192 & HFSA & 15 & 170 & 471 \\
(10,9) & 199 & 2192 & PC & 15 & 170 & 471 \\
(10,9) & 199 & 2192 & MLSA & 13 & 176 & 483 \\

\hline
(5,2) & 210 & 3008 & HC & 13 & 156 & 495 \\
(5,2) & 210 & 3008 & HFSA & 16 & 163 & 510 \\
(5,2) & 210 & 3008 & PC & 16 & 163 & 510 \\
(5,2) & 210 & 3008 & MLSA & 20 & 165 & 509 \\

\hline
(5,3) & 210 & 3024 & HC & 14 & 186 & 585 \\
(5,3) & 210 & 3024 & HFSA & 17 & 196 & 588 \\
(5,3) & 210 & 3024 & PC & 17 & 196 & 588 \\
(5,3) & 210 & 3024 & MLSA & 15 & 190 & 594 \\

\hline
(12,5) & 266 & 3568 & HC & 18 & 219 & 671 \\
(12,5) & 266 & 3568 & HFSA & 18 & 219 & 671 \\
(12,5) & 266 & 3568 & PC & 18 & 219 & 671 \\
(12,5) & 266 & 3568 & MLSA & 15 & 255 & 718 \\

\hline
(12,7) & 309 & 5504 & HC & 25 & 283 & 823 \\
(12,7) & 309 & 5504 & HFSA & 25 & 322 & 897 \\
(12,7) & 309 & 5504 & PC & 25 & 322 & 897 \\
(12,7) & 309 & 5504 & MLSA & 16 & 348 & 881 \\

\hline
(9,8) & 332 & 5824 & HC & 27 & 308 & 894 \\
(9,8) & 332 & 5824 & HFSA & 27 & 308 & 894 \\
(9,8) & 332 & 5824 & PC & 27 & 308 & 894 \\
(9,8) & 332 & 5824 & MLSA & 24 & 338 & 891 \\

\hline
(12,11) & 365 & 6832 & HC & 24 & 354 & 1029 \\
(12,11) & 365 & 6832 & HFSA & 25 & 366 & 1031 \\
(12,11) & 365 & 6832 & PC & 25 & 366 & 1031 \\
(12,11) & 365 & 6832 & MLSA & 24 & 379 & 996 \\

\bottomrule
\end{tabular}
\end{minipage}

\end{table*}
% \subsection{RHG Lattices}
\begin{table*}[t]
% \tiny
\setlength{\tabcolsep}{2.6pt}
\renewcommand{\arraystretch}{1.2}
\caption{Testing our algorithms on RHG lattices up to and including dimension $(3,4,4)$. Here HC denotes \texttt{hill\_climbing}, HFSA denotes \texttt{height\_function\_sa}, PC denotes \texttt{path\_clustering}, and MLSA denotes \texttt{min\_la\_sa}. We report the number of emitters, emitter CNOTs, and total predicted gate count. The CNOT count is calculated using the emission algorithm presented by \cite{li_algo}.}
\label{tab:rhg_results-total}

\begin{minipage}[t]{0.49\textwidth}
\centering
\begin{tabular}{@{}ccccccc@{}}
\toprule
Dims & $|V|$ & $|E|$ & Alg. & Emit. & CNOTs & Gates \\
\midrule

$(1,1,1)$ & 18 & 24 & HC & 4 & 14 & 40 \\
$(1,1,1)$ & 18 & 24 & HFSA & 4 & 14 & 40 \\
$(1,1,1)$ & 18 & 24 & PC & 4 & 14 & 40 \\
$(1,1,1)$ & 18 & 24 & MLSA & 4 & 15 & 45 \\

\hline
$(1,1,2)$ & 31 & 44 & HC & 4 & 30 & 73 \\
$(1,1,2)$ & 31 & 44 & HFSA & 4 & 30 & 73 \\
$(1,1,2)$ & 31 & 44 & PC & 4 & 30 & 73 \\
$(1,1,2)$ & 31 & 44 & MLSA & 5 & 30 & 70 \\

\hline
$(1,1,3)$ & 44 & 64 & HC & 4 & 39 & 93 \\
$(1,1,3)$ & 44 & 64 & HFSA & 4 & 39 & 95 \\
$(1,1,3)$ & 44 & 64 & PC & 4 & 39 & 95 \\
$(1,1,3)$ & 44 & 64 & MLSA & 4 & 39 & 93 \\

\hline
$(1,2,2)$ & 53 & 80 & HC & 7 & 50 & 126 \\
$(1,2,2)$ & 53 & 80 & HFSA & 7 & 50 & 126 \\
$(1,2,2)$ & 53 & 80 & PC & 7 & 50 & 126 \\
$(1,2,2)$ & 53 & 80 & MLSA & 7 & 51 & 115 \\

\hline
$(1,1,4)$ & 57 & 84 & HC & 4 & 51 & 118 \\
$(1,1,4)$ & 57 & 84 & HFSA & 4 & 59 & 126 \\
$(1,1,4)$ & 57 & 84 & PC & 4 & 59 & 126 \\
$(1,1,4)$ & 57 & 84 & MLSA & 4 & 54 & 125 \\

\hline
$(1,2,3)$ & 75 & 116 & HC & 7 & 79 & 171 \\
$(1,2,3)$ & 75 & 116 & HFSA & 7 & 85 & 181 \\
$(1,2,3)$ & 75 & 116 & PC & 7 & 85 & 181 \\
$(1,2,3)$ & 75 & 116 & MLSA & 11 & 85 & 232 \\

\hline
$(2,2,2)$ & 90 & 144 & HC & 11 & 101 & 282 \\
$(2,2,2)$ & 90 & 144 & HFSA & 11 & 101 & 282 \\
$(2,2,2)$ & 90 & 144 & PC & 11 & 101 & 282 \\
$(2,2,2)$ & 90 & 144 & MLSA & 12 & 114 & 242 \\

\hline
$(1,2,4)$ & 97 & 152 & HC & 7 & 111 & 227 \\
$(1,2,4)$ & 97 & 152 & HFSA & 7 & 123 & 237 \\
$(1,2,4)$ & 97 & 152 & PC & 8 & 112 & 237 \\
$(1,2,4)$ & 97 & 152 & MLSA & 10 & 102 & 258 \\

\hline
$(1,3,3)$ & 106 & 168 & HC & 10 & 110 & 254 \\
$(1,3,3)$ & 106 & 168 & HFSA & 10 & 112 & 256 \\
$(1,3,3)$ & 106 & 168 & PC & 10 & 112 & 256 \\
$(1,3,3)$ & 106 & 168 & MLSA & 11 & 131 & 292 \\

\hline
$(2,2,3)$ & 127 & 208 & HC & 12 & 188 & 357 \\
$(2,2,3)$ & 127 & 208 & HFSA & 12 & 174 & 337 \\
$(2,2,3)$ & 127 & 208 & PC & 12 & 174 & 337 \\
$(2,2,3)$ & 127 & 208 & MLSA & 17 & 155 & 325 \\

\bottomrule
\end{tabular}
\end{minipage}
\hfill
\begin{minipage}[t]{0.49\textwidth}
\centering
\begin{tabular}{@{}ccccccc@{}}
\toprule
Dims & $|V|$ & $|E|$ & Alg. & Emit. & CNOTs & Gates \\
\midrule

$(1,3,4)$ & 137 & 220 & HC & 10 & 176 & 347 \\
$(1,3,4)$ & 137 & 220 & HFSA & 10 & 179 & 342 \\
$(1,3,4)$ & 137 & 220 & PC & 10 & 179 & 342 \\
$(1,3,4)$ & 137 & 220 & MLSA & 14 & 179 & 418 \\

\hline
$(2,2,4)$ & 164 & 272 & HC & 12 & 236 & 434 \\
$(2,2,4)$ & 164 & 272 & HFSA & 12 & 237 & 441 \\
$(2,2,4)$ & 164 & 272 & PC & 12 & 237 & 441 \\
$(2,2,4)$ & 164 & 272 & MLSA & 22 & 238 & 466 \\

\hline
$(1,4,4)$ & 177 & 288 & HC & 13 & 203 & 465 \\
$(1,4,4)$ & 177 & 288 & HFSA & 13 & 200 & 450 \\
$(1,4,4)$ & 177 & 288 & PC & 21 & 235 & 503 \\
$(1,4,4)$ & 177 & 288 & MLSA & 25 & 213 & 583 \\

\hline
$(2,3,3)$ & 179 & 300 & HC & 17 & 227 & 479 \\
$(2,3,3)$ & 179 & 300 & HFSA & 17 & 226 & 478 \\
$(2,3,3)$ & 179 & 300 & PC & 19 & 240 & 534 \\
$(2,3,3)$ & 179 & 300 & MLSA & 24 & 231 & 468 \\

\hline
$(2,3,4)$ & 231 & 392 & HC & 17 & 360 & 644 \\
$(2,3,4)$ & 231 & 392 & HFSA & 17 & 368 & 650 \\
$(2,3,4)$ & 231 & 392 & PC & 24 & 348 & 665 \\
$(2,3,4)$ & 231 & 392 & MLSA & 31 & 349 & 661 \\

\hline
$(3,3,3)$ & 252 & 432 & HC & 21 & 303 & 810 \\
$(3,3,3)$ & 252 & 432 & HFSA & 21 & 303 & 810 \\
$(3,3,3)$ & 252 & 432 & PC & 31 & 363 & 736 \\
$(3,3,3)$ & 252 & 432 & MLSA & 25 & 345 & 664 \\

\hline
$(2,4,4)$ & 298 & 512 & HC & 22 & 419 & 849 \\
$(2,4,4)$ & 298 & 512 & HFSA & 22 & 415 & 851 \\
$(2,4,4)$ & 298 & 512 & PC & 29 & 417 & 811 \\
$(2,4,4)$ & 298 & 512 & MLSA & 40 & 425 & 831 \\

\hline
$(3,3,4)$ & 325 & 564 & HC & 24 & 574 & 989 \\
$(3,3,4)$ & 325 & 564 & HFSA & 24 & 560 & 959 \\
$(3,3,4)$ & 325 & 564 & PC & 31 & 486 & 908 \\
$(3,3,4)$ & 325 & 564 & MLSA & 32 & 450 & 853 \\

\hline
$(3,4,4)$ & 419 & 736 & HC & 28 & 519 & 1365 \\
$(3,4,4)$ & 419 & 736 & HFSA & 29 & 528 & 1380 \\
$(3,4,4)$ & 419 & 736 & PC & 54 & 700 & 1432 \\
$(3,4,4)$ & 419 & 736 & MLSA & 40 & 602 & 1125 \\

\bottomrule
\end{tabular}
\end{minipage}

\end{table*}

% \subsection{QECC}\label{app:qecc}

\begin{table}[t]
\centering
\scriptsize
\setlength{\tabcolsep}{2.5pt}
\renewcommand{\arraystretch}{1.4}
\caption{Testing our algorithms on several QECC families. HC denotes \texttt{hill\_climbing}, HFSA denotes \texttt{height\_function\_sa}, PC denotes \texttt{path\_clustering}, and MLSA denotes \texttt{min\_la\_sa}.}
\label{tab:qecc-init}

\resizebox{\columnwidth}{!}{%
\begin{tabular}{@{}ccccccc@{}}
\toprule
QECC & $|V|$ & $|E|$ & Alg. & Emit. & CNOTs & Gates \\
\midrule

Steane & 7 & 12 & HC & 3 & 5 & 19 \\
Steane & 7 & 12 & HFSA & 3 & 5 & 19 \\
Steane & 7 & 12 & PC & 3 & 5 & 19 \\
Steane & 7 & 12 & MLSA & 3 & 5 & 19 \\
\hline
Shor & 9 & 6 & HC & 1 & 0 & 20 \\
Shor & 9 & 6 & HFSA & 1 & 0 & 20 \\
Shor & 9 & 6 & PC & 1 & 0 & 20 \\
Shor & 9 & 6 & MLSA & 1 & 0 & 14 \\
\hline
$\mathrm{HGP}(34,4)$ & 34 & 81 & HC & 7 & 38 & 91 \\
$\mathrm{HGP}(34,4)$ & 34 & 81 & HFSA & 6 & 31 & 95 \\
$\mathrm{HGP}(34,4)$ & 34 & 81 & PC & 6 & 31 & 95 \\
$\mathrm{HGP}(34,4)$ & 34 & 81 & MLSA & 7 & 35 & 107 \\
\hline
$\mathrm{HGP}(52,4)$ & 52 & 164 & HC & 10 & 88 & 177 \\
$\mathrm{HGP}(52,4)$ & 52 & 164 & HFSA & 11 & 95 & 190 \\
$\mathrm{HGP}(52,4)$ & 52 & 164 & PC & 9 & 81 & 182 \\
$\mathrm{HGP}(52,4)$ & 52 & 164 & MLSA & 11 & 87 & 183 \\
\hline
$\mathrm{BB}[[72,12,6]]$ & 72 & 1283 & HC & 33 & 1127 & 2191 \\
$\mathrm{BB}[[72,12,6]]$ & 72 & 1283 & HFSA & 32 & 1075 & 2044 \\
$\mathrm{BB}[[72,12,6]]$ & 72 & 1283 & PC & 32 & 1206 & 2381 \\
$\mathrm{BB}[[72,12,6]]$ & 72 & 1283 & MLSA & 35 & 1179 & 2212 \\
\hline
$\mathrm{HGP}(80,16)$ & 80 & 453 & HC & 21 & 240 & 531 \\
$\mathrm{HGP}(80,16)$ & 80 & 453 & HFSA & 17 & 255 & 594 \\
$\mathrm{HGP}(80,16)$ & 80 & 453 & PC & 16 & 270 & 627 \\
$\mathrm{HGP}(80,16)$ & 80 & 453 & MLSA & 19 & 243 & 542 \\
\hline
$\mathrm{BB}[[90,8,10]]$ & 90 & 1347 & HC & 40 & 1216 & 2177 \\
$\mathrm{BB}[[90,8,10]]$ & 90 & 1347 & HFSA & 35 & 1260 & 2434 \\
$\mathrm{BB}[[90,8,10]]$ & 90 & 1347 & PC & 33 & 1213 & 2345 \\
$\mathrm{BB}[[90,8,10]]$ & 90 & 1347 & MLSA & 42 & 1288 & 2380 \\
\hline
$\mathrm{HGP}(106,16)$ & 106 & 542 & HC & 22 & 382 & 769 \\
$\mathrm{HGP}(106,16)$ & 106 & 542 & HFSA & 24 & 408 & 817 \\
$\mathrm{HGP}(106,16)$ & 106 & 542 & PC & 24 & 408 & 817 \\
$\mathrm{HGP}(106,16)$ & 106 & 542 & MLSA & 24 & 326 & 707 \\
\hline
$\mathrm{BB}[[108,8,10]]$ & 108 & 2281 & HC & 50 & 2474 & 4535 \\
$\mathrm{BB}[[108,8,10]]$ & 108 & 2281 & HFSA & 39 & 1905 & 3742 \\
$\mathrm{BB}[[108,8,10]]$ & 108 & 2281 & PC & 43 & 2291 & 4507 \\
$\mathrm{BB}[[108,8,10]]$ & 108 & 2281 & MLSA & 52 & 2465 & 4512 \\
\hline
$\mathrm{HGP}(130,4)$ & 130 & 1080 & HC & 17 & 341 & 687 \\
$\mathrm{HGP}(130,4)$ & 130 & 1080 & HFSA & 21 & 337 & 685 \\
$\mathrm{HGP}(130,4)$ & 130 & 1080 & PC & 21 & 337 & 685 \\
$\mathrm{HGP}(130,4)$ & 130 & 1080 & MLSA & 24 & 346 & 706 \\
\hline
$\mathrm{BB}[[144,12,12]]$ & 144 & 4928 & HC & 68 & 4255 & 7644 \\
$\mathrm{BB}[[144,12,12]]$ & 144 & 4928 & HFSA & 53 & 3268 & 6174 \\
$\mathrm{BB}[[144,12,12]]$ & 144 & 4928 & PC & 55 & 3684 & 6867 \\
$\mathrm{BB}[[144,12,12]]$ & 144 & 4928 & MLSA & 69 & 4335 & 7870 \\
\bottomrule
\end{tabular}%
}
\end{table}

\begin{table}[t]
\centering
\scriptsize
\setlength{\tabcolsep}{2.5pt}
\renewcommand{\arraystretch}{1.4}
\caption{Testing our algorithms on several QECC families after a logical Bell pair is imposed. HC denotes \texttt{hill\_climbing}, HFSA denotes \texttt{height\_function\_sa}, PC denotes \texttt{path\_clustering}, and MLSA denotes \texttt{min\_la\_sa}.}
\label{tab:qecc-bell}

\resizebox{\columnwidth}{!}{%
\begin{tabular}{@{}ccccccc@{}}
\toprule
Bell QECC & $|V|$ & $|E|$ & Alg. & Emit. & CNOTs & Gates \\
\midrule

Steane & 14 & 33 & HC & 3 & 17 & 54 \\
Steane & 14 & 33 & HFSA & 3 & 19 & 60 \\
Steane & 14 & 33 & PC & 3 & 19 & 60 \\
Steane & 14 & 33 & MLSA & 3 & 16 & 53 \\
\hline
Shor & 18 & 21 & HC & 3 & 3 & 29 \\
Shor & 18 & 21 & HFSA & 2 & 4 & 35 \\
Shor & 18 & 21 & PC & 3 & 3 & 29 \\
Shor & 18 & 21 & MLSA & 3 & 4 & 32 \\
\hline
$\mathrm{HGP}(34,4)$ & 68 & 306 & HC & 14 & 145 & 324 \\
$\mathrm{HGP}(34,4)$ & 68 & 306 & HFSA & 14 & 174 & 444 \\
$\mathrm{HGP}(34,4)$ & 68 & 306 & PC & 15 & 167 & 410 \\
$\mathrm{HGP}(34,4)$ & 68 & 306 & MLSA & 23 & 265 & 533 \\
\hline
$\mathrm{HGP}(52,4)$ & 104 & 466 & HC & 16 & 267 & 549 \\
$\mathrm{HGP}(52,4)$ & 104 & 466 & HFSA & 16 & 267 & 549 \\
$\mathrm{HGP}(52,4)$ & 104 & 466 & PC & 16 & 267 & 549 \\
$\mathrm{HGP}(52,4)$ & 104 & 466 & MLSA & 17 & 304 & 623 \\
\hline
$\mathrm{BB}[[72,12,6]]$ & 144 & 7164 & HC & 49 & 3474 & 6211 \\
$\mathrm{BB}[[72,12,6]]$ & 144 & 7164 & HFSA & 51 & 3711 & 6921 \\
$\mathrm{BB}[[72,12,6]]$ & 144 & 7164 & PC & 51 & 3711 & 6921 \\
$\mathrm{BB}[[72,12,6]]$ & 144 & 7164 & MLSA & 49 & 3510 & 6404 \\
\hline
$\mathrm{HGP}(80,16)$ & 160 & 1874 & HC & 54 & 1414 & 2416 \\
$\mathrm{HGP}(80,16)$ & 160 & 1874 & HFSA & 41 & 1392 & 2575 \\
$\mathrm{HGP}(80,16)$ & 160 & 1874 & PC & 40 & 1147 & 2103 \\
$\mathrm{HGP}(80,16)$ & 160 & 1874 & MLSA & 48 & 1536 & 2727 \\
\hline
$\mathrm{BB}[[90,8,10]]$ & 180 & 7266 & HC & 56 & 4433 & 7875 \\
$\mathrm{BB}[[90,8,10]]$ & 180 & 7266 & HFSA & 54 & 4728 & 8985 \\
$\mathrm{BB}[[90,8,10]]$ & 180 & 7266 & PC & 53 & 4308 & 7901 \\
$\mathrm{BB}[[90,8,10]]$ & 180 & 7266 & MLSA & 55 & 4242 & 7753 \\
\hline
$\mathrm{HGP}(106,16)$ & 212 & 1741 & HC & 55 & 1393 & 2405 \\
$\mathrm{HGP}(106,16)$ & 212 & 1741 & HFSA & 50 & 1722 & 2943 \\
$\mathrm{HGP}(106,16)$ & 212 & 1741 & PC & 50 & 1722 & 2943 \\
$\mathrm{HGP}(106,16)$ & 212 & 1741 & MLSA & 48 & 1271 & 2392 \\
\hline
$\mathrm{BB}[[108,8,10]]$ & 216 & 15014 & HC & 92 & 9085 & 16323 \\
$\mathrm{BB}[[108,8,10]]$ & 216 & 15014 & HFSA & 83 & 9165 & 16980 \\
$\mathrm{BB}[[108,8,10]]$ & 216 & 15014 & PC & 83 & 9165 & 16980 \\
$\mathrm{BB}[[108,8,10]]$ & 216 & 15014 & MLSA & 84 & 8713 & 15530 \\
\hline
$\mathrm{HGP}(130,4)$ & 260 & 2416 & HC & 35 & 1354 & 2700 \\
$\mathrm{HGP}(130,4)$ & 260 & 2416 & HFSA & 35 & 923 & 1822 \\
$\mathrm{HGP}(130,4)$ & 260 & 2416 & PC & 35 & 923 & 1822 \\
$\mathrm{HGP}(130,4)$ & 260 & 2416 & MLSA & 40 & 1270 & 2504 \\
\hline
$\mathrm{BB}[[144,12,12]]$ & 288 & 28888 & HC & 96 & 13941 & 25242 \\
$\mathrm{BB}[[144,12,12]]$ & 288 & 28888 & HFSA & 102 & 14617 & 26590 \\
$\mathrm{BB}[[144,12,12]]$ & 288 & 28888 & PC & 102 & 14617 & 26590 \\
$\mathrm{BB}[[144,12,12]]$ & 288 & 28888 & MLSA & 98 & 13598 & 23192 \\
\bottomrule
\end{tabular}%
}
\end{table}

\begin{table*}[t]
% \tiny
\setlength{\tabcolsep}{2.6pt}
\renewcommand{\arraystretch}{1.2}
\caption{Testing our algorithms on several QECC families after $l_f$ foliations imposed. HC denotes \texttt{hill\_climbing}, HFSA denotes \texttt{height\_function\_sa}, PC denotes \texttt{path\_clustering}, and MLSA denotes \texttt{min\_la\_sa}.}

\label{tab:qecc-fol}

\begin{minipage}[t]{0.49\textwidth}
\centering
\begin{tabular}{@{}cccccccc@{}}
\toprule
$l_f$ QECC & $|V|$ & $|E|$ & Alg. & Emit. & CNOTs & Gates \\
\midrule

1 Steane & 20 & 31 & HC & 5 & 22 & 63 \\

1 Steane & 20 & 31 & HFSA & 5 & 22 & 63 \\

1 Steane & 20 & 31 & PC & 5 & 22 & 63 \\

1 Steane & 20 & 31 & MLSA & 6 & 20 & 67 \\

\hline

1 Shor & 26 & 33 & HC & 3 & 13 & 64 \\

1 Shor & 26 & 33 & HFSA & 4 & 15 & 63 \\

1 Shor & 26 & 33 & PC & 4 & 15 & 68 \\

1 Shor & 26 & 33 & MLSA & 3 & 15 & 69 \\

\hline

3 Steane & 60 & 107 & HC & 7 & 98 & 175 \\

3 Steane & 60 & 107 & HFSA & 7 & 92 & 173 \\

3 Steane & 60 & 107 & PC & 7 & 92 & 173 \\

3 Steane & 60 & 107 & MLSA & 7 & 90 & 173 \\

\hline

3 Shor & 78 & 117 & HC & 7 & 84 & 240 \\

3 Shor & 78 & 117 & HFSA & 8 & 82 & 238 \\

3 Shor & 78 & 117 & PC & 13 & 98 & 285 \\

3 Shor & 78 & 117 & MLSA & 10 & 68 & 173 \\

\hline

1 $\mathrm{HGP}(34,4)$ & 98 & 178 & HC & 15 & 136 & 290 \\

1 $\mathrm{HGP}(34,4)$ & 98 & 178 & HFSA & 15 & 136 & 290 \\

1 $\mathrm{HGP}(34,4)$ & 98 & 178 & PC & 15 & 168 & 373 \\

1 $\mathrm{HGP}(34,4)$ & 98 & 178 & MLSA & 15 & 147 & 314 \\

\hline

5 Steane & 100 & 183 & HC & 7 & 175 & 294 \\

5 Steane & 100 & 183 & HFSA & 7 & 173 & 294 \\

5 Steane & 100 & 183 & PC & 7 & 173 & 294 \\

5 Steane & 100 & 183 & MLSA & 8 & 150 & 274 \\

\hline

5 Shor & 130 & 201 & HC & 9 & 139 & 292 \\

5 Shor & 130 & 201 & HFSA & 9 & 140 & 293 \\

5 Shor & 130 & 201 & PC & 22 & 155 & 459 \\

5 Shor & 130 & 201 & MLSA & 9 & 133 & 284 \\

\hline

1 $\mathrm{HGP}(52,4)$ & 152 & 292 & HC & 18 & 259 & 589 \\

1 $\mathrm{HGP}(52,4)$ & 152 & 292 & HFSA & 18 & 259 & 589 \\

1 $\mathrm{HGP}(52,4)$ & 152 & 292 & PC & 18 & 259 & 589 \\

1 $\mathrm{HGP}(52,4)$ & 152 & 292 & MLSA & 28 & 256 & 525 \\

\hline

1 $\mathrm{BB}[[72,12,6]]$ & 216 & 504 & HC & 67 & 1009 & 1674 \\

1 $\mathrm{BB}[[72,12,6]]$ & 216 & 504 & HFSA & 64 & 986 & 1719 \\

1 $\mathrm{BB}[[72,12,6]]$ & 216 & 504 & PC & 64 & 1112 & 1985 \\

1 $\mathrm{BB}[[72,12,6]]$ & 216 & 504 & MLSA & 72 & 774 & 1288 \\

\bottomrule
\end{tabular}
\end{minipage}
\hfill
\begin{minipage}[t]{0.49\textwidth}
\centering
\begin{tabular}{@{}cccccccc@{}}
\toprule
$l_f$ QECC & $|V|$ & $|E|$ & Alg. & Emit. & CNOTs & Gates \\
\midrule

1 $\mathrm{HGP}(80,16)$ & 224 & 368 & HC & 26 & 240 & 694 \\

1 $\mathrm{HGP}(80,16)$ & 224 & 368 & HFSA & 26 & 240 & 694 \\

1 $\mathrm{HGP}(80,16)$ & 224 & 368 & PC & 43 & 387 & 948 \\

1 $\mathrm{HGP}(80,16)$ & 224 & 368 & MLSA & 31 & 289 & 660 \\

\hline

1 $\mathrm{BB}[[90,8,10]]$ & 270 & 630 & HC & 65 & 1059 & 1754 \\

1 $\mathrm{BB}[[90,8,10]]$ & 270 & 630 & HFSA & 67 & 1012 & 1799 \\

1 $\mathrm{BB}[[90,8,10]]$ & 270 & 630 & PC & 67 & 1012 & 1799 \\

1 $\mathrm{BB}[[90,8,10]]$ & 270 & 630 & MLSA & 90 & 1139 & 1787 \\

\hline

3 $\mathrm{HGP}(34,4)$ & 294 & 602 & HC & 34 & 787 & 1211 \\

3 $\mathrm{HGP}(34,4)$ & 294 & 602 & HFSA & 34 & 780 & 1194 \\

3 $\mathrm{HGP}(34,4)$ & 294 & 602 & PC & 42 & 648 & 1126 \\

3 $\mathrm{HGP}(34,4)$ & 294 & 602 & MLSA & 34 & 673 & 1077 \\

\hline

1 $\mathrm{HGP}(106,16)$ & 302 & 526 & HC & 36 & 455 & 978 \\

1 $\mathrm{HGP}(106,16)$ & 302 & 526 & HFSA & 36 & 449 & 966 \\

1 $\mathrm{HGP}(106,16)$ & 302 & 526 & PC & 44 & 524 & 1050 \\

1 $\mathrm{HGP}(106,16)$ & 302 & 526 & MLSA & 78 & 492 & 1189 \\

\hline

1 $\mathrm{BB}[[108,8,10]]$ & 324 & 756 & HC & 69 & 1572 & 2714 \\

1 $\mathrm{BB}[[108,8,10]]$ & 324 & 756 & HFSA & 72 & 1550 & 2686 \\

1 $\mathrm{BB}[[108,8,10]]$ & 324 & 756 & PC & 72 & 1550 & 2686 \\

1 $\mathrm{BB}[[108,8,10]]$ & 324 & 756 & MLSA & 108 & 1151 & 1875 \\

\hline

1 $\mathrm{HGP}(130,4)$ & 386 & 802 & HC & 32 & 719 & 1389 \\

1 $\mathrm{HGP}(130,4)$ & 386 & 802 & HFSA & 32 & 742 & 1422 \\

1 $\mathrm{HGP}(130,4)$ & 386 & 802 & PC & 32 & 742 & 1422 \\

1 $\mathrm{HGP}(130,4)$ & 386 & 802 & MLSA & 114 & 828 & 1669 \\

\hline

1 $\mathrm{BB}[[144,12,12]]$ & 432 & 1008 & HC & 92 & 1632 & 2827 \\

1 $\mathrm{BB}[[144,12,12]]$ & 432 & 1008 & HFSA & 84 & 1440 & 2544 \\

1 $\mathrm{BB}[[144,12,12]]$ & 432 & 1008 & PC & 84 & 1440 & 2544 \\

1 $\mathrm{BB}[[144,12,12]]$ & 432 & 1008 & MLSA & 144 & 1796 & 2836 \\

\hline

3 $\mathrm{HGP}(52,4)$ & 456 & 980 & HC & 45 & 1090 & 2084 \\

3 $\mathrm{HGP}(52,4)$ & 456 & 980 & HFSA & 46 & 1015 & 2009 \\

3 $\mathrm{HGP}(52,4)$ & 456 & 980 & PC & 50 & 1014 & 2089 \\

3 $\mathrm{HGP}(52,4)$ & 456 & 980 & MLSA & 53 & 1030 & 1649 \\

\hline

5 $\mathrm{HGP}(34,4)$ & 490 & 1026 & HC & 34 & 1311 & 1921 \\

5 $\mathrm{HGP}(34,4)$ & 490 & 1026 & HFSA & 34 & 1219 & 1831 \\

5 $\mathrm{HGP}(34,4)$ & 490 & 1026 & PC & 87 & 1323 & 2178 \\

5 $\mathrm{HGP}(34,4)$ & 490 & 1026 & MLSA & 35 & 1055 & 1678 \\

\bottomrule
\end{tabular}
\end{minipage}
\end{table*}

\end{document}